\documentclass[aps,amsfonts,nofootinbib,onecolumn,floatfix]{revtex4}
\usepackage{hyperref}
\usepackage{bm}
\usepackage{times}
\usepackage{graphicx}
\usepackage{amssymb}
\usepackage{url,hyperref}

\input epsf
\usepackage{graphics}
\usepackage{amsmath}
\usepackage{color}
\usepackage{dcolumn}
\usepackage{graphicx}

\newcommand{\be}{\begin{equation}}
\newcommand{\ee}{\end{equation}}
\newcommand{\bea}{\begin{eqnarray}}
\newcommand{\eea}{\end{eqnarray}}
\newcommand{\fnl}{f_{\mathrm{NL}}}

\newcommand{\calb}{a}
\newcommand{\rot}{\omega}
\newcommand{\bn}{\hat{\bf n}}

\newcommand{\aap}{Astron.\ Astrophys.\ }
\newcommand{\apjl}{Astrophys.\ J.\ Lett.\  }

\newcommand{\mnras}{Mon.\ Not.\ R.\ Astron.\ Soc.\ }

\newcommand{\kv}{{\bf k}}

\newcommand{\skyp}[1]{}

\newcommand{\bi}{B_{\ellt}}
\def\be{\begin{equation}}
\def\ee{\end{equation}}
\def\bea{\begin{eqnarray}}
\def\eea{\end{eqnarray}}

\def\nng{n_{\rm NG}}

\def\fnl{f_{\rm NL}}

\def\ellt{\ell_1\ell_2\ell_3}

\newcommand\eqn[1]{Eq.~(\ref{#1})}
\begin{document}

\title{Primordial Non-Gaussianity in the Cosmic Microwave Background}
\author{Amit P.S. Yadav$^{1}$}\email{ayadav@ias.edu}
\author{Benjamin D. Wandelt$^{2,3}$}
\affiliation{$^1$ Institute for Advanced Study, School of Natural Sciences, Einstein Drive, Princeton, NJ 08540, USA}
\affiliation{$^2$ UPMC Univ Paris 06, Institut d'Astrophysique de Paris, 98 bis, boulevard Arago 75014 Paris, France}
\affiliation{$^3$ Department of Astronomy and Physics, University of Illinois at Urbana-Champaign, Urbana IL 61801, USA}

\begin{abstract}
In the last few decades, advances in observational cosmology have given us a standard model of cosmology. We know the content of the universe to within a few percent. With more ambitious experiments on the way, we hope to move beyond the knowledge of what the universe is made of, to why the universe is the way it is. In this review paper we focus on primordial non-Gaussianity as a probe of the physics of the dynamics of the universe at the very earliest moments. We discuss 1) theoretical predictions from inflationary models and their observational consequences in the cosmic microwave background (CMB) anisotropies; 2) CMB--based estimators for constraining primordial non-Gaussianity with an emphasis on bispectrum templates; 3) current constraints on non-Gaussianity and what we can hope to achieve in the near future; and 4) non-primordial sources of non-Gaussianities in the CMB such as bispectrum due to second order effects, three way cross-correlation between primary-lensing-secondary CMB, and possible instrumental effects.   

\end{abstract}

\maketitle

\section{Motivation} 
\label{introduction}

In the  last few decades  the advances in observational cosmology have led the field to its ``golden age.'' Cosmologists are beginning to
nail down  the basic cosmological parameters. We now know that we live in a Universe which is $13.7\pm 0.1$ Gyr old and is spatially flat to about $1\%$, and is made of $4.6\pm 0.1 \%$ baryons, $22.8\pm 1.3\%$ dark matter, and remaining $72.6 \pm 1.5 \%$ in the form of dark energy. Although we know the constituents to high accuracy, we still do not completely understand the physics of the beginning, the nature of dark energy and dark matter. Many upcoming CMB experiments complimented with observational campaign to map 3D structure of the Universe and new particle physics constraints from the Large Hadron Collider will enable us to move beyond the knowledge of what the universe is made of, to why the universe is the way it is. In this paper we focus on learning about the physics responsible for the initial conditions for the universe.

Inflation~\cite{Guth81, Sato81, Linde81,Albrecht_Steinhardt82} is perhaps one of the most promising paradigms for the early universe, which apart
from solving some of the problems of the Big Bang model like the flatness and  horizon problem, also gives a mechanism
for  producing the  seed  perturbations for  structure formation~\cite{Guth_Pi82,Starobinsky82,
Hawking82,Bardeen_etal83,Mukhanov_et92}, and other  testable  predictions

Most observational probes based on 2-point statistics like CMB power spectrum still allow vast number of inflationary models. Moreover, the alternatives to inflation such as cyclic models are also compatible with the data. Characterizing the non-Gaussianity in the primordial perturbations has emerged as powerful probe of the early universe. The amplitude of non-Gaussianity is described in terms of dimensionless non-linearity parameter
$\fnl$ (defined in Sec.~\ref{model}). Different models  of
inflation predict different amounts  of $\fnl$, starting from $O(1)$
to $\fnl\sim 100$, above which values have been excluded by the WMAP
data already. Non-Gaussianity from the simplest inflation models that are based on a
slowly rolling scalar field   is very
small~\cite{Salopek_Bond90,Salopek_Bond91,Falk_et93,Gangui_etal94,Acquaviva02,Maldacena03};
however, a very large class of more general models with, e.g., multiple
scalar fields, features in inflaton potential, non-adiabatic fluctuations, non-canonical kinetic terms, deviations from Bunch-Davies
vacuum, among others \cite[for a review and references therein]{BKMR_04} 
generates substantially higher
amounts of non-Gaussianity.

The measurement of the bispectrum of the CMB anisotropies is one of the most promising and ``clean'' way of constraining $\fnl$.
Many efficient methods for evaluating bispectrum of CMB temperature anisotropies 
exist ~\cite{KSW05,IntegratedBispectrum,Smith_Zaldarriaga06,YKW07,Yadav_etal08a}. 
So far, the bispectrum tests of non-Gaussianity have
not detected any significant $\fnl$ in temperature fluctuations mapped by
COBE~\cite{nong_bdw} and
WMAP~\cite{nong_wmap,wmap_2nd_spergel,Creminelli_wmap1,creminelli_wmap2,IntegratedBispectrum,szapudi06,YW08}. 
 On the other hand,
some authors have claimed non-Gaussian signatures in the WMAP
temperature
data~\cite{hotcold,larsonwandelt,nong_cmb1,nong_cmb2,nong_cmb3}. These
signatures cannot be characterized by $\fnl$ and are consistent with
non-detection of $\fnl$.

Currently  the constraints on  the $\fnl$  come from  
temperature anisotropy data alone. By  also having  the polarization
information in the  cosmic microwave background, one can improve
sensitivity to  primordial fluctuations \cite{BZ04,YW05}. Although
the  experiments have already  started characterizing  polarization
anisotropies \cite{dasi_pol_02, wmap_1st_pol, wmap_2nd_pol,boom_ee}, the errors  are large
in comparison to temperature anisotropy. The upcoming experiments such
as Planck will characterize polarization anisotropy to high
accuracy. 

The organization of the paper is as following: In Section~\ref{sec:intro} we review the inflationary cosmology focusing on how the microscopic quantum fluctuations during inflation gets converted into macroscopic sees perturbations for structure formation, and as CMB anisotropies. In Section~\ref{model} we discuss theoretical predictions for non-Gaussianity from the inflationary cosmology. In Section~\ref{sec:cmbbispectrum} we show how the primordial non-Gaussianity is connected to the CMB bispectrum, and describe/review CMB bispectrum based estimators to constrain primordial non-Gaussianity ($\fnl$). In Section~\ref{sec:constraints} we discuss the current constraints on $\fnl$ by CMB bispectrum and what we can hope to achieve in near future. We also discuss  non-primordial sources of non-Gaussianity which contaminate primordial bispectrum signal. In section~\ref{sec:otherprobes} we discuss other methods for constraining $\fnl$ besides CMB bispectrum. Finally in Section~\ref{sec:summary} we summarize with concluding remarks. 

\section{Introduction: The Early Universe}
\label{sec:intro}
One of the most promising paradigms of the early universe is inflation~\cite{Guth81, Sato81, Albrecht_Steinhardt82}, which apart from solving the flatness, homogeneity and isotropy problem, also gives a mechanism for producing the seed perturbations for structure formation, and other testable predictions 
\footnote{Although inflation is the most popular theory for the early universe, other mechanisms, for example, ekpyrotic models~\cite{Khoury_et_01} and cyclic models~\cite{Steinhardt_Turok_02a, Steinhardt_Turok_02b} have been proposed for generating nearly scale invariant Gaussian perturbations, while retaining homogeneity and flatness.  In the cyclic universe, there is no beginning of time, and our expansion of the universe is one out of the infinite number of such cycles. Each cycle consists of the following phases: (1) A hot big bang phase, during which a structure formation takes place. (2) An accelerated expansion phase which dilutes the matter and radiation energy density. Since observations suggest that our universe is going through an accelerated expansion phase, in the cyclic model interpretation, we are presently going through this phase. (3) A decelerating phase, which makes the universe flat, and generates nearly Gaussian and scale invariant density perturbations. (4) A big crunch/bang transition phase during which matter and radiation is created. Although the mechanism is different, the outcome of phase (3) of the cyclic model is in some sense analogous to a slow-roll expansion phase of inflation; and phase (4) will correspond with the reheating phase in the inflationary scenario. As we will discuss in the next section these two scenarios can be distinguished by their different predictions about the gravitational waves, and non-Gaussianity. Cyclic models predict negligible contribution of gravitational waves while inflationary models can produce large gravitational wave contribution, which can be detected by next generation experiments. Second, cyclic models produce much larger non-Gaussianity (of local type) in comparison to the standard slow-roll inflationary scenario.} (for a recent review of inflationary cosmology see~\cite{2009arXiv0907.5424B}). During inflation, the universe goes through an exponentially expanding phase. From the Friedman equation, the condition for the accelerated expansion is 
\be
\rho+3p<0. 
\ee
For both matter and radiation this condition is not satisfied. But it turns out that for a scalar field, the above condition can be achieved. 
For a spatially homogeneous scalar field, $\phi$, moving in a potential, $V(\phi)$, the energy density is given by
\be
\rho_\phi= \frac12(d\phi/dt)^2+V(\phi),
\ee 
and the pressure is given by
\be 
p_\phi= \frac12(d\phi/dt)^2-V(\phi).
\ee 
Hence the condition for accelerated expansion of the universe dominated with scalar field $\phi$ is 
\be
\label{eq:sr_condition}
(d\phi/dt)^2 < V(\phi).
\ee 
Physically this condition corresponds to situations where kinetic energy of the field is much smaller than its potential energy. This condition is referred to {\it slowly-rolling} of the scalar field.
During such {\it slow-roll}, the Hubble parameter, $H(t)=d\ln a/dt$, is nearly constant in time, and the
expansion scale factor, $a(t)$ is given by
\begin{equation}
 \label{eq:accel}
  a(t)= a(t_0)\exp\left(\int_{t_0}^t H(t') dt'\right)
  \approx
  a(t_0)\exp\left[H(t)\left(t-t_0\right)\right].
\end{equation}
This exponential expansion drives the observable universe spatially flat, homogeneous and isotropic.

\begin{figure}[t]
\includegraphics[height=8cm, angle=0]{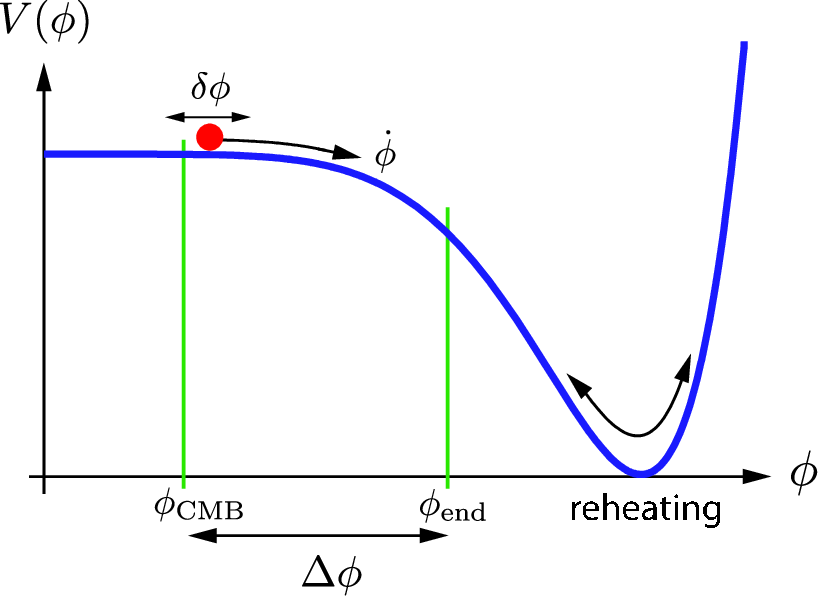}
\caption{A toy scenario for the dynamics of the scalar field during inflation. During the flat part of potential, universe expand exponentially. When field reaches near the minima of the potential, the field oscillates and the radiation is generated.
  \label{fig:potential}}
\end{figure}
A toy model is shown in Fig.~\ref{fig:potential}. 
In the {\it slow-roll} phase, $\phi$ rolls down on $V(\phi)$ slowly, satisfying Eq.~(\ref{eq:sr_condition}) and hence driving the 
universe to expand exponentially.
Near the minima of the potential, $\phi$ oscillates rapidly and inflation ends.
After inflation ends, interactions of $\phi$ with other particles 
lead $\phi$ to decay with a decay rate of $\Gamma_\phi$, producing 
particles and radiation. 
This is called a {\it reheating} phase of the universe,
as $\phi$ converts its energy density into heat by the particle
production.  

\begin{figure}[t]
\includegraphics[height=9cm, angle=0]{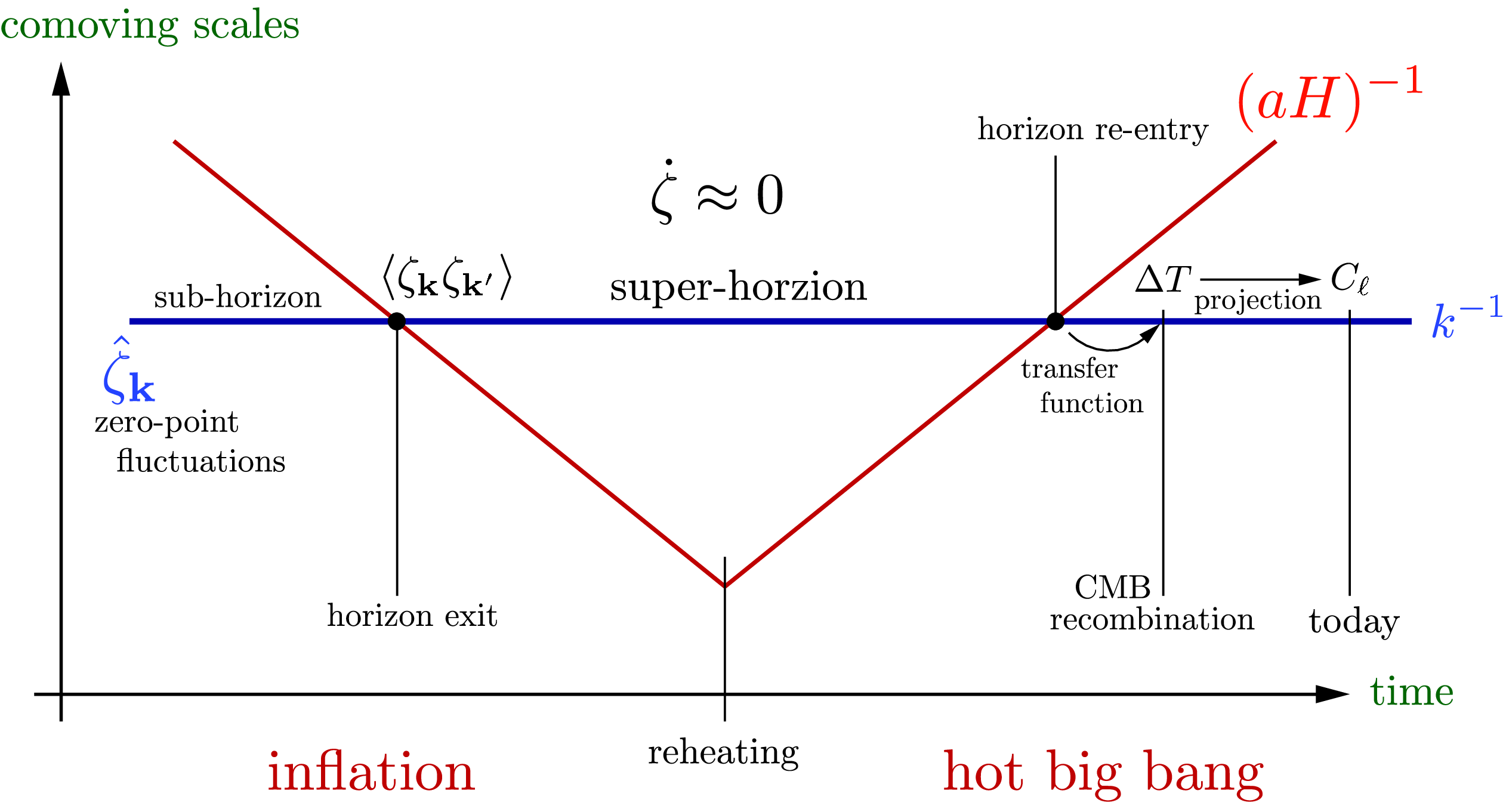}
\caption{Evolution of comoving horizon and generation of perturbations in the inflationary universe. Figure from Ref.~\cite{2009arXiv0907.5424B}.
  \label{fig:scales}}
\end{figure}

Not only inflation solves the flatness, homogeneity and isotropy problem, it also gives a mechanism for generating seed perturbations. During inflation the quantum fluctuation in the field $\phi$ are exponentially stretched due to the rapid expansion phase. The proper wavelength of the fluctuations are stretched out of the 
Hubble-horizon scale to that time, $H^{-1}$. Once outside the horizon, the characteristic {\it r.m.s.} amplitude of these fluctuations is $\left|\phi\right|_{\rm rms}\sim H/(2\pi)$. These fluctuations do not change in time while outside the horizon. After inflation, and reheating, the standard hot-big scenario starts. As the universe decelerates, at some point the fluctuations re-enter 
the Hubble horizon, seeding matter and radiation fluctuations in the 
universe.
Figure~\ref{fig:scales} summarizes the evolution of characteristic
length scales.

\subsubsection*{Primordial Perturbations}
 We use linearly perturbed conformal Friedmann Lemaître Robertson Walker (FLRW) metric of the form,
\begin{equation}
 \label{eq:metric}
  ds^2  = a^2(\tau)\left\{
		    -(1+2AQ) d\tau^2 
		    -2 B Q_i d\tau dx^i 
		    + \left[\left(1+2 H_{\rm L}Q\right)\delta_{ij} 
		       + 2 H_{\rm T}Q_{ij}\right]dx^idx^j
		  \right\}.
\end{equation}
where all the metric perturbations, $A$, $B$, $H_{\rm L}$, and $H_{\rm T}$,
are $\ll 1$, and functions of conformal time $\tau$.
The spatial coordinate dependence of the perturbations is described
by the scalar harmonic eigenfunctions, $Q$, $Q_i$, and $Q_{ij}$, that 
satisfy
$\delta^{ij}Q_{,ij} = -k^2 Q$, $Q_i= -k^{-1}Q_{,i}$, and
$Q_{ij} =  k^{-2}Q_{,ij} + \frac13\delta_{ij}Q$.
Note that $Q_{ij}$ is traceless: $\delta^{ij}Q_{ij}=0$.

Lets consider two new perturbation variables \cite{MFB92, Bardeen_etal83},
\begin{equation}
 \label{eq:u}
 u\equiv \delta\phi 
  - \frac{\dot{\phi}}{aH}{\cal R},
\end{equation}
and 
\begin{equation}
 \zeta\equiv -\frac{aH}{\dot{\phi}}u = {\cal R}-\frac{aH}{\dot{\phi}}
  \delta\phi,
\end{equation} 
which are Gauge invariant.
Here ${\cal R}\equiv H_{\rm L}+\frac13H_{\rm T}$, is perturbations in the intrinsic spatial curvature.
While $u$ reduces to $\delta\phi$ in the spatially flat gauge
(${\cal R}\equiv 0$), or to $-(\dot{\phi}/aH){\cal R}$ in the 
comoving gauge ($\delta\phi\equiv 0$), its value is invariant 
under any gauge transformation.
 Similarly $\zeta$, which reduces to ${\cal R}$ in the comoving gauge, and to 
$-(aH/\dot{\phi})\delta\phi$ in the spatially flat gauge, is also gauge invariant.
The perturbation variable $\zeta$ helps the perturbation analysis not only because of
being gauge invariant, but also because it is {\it conserved} on super-horizon
scales throughout the cosmic evolution.

The quantum fluctuations generate the gauge-invariant perturbation, $u$, that reduces to either $\delta\phi$ or $(\dot{\phi}/aH){\cal R}$ depending on which gauge we use, 
either the spatially flat gauge or the comoving gauge.
Hence, $\delta\phi_{\rm flat}$ and $(\dot{\phi}/aH){\cal R}_{\rm com}$ 
are equivalent to each other at linear order.
The benefit of using $u$ is that it relates these two variables unambiguously,
simplifying the transformation between 
$\delta\phi_{\rm flat}$ and ${\cal R}_{\rm com}$.

%

The solution for $\zeta$ is valid throughout the cosmic
history regardless of whether a scalar field, radiation, or matter
dominates the universe; thus, once created and leaving
the Hubble horizon during inflation, $\zeta$ remains constant in time 
throughout the subsequent cosmic evolution until reentering the horizon.
The amplitude of $\zeta$ is fixed by the 
quantum-fluctuation amplitude in $u$
\begin{equation}
 \Delta^2_{\zeta}(k)= \left(\frac{aH}{\dot{\phi}}\right)^2
  \Delta_\phi^2(k)
  \approx \left(\frac{aH^2}{2\pi\dot{\phi}}\right)^2
  = \left[\frac{H^2}{2\pi (d\phi/dt)}\right]^2.
\end{equation}

This is the spectrum of $\zeta$ on super-horizon scales.

\subsubsection*{From Primordial Perturbations to CMB Anisotropies}

The metric perturbations perturb CMB, producing the CMB anisotropy
on the sky.
Among the metric perturbation variables, the curvature perturbations
play a central role in producing the CMB anisotropy.

As we have shown in the previous subsection, 
the gauge-invariant perturbation, $\zeta$, does not change in time 
on super-horizon scales throughout the cosmic evolution regardless 
of whether a scalar field, radiation, or matter dominates the universe.
The intrinsic spatial curvature perturbation, ${\cal R}$, 
however, does change when equation of state of the universe,
$w\equiv p/\rho$, changes.
Since $\zeta$ remains constant, it is useful to write the evolution
of ${\cal R}$ in terms of $\zeta$ and $w$; however,
${\cal R}$ is {\it not} gauge invariant itself, but $\zeta$ is gauge
invariant, so that the relation between ${\cal R}$ and $\zeta$ may
look misleading. In 1980, Bardeen~\cite{Bardeen80} introduced another gauge-invariant variable,
$\Phi$ (or $\Phi_{\rm H}$ in the original notation),
which reduces to ${\cal R}$ in the zero-shear gauge, or the Newtonian
gauge, in which $B\equiv 0\equiv H_{\rm T}$.
$\Phi$ is given by
\begin{equation}
 \Phi\equiv {\cal R}-\frac{aH}k\left(-B+\frac{\dot{H}_{\rm T}}{k}\right).
\end{equation}
Here, the terms in the parenthesis represent the shear, or the anisotropic
expansion rate, of the $\tau={\rm constant}$ hypersurfaces.
While $\Phi$ represents the curvature perturbations in the zero-shear gauge,
it also represents the shear in the spatially flat gauge in which 
${\cal R}\equiv 0$.
Using $\Phi$, we may write $\zeta$ as
\begin{equation}
 \label{eq:zetaphi}
 \zeta= {\cal R} - \frac{aH}{\dot{\phi}}\delta\phi 
      = \Phi - \frac{aH}k \left(v_\phi-\frac{\dot{H}_{\rm T}}{k}\right),
\end{equation}
where the terms in the parenthesis represent the 
gauge-invariant fluid velocity.

We use $\Phi$ in rest of the paper because it gives the closest analogy to the Newtonian
potential, which we have some intuition of. $\Phi$ reduces to ${\cal R}$ in the zero-shear gauge
(or the Newtonian gauge) in which the metric (Eq.(\ref{eq:metric})) 
becomes just like the Newtonian limit of the general relativity.

The gauge-invariant velocity term, $v-k^{-1}\dot{H}_{\rm T}$,
differentiates $\zeta$ from $\Phi$.
Since this velocity term depends on the equation of state of the universe,
$w=p/\rho$, the velocity and $\Phi$ change as $w$ changes, while 
$\zeta$ is independent of $w$. The evolution of $\Phi$ on super-horizon scales in cosmological linear perturbation theory gives the following~\citep{KS84},
\begin{equation}
 \Phi
  = \frac{3+3w}{5+3w}\zeta,
\label{PhiZeta}
\end{equation}
for adiabatic fluctuations, and hence $\Phi=\frac23\zeta$ in the radiation era
($w=1/3$), and $\Phi=\frac35\zeta$ in the matter era ($w=0$).
$\Phi$ then perturbs CMB through the so-called (static) Sachs--Wolfe 
effect \citep{SW}
\begin{equation}
 \label{eq:sweffect}
 \frac{\Delta T}T = -\frac{1+w}{5+3w}\zeta.
\end{equation}

At the decoupling epoch, the universe has already been in the matter era
in which $w=0$,
so that we observe adiabatic temperature fluctuations of
$\Delta T/T= -\frac13\Phi= -\frac15\zeta$,
and the CMB fluctuation spectrum of the Sachs--Wolfe effect,
$\Delta_{\rm SW}^2(k)$, is
\begin{equation}
 \Delta_{\rm SW}^2(k)
  = \frac19\Delta_{\Phi}^2(k)
  = \frac1{25}\Delta_{\zeta}^2(k)
\end{equation}
By projecting the 3-dimension CMB fluctuation spectrum,
$\Delta_{\rm SW}^2(k)$, on the sky, we obtain
the angular power spectrum\footnote{For the scale invariant ($n=1$) 
case, $C_l^{\rm SW}=\left[l(l+1)\right]^{-1}6C_2^{\rm SW}$.}, $C_l$ \citep{Bond_Efstathiou87},
\begin{equation}
 \label{eq:sw}
 C_l^{\rm SW} = 4\pi \int_0^\infty \frac{dk}k~
  \Delta_{\rm SW}^2(k) 
  j^2_l\left[k(\tau_0-\tau_{\rm dec})\right]
  =
  C^{\rm SW}_2
  \frac{\Gamma\left[(9-n_s)/2\right]\Gamma\left[l+(n_s-1)/2\right]}
       {\Gamma\left[(n_s+3)/2\right]\Gamma\left[l+(5-n_s)/2\right]},
\end{equation}
where $\tau_0$ and $\tau_{\rm dec}$ denote the conformal time at the present epoch and at the 
decoupling epoch, respectively, and 
$n_s\equiv 1+\left[d\ln \Delta^2(k)/d\ln k\right]$ 
is a spectral index which is conventionally used in the literature.

On small angular scales ($\ell > 10$), the Sachs--Wolfe
approximation breaks down, and the acoustic physics in the photon-baryon
fluid system modifies the primordial radiation spectrum
\citep{Peebles_Yu70}.
To calculate the anisotropies at all the scales, one has to solve the Boltzmann
photon transfer equation together with the Einstein equations. These equations can be solved numerically with
the Boltzmann code such as CMBFAST \citep{cmbfast}. The CMB power spectrum then can be written as

\begin{equation}
 C_\ell= 4\pi \int_0^\infty \frac{dk}k~ \Delta_{\Phi}^2(k) g^2_{{\rm T}\ell}(k).
\end{equation}

Here $g_{{\rm T}\ell}(k)$ is called the radiation transfer function, and it contains all the physics which modifies the primordial power spectrum $\Delta_\Phi$ to generate CMB power spectrum $C_\ell$. For the adiabatic initial conditions, in the Sachs--Wolfe limit,
$g_{{\rm T}l}(k)=-\frac13j_l\left[k(\tau_0-\tau_{\rm dec})\right]$.
Often in the literature power spectrum, $P_\Phi(k)$, is used instead of 
$\Delta_{\Phi}^2(k)$. The two are related as 
$\Delta_{\Phi}^2(k)= (2\pi^2)^{-1} k^3P_\Phi(k)$. 
$\Delta_{\Phi}^2(k)$ is called the dimensionless power spectrum.

If $\Phi$ were exactly Gaussian, all the statistical properties of $\Phi$ would be encoded in the {\it two-point function} or in $C_\ell$ in the spherical harmonic space. Since $\Phi$ is directly related to $\zeta $ through Eq.~(\ref{PhiZeta}), all the information of $\zeta$ is also in-coded in $C_\ell$. Although $\zeta$ which is related to a Gaussian variable, $u$, through
$\zeta=-(aH/\dot{\phi})u$, in the linear order $\zeta$ also obeys 
Gaussian statistics; however the {\it non-linear} relation between $\zeta$ and $u$ makes $\zeta$ (and hence $\Phi$ and CMB anisotropies) slightly non-Gaussian. The {\it non-linear} relation between $\zeta$ and $\Phi$ is not the only source of non-Gaussianity in the CMB anisotropies. For example, at the second order, the relationship between $\Phi$ and $\Delta T/T$ is also {\it non-linear}.

\subsubsection*{Probes of the Cosmological Initial Conditions}
 The main predictions of a canonical inflation model are: 
\begin{itemize}
\item{spatial flatness of the observable universe,}
\item{homogeneity and isotropy  on large angular scales of the
observable universe,}
\item{seed scalar and tensor perturbation with primordial density perturbations being\\
\indent \indent(a) nearly scale invariant,\\
 \indent \indent(b) nearly adiabatic, and\\
\indent \indent(c) very close to Gaussian.}
\end{itemize}
At the time of writing, these predictions are consistent with all current observations. This represents a major success for the inflationary paradigm. On the other hand, the inflationary paradigm can be realized by a large {\it `zoo'}~\footnote{Example of some inflationary models are: eternal inflation, hybrid inflation, chaotic, Ghost inflation, Tilted Ghost inflation, DBI inflation, brane inflation, N-flation, bubble inflation, extended inflation, false vacuum inflation, power law inflation, k-inflation, hyperextended inflation, supersymmetric inflation, Quintessential inflation, Natural inflation, Super inflation, Supernatural inflation, D-term inflation, B -inflation, Thermal inflation, discrete inflation, Assisted inflation, Polar cap inflation, Open inflation, Topological inflation, Double inflation, Multiple inflation, Induced-gravity inflation, Warm inflation, stochastic inflation, Generalized assisted inflation, self-sustained inflation, Graduated inflation, Local inflation, Singular inflation, Slinky inflation, locked inflation, Elastic inflation, Mixed inflation, Phantom inflation, Boundary inflation,  Non-commutative inflation, Tachyonic inflation, Tsunami inflation, Lambda inflation, Steep inflation, Oscillating inflation, Mutated Hybrid inflation, intermediate inflation, Inhomogeneous inflation.} of models. In addition, somewhat surprisingly, there exist scenarios where the Universe first contracts and then expands (such as the ekpyrotic/cyclic model), which (up to theoretical uncertainties regarding the precise mechanics of the bounce) also reproduce Universes with the properties described above. What we would like to do is to find observables that allows us to distinguish between members of the inflationary zoo. The exciting fact is that upcoming experiments will have the sensitivity to achieve this goal. 
{\it Tilt and Running:} 
Inflationary models very generically predict a slight deviation from completely flat spectrum. If we write the primordial power spectrum as $\Delta_{\Phi}(k)=A(k_0)\big(\frac{k}{k_0}\big)^{n_s -1}$, then $n_s=1$ correspond to flat spectrum and the quantity $|n_s-1|$ is called a tilt, which characterizes the deviation from scale invariant spectrum. Although the deviations from the scale invariance are predicted to be small,  the exact amount of deviation depends on the details of the inflationary model. For example in most slow roll models $|n-1|$ is of order $1/N_e$, where $N_e \sim 60$ is a number of e-folds to the end of inflation. Ghost inflation, however, predicts negligible tilt. Hence characterizing the tilt of the scalar spectral index is a useful probe of the early universe. Currently the most stringent constraints on tilt come from the WMAP 5-year data, $n_s =0.960^{+0.014}_{-0.013}$~\cite{wmap5_cosmol}, which already disfavors inflationary models with 'blue spectral index' ($n_s>1$). The $1\sigma$ error on $n_s$ will reduce to $\Delta n_s=0.0036$ for upcoming Planck satellite and to $\Delta n_s=0.0016$ for futuristic CMBPol like satellite~\cite{cmbpol_Baumann08}.

Apart from the tilt in the primordial power spectrum, inflationary models also predict $n_s$ to be slightly scale dependent. This scale dependence is referred to as `running' of the spectral index $n_s$, and is defined as $dn_s/d\ln k$. The constraints on the running from the WMAP 5-year data are $-0.090<dn_s/d\ln k <0.0019$~\cite{wmap5_cosmol}. The $1\sigma$ error will reduce to $\Delta (dn_s/d\ln k)=0.0052$ for upcoming Planck satellite and to $\Delta  (dn_s/d\ln k)=0.0036$ for a fourth-generation satellite such as CMBPol~\cite{cmbpol_Baumann08}. 

{\it Primordial Gravitational Waves:}
Inflation also generates tensor perturbations (gravitational waves), which although small compared to scalar component, are still detectable, in principle. So far primordial gravitational waves have not been detected. There are upper limits on their amplitude; see Ref.~\cite{Collaboration:2009ws} for a current observational bounds on the level for primordial gravitational waves. Detection of these tensor perturbations or primordial gravitational waves is considered a `smoking gun' for the inflationary scenario. In contrast to inflation, ekpyrotic (cyclic) models predict an amount of gravitational waves that is much smaller than polarized foreground emission would allow us to see even for an ideal CMB experiment. Primordial scalar perturbations create only E-modes of the CMB\footnote{To first order in perturbations, primordial scalar perturbations do not generate B-modes of CMB. However at second (and higher) order in perturbations, scalar perturbations do produce B-modes~\cite{Bartolo:2006fj,Baumann:2007zm}. The B-modes generated from higher order perturbations are expected to be smaller than the tensor B-mode levels that the upcoming and future experiments (like CMBPol) are sensitive too.}, while primordial tensor perturbations generate both parity even E-modes and parity odd B-modes polarization~\cite{SeljakZaldarriaga97,1997PhRvD..55.7368K,1997PhRvL..78.2058K}. The detection of primordial tensor B-modes in the CMB would confirm the existence of tensor perturbations in the early universe. This primordial {\it B-mode} signal is directly related to the Hubble parameter $H$ during inflation, and thus a detection would establish the energy scale at which inflation happened. Various observational efforts are underway to detect such B-mode signal of the CMB~\cite{cmb_polarization_experiments}. Search for primordial B-modes is challenging. Apart from foreground subtraction challenges, and the challenge of reaching the instrumental sensitivity to detect primordial B-modes, there are several non-primordial sources such as weak lensing of CMB by the large scale structure~\cite{1996ApJ...463....1S,1998PhRvD..58b3003Z}, rotation of the CMB polarization~\cite{LueWangKamionkowski,Kamionkowski_rotation09,Yadav_etal_09_rotation,GluscevicKamionkowskiCooray09}, and instrumental systematics that generate B-modes which contaminate the inflationary signal~\cite{HHZ,Shimon_etal08,YSZ09}. The amplitude of gravitational waves is parametrized as the ratio of the amplitude of tensor and scalar perturbations, $r$. The limit from WMAP 5-year data is $r < 0.22$ ($2 \sigma$)~\cite{wmap5_cosmol}.

{\it Isocurvature Modes:}
Inflationary models with a single scalar field predict primordial perturbations to be adiabatic. Hence detection of isocurvature density perturbations is a "smoking gun" for multi-field models of inflation. A large number of inflationary models with multiple scalar fields predict some amount of isocurvature modes~\cite{Linde1984,Kofman1987,Polarski:1994rz,Gordon:2000hv,Linde1985,Efstathiou_Bond_1986,Peebles1987,Kodama_Sasaki_1986,Weinberg:2004kr,GarciaBellido:1995qq,Sasaki:1995aw,Sasaki:1998ug,Bartolo:2001rt}. For example, curvaton models predict the primordial 
perturbations to be a mixture of adiabatic and isocurvature perturbations. Isocurvature initial conditions specify perturbations in the energy densities of two (or more) species that add up to zero. It does not perturb the spatial curvature of comoving slice (i.e $\cal R $ is zero, hence the name isocurvature). In general, there can be four types of isocurvature modes, namely: 
baryon isocurvature modes, CDM 
isocurvature modes, neutrino density isocurvature modes and, neutrino velocity isocurvature modes. These perturbations imprint distinct signatures in the CMB temperature and E-polarization anisotropies~\cite{Bucher:2000kb}. The contribution of isocurvature modes is model 
dependent, and different models predict different amounts of it. There  exists an upper limit on the allowed isocurvature modes using CMB 
temperature anisotropies~\cite{Bucher_Moodley_Turok_01,Bean_etal_06}, a characterization (or detection of any) of isocurvature modes has a potential of discriminating between early Universe models.

{\it Primordial Non-Gaussianity:}  Canonical inflationary models predict primordials perturbations to be very close to
Gaussian~\cite{Guth_Pi82,Starobinsky82, Hawking82,Bardeen_etal83,Mukhanov_et92}, and any non-Gaussianity predicted by the canonical inflation models 
is very small~\cite{Acquaviva02,Maldacena03}. However
models with non-linearity~\cite{Salopek_Bond90, Gangui_etal94, Gupta_et02}, interacting scalar fields~\cite{Allen_et87, Falk_et93}, 
and deviation from ground state~\cite{Lesgourgues_et97, Martin_et2000} can generate large non-Gaussian perturbations. The amplitude of the non-Gaussian contribution to the perturbation is often referred to as $\fnl$ even if the nature of the non-Gaussianities can be quite different. Different models of inflation predict different amounts of $\fnl$, starting from very close to zero for almost Gaussian perturbations, to $\fnl\approx 100$ for large non-Gaussian perturbations. 
For example, the canonical inflation models with slow roll inflation, where only a couple of derivatives of potential are responsible for inflationary dynamics, predict $\fnl\sim 0.05$~\cite{Maldacena03}. In models where higher order derivatives of the potential are important, the value of $\fnl$ varies from $\fnl\sim 0.1$ where higher order derivatives are suppressed by a low UV cutoff~\cite{Creminelli03} to $\fnl\sim 100$  based on Dirac-Born-Infeld effective action. Ghost inflation, where during inflation, the background has a constant rate of change as opposed to the constant background in conventional inflation, is also capable of giving $\fnl\sim 100$~\cite{Arkani_et_04}. 
The additional field models generating inhomogeneities in non-thermal species~\cite{Dvali_Gruzinov_Zaldarriaga_03} can generate $\fnl\sim 5$~\cite{Zaldarriaga04}; while curvaton models, where isocurvature perturbations in second field during the inflation generate adiabatic perturbations after the inflation, can have $\fnl\sim 10$~\cite{Lyth_etal03}. 

In the following we will see that  non-Gaussianity, far from being merely a test of standard inflation, may reveal detailed information about the state and physics of the very early Universe, if it is present at the level suggested by the theoretical arguments above.

\begin{figure*}[t]
\includegraphics[height=9cm, angle=0]{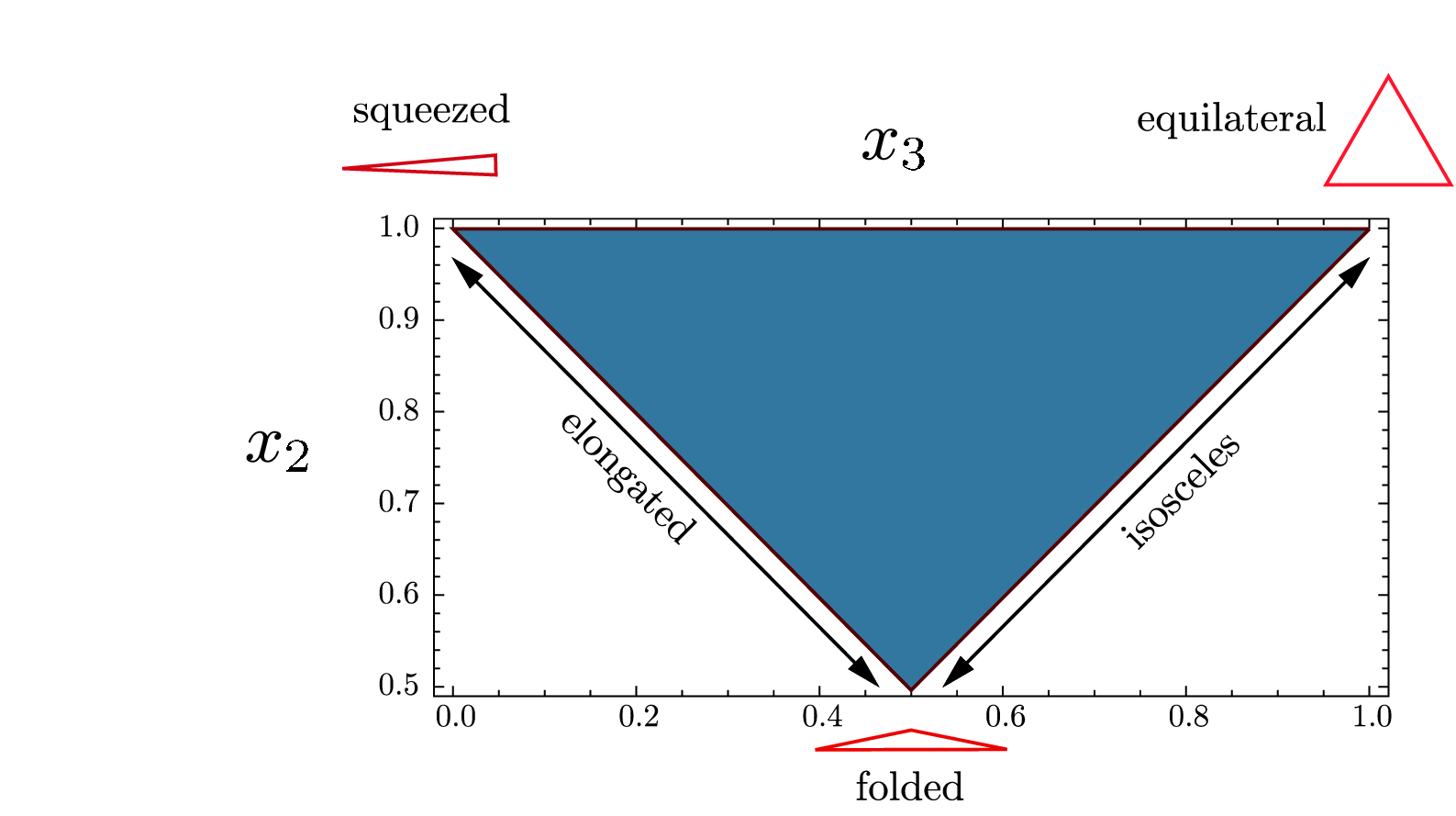} 
\caption{Shapes of Non-Gaussianity. The shape function $F(k_1, k_2, k_3)$ forms a triangle in Fourier space. The triangles are parametrized by rescaled Fourier modes, $x_2=k_2/k_1$ and $x_3=k_3/k_1$. Figure from Ref.~\cite{2009arXiv0907.5424B}
  \label{fig:ng_shapes}}
\end{figure*}

\section{Primordial Non-Gaussianity}
\label{model}
 Large primordial non-Gaussianity can be generated if any of the following condition is violated~\cite{Komatsu:2009kd}
\begin{itemize}
\item{{\it Single Field.} Only one scalar field is responsible for driving the inflation and the quantum fluctuations in the same field is responsible for generating the seed classical perturbations.}
\item{{\it Canonical Kinetic Energy.} The kinetic energy of the field is such that the perturbations travel at the speed of light.}
\item{{\it Slow Roll.} During inflation phase the field evolves much slowly than the Hubble time during inflation.}
\item{{\it Initial Vacuum State.} The quantum field was in the Bunch-Davies vacuum state before the quantum fluctuation were generated.}
\end{itemize}
To characterize the non-Gaussianity one has to consider the higher order moments beyond two-point function, which contains all the information for Gaussian perturbations. The 3-point function which is zero for Gaussian perturbations contains the information about non-Gaussianity.
The 3-point correlation function of Bardeen's curvature perturbations, $\Phi(k)$, can be simplified using the translational symmetry to give
\begin{eqnarray}
\langle \Phi(\mathbf{k_1})\Phi(\mathbf{k_2})\Phi(\mathbf{k_3})\rangle = (2\pi)^3\delta^3(\mathbf{k_1} + \mathbf{k_2} + \mathbf{k_3}) \fnl \cdot F(k_1, k_2, k_3).
\label{eq:3pt}
\end{eqnarray}
where $F(k_1, k_2, k_3)$ tells the shape of the bispectrum in momentum space while the amplitude of non-Gaussianity is captured dimensionless non-linearity parameter
$\fnl$. The shape function $F(k_1,k_2,k_3)$ correlates fluctuations with three wave-vectors and form a triangle in Fourier space. Depending on the physical mechanism responsible for the bispectrum, the shape of the 3-point function, $F(k_1, k_2,
k_3)$ can be broadly classified into three
classes~\citep{Babich_etal_04}.
The local, ``squeezed,'' non-Gaussianity where
$F(k_1, k_2, k_3)$ is large for the configurations in which $k_1 \ll k_2\approx k_3$. Most of the studied inflationary and Ekpyrotic models produce non-Gaussianity of local shape (eg.~\cite{Bartolo:2001cw,Bernardeau:2002jy,Bernardeau:2002jf,Sasaki:2008uc,Naruko:2008sq,Byrnes:2008wi,Byrnes:2006fr,Langlois:2008vk,Valiviita:2008zb,Assadullahi:2007uw,Valiviita:2006mz,Vernizzi:2006ve,Allen:2005ye,Linde_Mukhanov1997,Dvali_Gruzinov_Zaldarriaga_03,Lyth_etal03,Kofman:2003nx,Lehners:2007wc,Lehners:2008my,Koyama:2007ag}). Second, the non-local, ``equilateral,'' non-Gaussianity where $F(k_1, k_2, k_3)$ is
large for the configuration when $k_1 \approx k_2 \approx k_3$. Finally the
folded~\cite{Holman:2007na,Chen_etal07} shape where $F(k_1, k_2, k_3)$ is large for the configurations in which  $k_1\approx 2k_2 \approx 2k_3$. Figure~\ref{fig:ng_shapes} shows these three shapes.

{\it Non-Gaussianity of Local Type:}
The local form of non-Gaussianity may be parametrized in real space
as\footnote{Or equivalently $\label{eqn:phiNG}
\Phi(\mathbf{r}) = \Phi_L(\mathbf{r}) + \fnl \left( \Phi_L^2(\mathbf{r}) - \langle \Phi_L^2(\mathbf{r}) \rangle
\right)$}~\cite{Gangui_etal94,verde00,KS2001}: 
\begin{equation}
\label{phi_bispec}
\zeta(\mathbf{r}) = \zeta_L(\mathbf{r}) + \frac{3}{5}\fnl \left( \zeta_L^2(\mathbf{r}) - \langle \zeta_L^2(\mathbf{r}) \rangle
\right)
\end{equation}
where $\zeta_L(r)$  is the linear  Gaussian part of  the perturbations, and $\fnl$ characterizes the amplitude of primordial
non-Gaussianity. 
Different
inflationary  models predict  different amounts  of  $\fnl$, starting
from $O(1)$ to $\fnl\sim 100$, beyond which values have been excluded
by the Cosmic Microwave Background (CMB) bispectrum of WMAP temperature
data. The bispectrum in this model can be
written as
\be
F_{local}(k_1,k_2,k_3)=2\Delta^2_\Phi \fnl \left[ \frac{1}{k^{3-(n_s-1)}_1 k^{3-(n_s-1)}_2}+ \frac{1}{k^{3-(n_s-1)}_1 k^{3-(n_s-1)}_3}+ \frac{1}{k^{3-(n_s-1)}_2 k^{3-(n_s-1)}_3} \right]
\label{eq:Flocal}
\ee
where $\Delta_{\Phi}$ is the amplitude of the primordial power spectrum. 

The local form arises from a non-linear relation between
inflaton and curvature
perturbations~\cite{Salopek_Bond90,Salopek_Bond91,Gangui_etal94},
curvaton models~\cite{Lyth_etal03}, 
or the New Ekpyrotic models~\cite{Koyama_etal07,Buchbinder_etal07}.            Models with fluctuations in the reheating efficiency [9, 10] and
multi-field inflationary models [17] also generate non-Gaussianity of local type. 

Being local in real space, non-Gaussianity of local type describes correlations among Fourier modes of very different
$k$. In the limit in which one of the modes becomes of very long wavelength 
\cite{Maldacena02}, $k_3 \rightarrow 0$, (i.e. the other 
two $k$'s become equal and opposite), $\zeta_{\vec k_3}$ freezes out much before $k_1$ and $k_2$ and behaves as a background for 
their evolution.  
In this limit $F_{\rm local}$ is proportional to the power spectrum of the short and long wavelength modes
\begin{equation}
\label{eq:local_limit}
F_{\rm local} \propto \frac1{k_3^3} \frac1{k_1^3} \;.
\end{equation}  

\begin{figure*}[]
\begin{center}
\includegraphics[height=60mm]{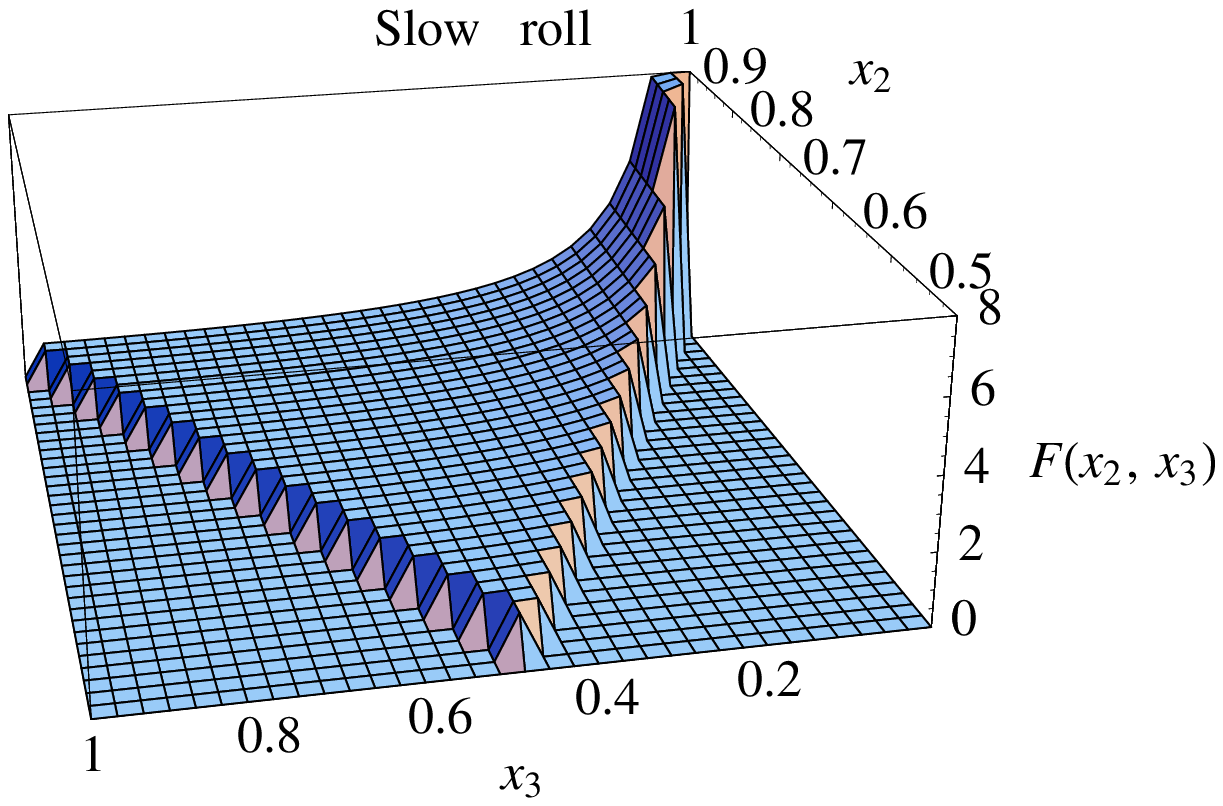} 
\includegraphics[height=60mm]{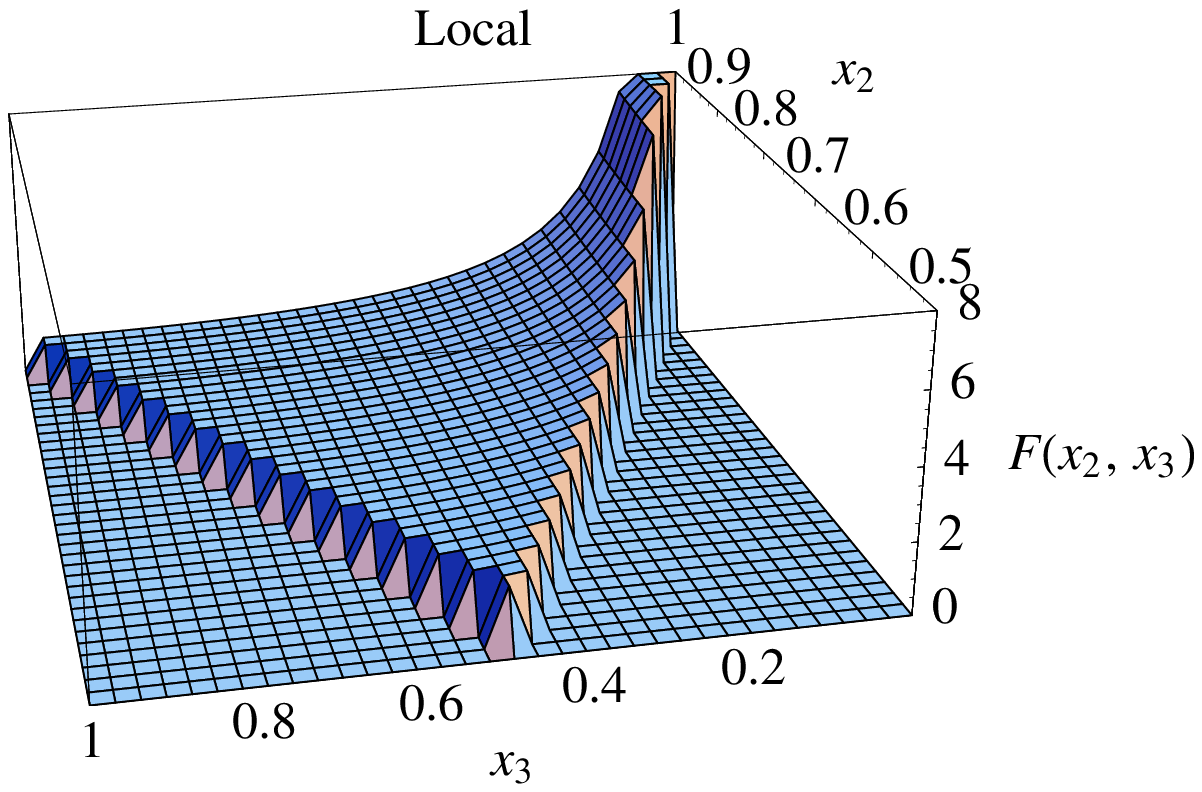}
\caption{Plot of the function $F (1, x_2 , x_3 ) x^2_2 x^2_3$ for the Slow-Roll inflation as given by Eq.~(21) (left panel) and the local distribution as given by Eq.~(19) (right panel). The figures are normalized to have value 1 for equilateral configurations x2 = x3 = 1 and set to zero outside the region $1 - x_2 \leq x_3 \leq x_2$. Here $x3 \equiv k3 /k1$, $x2 \equiv k2 /k1$ and $\epsilon =\eta =1/30$. The figures are taken from Babich et al. 2003~\cite{Babich_etal_04}.}
\label{fig:Flocalslow}
\end{center}
\end{figure*}
\begin{figure*}[]
\begin{center}
\includegraphics[height=60mm]{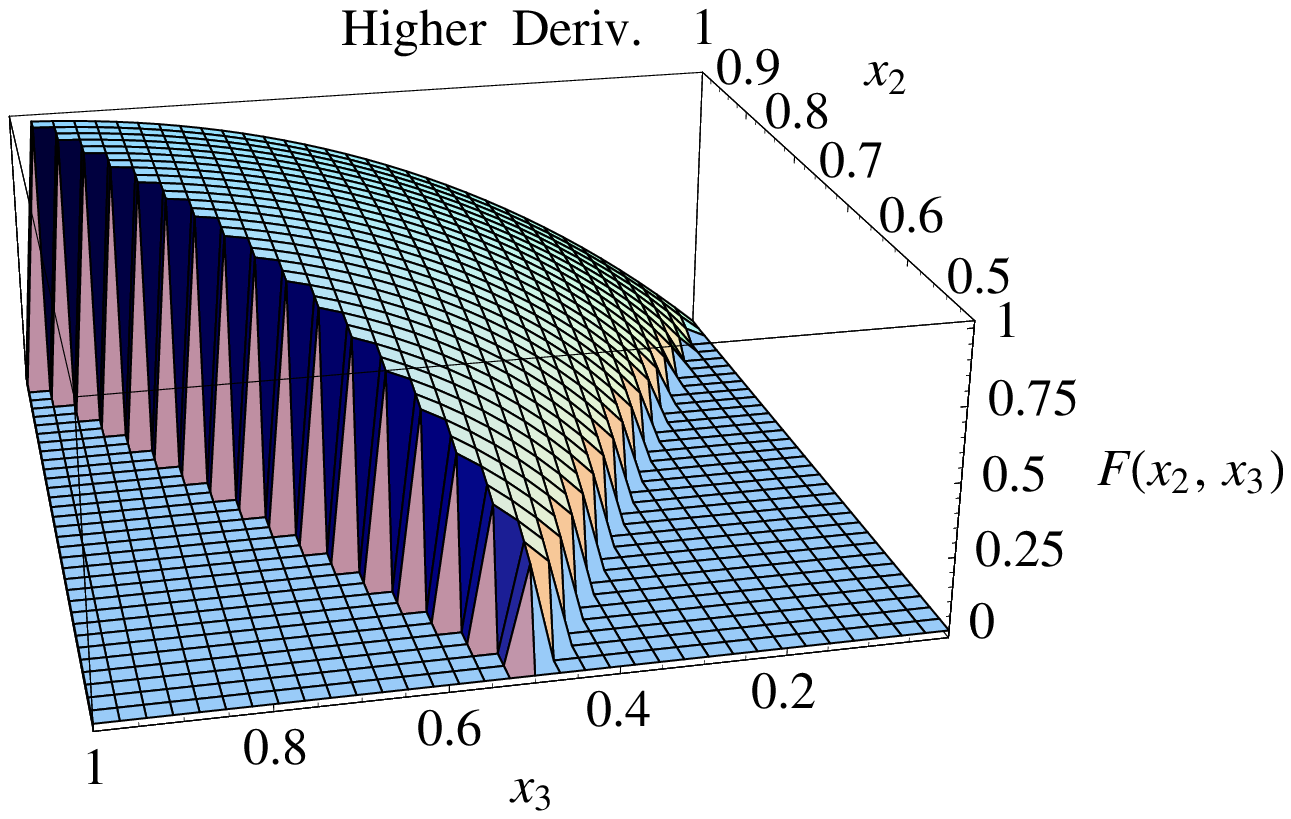}
\includegraphics[height=60mm]{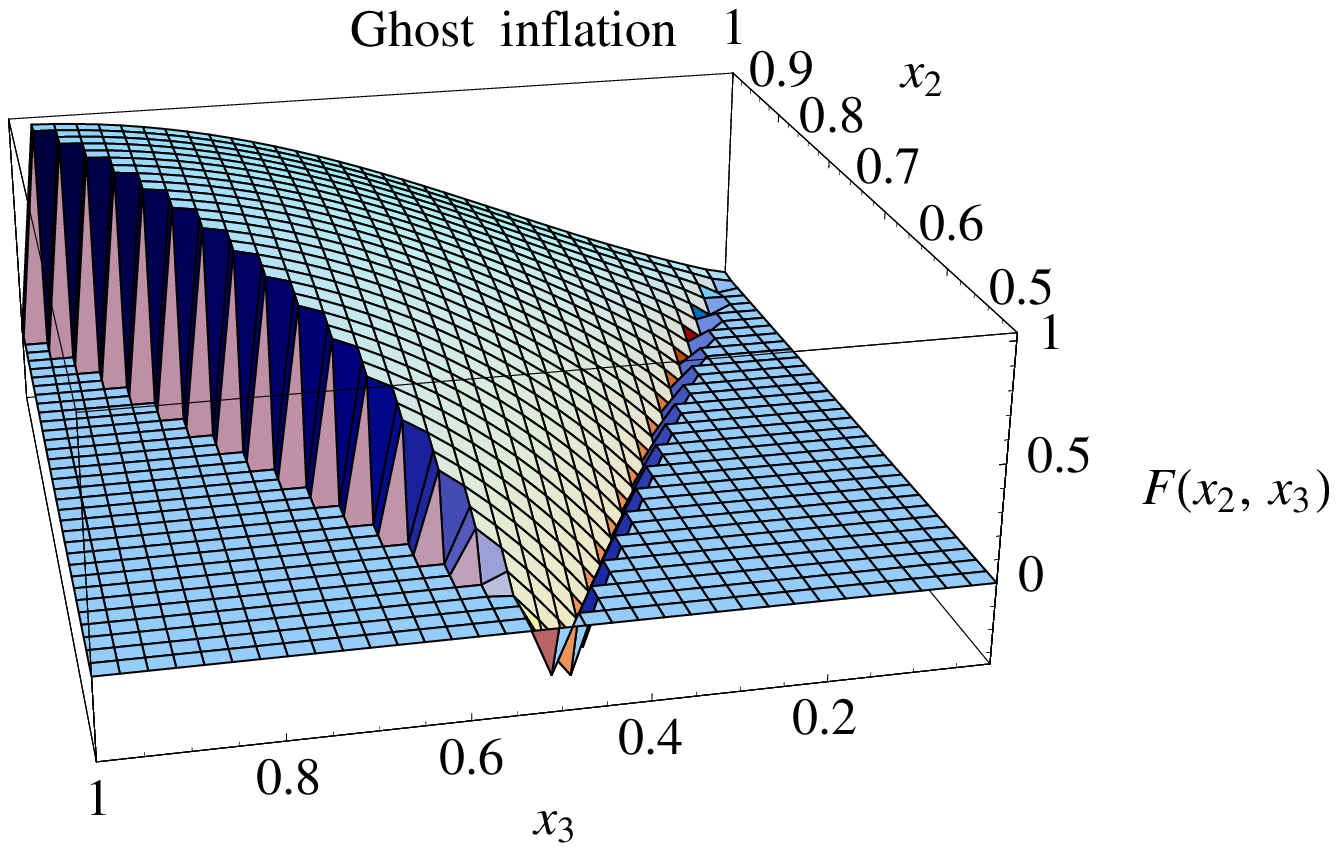}
\caption{Plot of the function $F (1, x_2 , x_3 ) x^2_2 x^2_3$ for the inflation with higher derivatives as given by Eq.~(23) (left panel) and the ghost inflation as given by Eq.~(24) (right panel). The figures are normalized to have value 1 for equilateral configurations x2 = x3 = 1 and set to zero outside the region $1 - x_2 \leq x_3 \leq x_2$. Here $x3 \equiv k3 /k1$ and $x2 \equiv k2 /k1$. The figures are taken from Babich et al. 2003~\cite{Babich_etal_04}}
\end{center}
\label{fig:FhdGhost}
\end{figure*}

As an example, for canonical single field slow-roll inflationary models, the three point function is given by~\cite{Maldacena03}
\begin{equation}
\label{eq:Fslow-roll}
F_\text{slow-roll}(\vec k_1,\vec k_2, \vec k 3)=\frac{1}{8}(3/5) \Delta^2_{\Phi} \frac{1}{\Pi k^3_i}\left [ (3\epsilon -2\eta)\sum_{\vec k_i} k^3_i+ \epsilon \sum_{i\ne j} k_i k^2_j+ 8\epsilon \frac{\sum_{i > j} k^2_i k^2_j}{k_1 + k_2 +k_3}\right ],
\end{equation}
where $\epsilon$ and $\eta$ are the usual slow-roll parameters and are assumed to be much smaller than unity. Taking the limit $k_3 \rightarrow 0$ gives the local form as in Eq.~(\ref{eq:local_limit}). To show this point, Figure.~\ref{fig:Flocalslow} compares the non-Gaussianity shape for local type and for slow-roll model. Although in this limit, slow-roll models do predict no-Gaussianity of local type but as evident from Eq.~(\ref{eq:Fslow-roll}), the bispectrum of {\it inflaton perturbations} yields a non-trivial
scale dependence of $\fnl$ \cite{Falk_et93,Maldacena03}. However in the slow roll limit $\eta, \epsilon <<1$ and hence the amplitude is too small to detect.

{\it Non-Gaussianity of equilateral type:} While vast number of inflationary models predict non-Gaussianity of local type, this model, for instance, fails completely when non-Gaussianity is
localized in a specific range in $k$ space, the case that is predicted
from inflation models with higher derivative kinetic terms~\cite{2003JCAP...10..003C,Alishahiha_etal04,Arkani_et_04,2005JCAP...06..003S,Chen_etal07,Cheung_Creminelli_etal07}. In these models the correlation
is among modes with comparable wavelengths which go out of the horizon
nearly at the same time.  The shape function for the equilateral shape can be written as~\cite{Creminelli_wmap1}
\begin{eqnarray}
F_\text{equil.}(k_1,k_2,k_3) &=& f_{\rm NL}^{\rm equil.} \cdot 6  \Delta_\Phi^2 \cdot \Bigg[-\frac1{k_1^{3-(n_s-1)} k_2^{3-(n_s-1)}}+ (\text{2 perm.}) - \frac2{(k_1 k_2 k_3)^{2-2(n_s-1)/3}} \nonumber \\
&& + \frac1{k^{1-(n_s-1)/3}_1 k^{2-2(n_s-1)/3}_2 k^{3-(n_s-1)}_3}
 + (5 \; \text{perm.}) \Bigg] \;,
\label{eq:Fequil}
\end{eqnarray}

The models of this kind have large $F(k_1,k_2,k_3)$ for the configurations where $k_1\approx k_2 \approx k_3$. The equilateral form arises from non-canonical kinetic terms such 
as the Dirac-Born-Infeld (DBI)
action~\cite{Alishahiha_etal04}, the ghost
condensation~\cite{Arkani_et_04}, or any other single-field models in
which the scalar field acquires a low speed of sound
~\cite{Chen_etal07,Cheung_Creminelli_etal07}. 

As an example, models with higher derivative operators in the usual inflation scenario and a model of inflation based on the Dirac-Born-Infeld (DBI) action produce  a bispectrum of the form
\be
F_{hd}(\vec k_1,\vec k_2, \vec k 3)=\frac{3}{40} \Delta^2_{\Phi} \frac{\dot \phi^2}{\Lambda^4}\frac{1}{\Pi k^3_i}\left [ \frac{1}{k_1+k_2+k_3}\left( \sum_i k^5_i + \sum_{i\ne j} (2k^4_i k_j- 3k^3_i k^2_j)+ \sum_{i\ne j\ne\ell} (k^3_i k_j k_\ell- 4k^2_i k^2_j k_\ell)\right) \right ]
\label{eq:Fhd}
\ee
The above model uses $\frac{1}{8\Lambda^4}(\nabla \phi)^2(\nabla \phi)^2$ as a leading order operator. DBI inflation, which can produce large non-Gaussianity, $\fnl\sim 100$, also has $F(k_1,k_2,k_3)$ of a similar form.  

Ghost inflation, where an inflationary de Sitter phase is obtained with a ghost condensate, produces a bispectrum of the following form~\cite{Arkani_et_04}
\begin{align}
\label{eq:Fghost}
F_{ghost}(k_1,k_2, k_3)  & = 
-\big(3/5\big)^3 \cdot \frac{2 \sqrt{2} \pi^{3/2}}{\Gamma(1/4)^3} \frac{H^5 \beta}{M^2}
\left(\frac{M}{\alpha H}\right)^4 \\ \nonumber & \frac1{\prod_i k_i^3}
\int_{-\infty}^0 d\eta \;\eta^{-1} F^*(\eta)
F^*\left(\frac{k_2}{k_1} \eta\right) F^{\prime
*}\left(\frac{k_3}{k_1}\eta\right) k_3 (\vec k_1 \cdot \vec k_2) +
{\rm symm.},
\end{align}
where $\alpha$ and $\beta$ are free parameters of order unity, and 
\begin{equation} \label{eq:F} F(x) = \sqrt{\frac{\pi}{8}} (-x)^{3/2}
H^{(1)}_{3/4}(x^2/2).
\end{equation} 
Ghost inflation also produces large non-Gaussianity, $\fnl \sim 100$. Figure~\ref{fig:FhdGhost} shows the shape of non-Gaussianity of equilateral type by showing $F(k_1,k_2,k_3)$ for ghost inflation and for a model with  a higher derivative term.

{\it Folded Shape:} So far the 3-point functions were calculated assuming the regular Bunch-Davis vacuum state, giving rise to either local or equilateral type non-Gaussianity. However if the bispectrum is calculated by dropping the assumption of Bunch-Davis initial state give rise to bispectrum shape which peaks for the folded shape, $k_1\approx 2k_2 \approx 2k_3$, with shape function given as~\cite{Holman:2007na,2009JCAP...05..018M,Chen_etal07} 
\begin{equation}
F_\text{non-BD}(k_1,k_2,k_3)=\big(3/5\big)^3 \cdot \frac{1}{M_{p}^{2}}\frac{4}{\prod(2k_{i}^{3})}\frac{H^{6}}{\dot{\phi}^{2}}\sum_{j}\frac{3k_{1}^{2}k_{2}^{2}k_{3}^{2}}{k_{j}^{2}\tilde{k}_{j}}\mathcal{R}e(\beta_{k_{j}})\left(\mathrm{cos}(\tilde{k}_{j}\eta_{0})-1\right)\label{modin2}
\end{equation}
where $\beta_{k_j}$ are the Bogoliubov coefficients which encode information about the initial conditions, $\eta_0$ is the initial conformal time and $\tilde k_j=\sum_i k_i-2k_j$.

\section {The Cosmic Microwave Background Bispectrum}
\label{sec:cmbbispectrum}
Since the discovery of CMB by Penzias and Wilson in 1965~\cite{PW69} and the first detection of CMB temperature anisotropies on large scales by the COBE DMR~\cite{COBE_first_det_1992}, the space satellite WMAP and over a dozens of balloon and ground based experiments have characterized the CMB temperature anisotropies to a high accuracy and over a wide range of angular scales. The space satellite Planck which launched in 2009 will soon characterize the temperature
anisotropies to even higher accuracy up to angular scales of $\ell_{max} \approx 2500$. The CMB power spectrum is obtained by reducing all the information of $N_{pix}$ ($\sim 10^{6}$ for WMAP and $\sim 10^7$ for Planck). Such reduction is justified to obtain a fiducial model, given the non-Gaussianities are expected to be small. With high quality data on the way, the field of non-Gaussianity is taking off. CMB bispectrum contains information which is not present in the power-spectrum and as we say in the previous section, is a unique probe of the early universe.

The harmonic coefficients of the CMB anisotropy $a_{lm}=T^{-1}\int d^2{\mathbf {\hat n}}\Delta T({\mathbf {\hat {\bf n}}}) Y^{*}_{\ell m}$ can
be related to the primordial fluctuation $\Phi$ as
\begin{eqnarray} 
\label{phi_alm}
a_{\ell    m}^p=b_{\ell}\,4\pi (-i)^\ell\int  \frac{d^3k}{(2 \pi)^3} \Phi(\mathbf{k})   \,    g^p_{\ell}(k)   Y^*_{\ell m}(\hat \kv)+ n_{\ell m},
\end{eqnarray} 
where  $\Phi(\kv)$  is the primordial curvature perturbations, for a
comoving wavevector $\mathbf  {k}$, $g^p_{\ell}(r)$ is the radiation
transfer  function where the index $p$ refers to either temperature
($T$) or {\it E}-polarization ($E$) of the CMB. A beam function
$b_{\ell}$ and the harmonic coefficient of noise $n_{\ell  m}$ are
instrumental effects. Eq.~(\ref{phi_alm}) is written for a flat
background, but can easily be generalized. 

Any non-Gaussianity present in the primordial perturbations $\Phi(\mathbf{k})$ gets transferred to the observed CMB via Eq.~(\ref{phi_alm}). The most common way to look for non-Gaussianity in the CMB is to study the {\it bispectrum}, the  three-point function of temperature and polarization anisotropies in harmonic space. The CMB angular bispectrum is defined as 
\be
B^{pqr}_{\ell_1 \ell_2 \ell_3, m_1m_2m_3}\equiv \langle a^p_{\ell_1 m_1}a^q_{\ell_2 m_2}a^r_{\ell_3 m_3} \rangle\, ,
\ee   
and the angular-averaged bispectrum is 
\be
\bi^{pqr}=\sum_{m_1m_2m_3}  \left(\begin{array}{ccc} \ell_1 & \ell_2 & \ell_3 
\\ m_1 & m_2 & m_3 \end{array}\right)   B^{pqr}_{\ell_1 \ell_2 \ell_3, m_1m_2m_3}\, ,
\ee
where the matrix is the Wigner 3J symbol imposing selection rules which makes bispectrum zero unless \\
 (i) $\ell_1 +\ell_2 + \ell_3 =$integer \\
(ii) $m_1 + m_2 + m_3=0$ \\
(iii) $\vert \ell_i -\ell_j \vert \le \ell_k \le \ell_i + \ell_j$ for $i,j,k=1,2,3$.\\
Using Eq.(\ref{phi_alm}) the bispectrum can be written as
 \begin{eqnarray}
 \label{eq:cmb_bi}
 \bi^{pqr} & = & (4\pi)^3 (-i)^{\ell_1+\ell_2+\ell_3} \sum_{m_1m_2m_3} \left(\begin{array}{ccc} \ell_1 & \ell_2 & \ell_3 \\ m_1 & m_2 & m_3 \end{array} \right)  \int \frac{d^3k_1}{(2 \pi)^3}\frac{d^3k_2}{(2 \pi)^3}\frac{d^3k_3}{(2 \pi)^3} \; Y^*_{\ell_1 m_1}(\hat{\kv}_1) Y^*_{\ell_2 m_2}(\hat{\kv}_2)Y^*_{\ell_3 m_3}(\hat{\kv}_3) 
 \nonumber \\ & & 
 \times g^p_{\ell_1}(k_1)
 g^q_{\ell_2}(k_2) g^r_{\ell_3}(k_3) \; \langle\Phi(\mathbf{k}_1)\Phi(\mathbf{k}_2)\Phi(\mathbf{k}_3) \rangle \;,
 \end{eqnarray}
where $\langle\Phi(\mathbf{k}_1)\Phi(\mathbf{k}_2)\Phi(\mathbf{k}_3) \rangle$ is the primordial curvature three-point function as defined in Eq. (\ref{eq:3pt}). 

To forecast constraints on non-Gaussianity using CMB data, we will perform a Fisher matrix analysis. The Fisher matrix for the parameters $p_a$ can be written as ~\cite{KS2001,BZ04,Yadav_etal08a} 
\be
{\mathcal F}_{ab} = 
\sum_{\left\{i\!jk,~pqr\right\}}\sum_{\ell_1\le\ell_2\le\ell_3}
 \frac{1}{\Delta_{\ellt}} 
 \frac{\partial B^{pqr}_{\ellt}}{\partial p_a}
 \left({\bf Cov}^{-1}\right)_{i\!jk,~pqr}
 \frac{\partial B^{\,i\!jk}_{\ellt}}{\partial p_b}\,.
\label{eq:fisher}
\ee
The indices $a$ and $b$ run over all the parameters bispectrum depends on, we will assume all the cosmological parameters except $\fnl$ to be known.  Indices $ijk$  and $pqr$ run over all the eight possible ordered combinations of temperature and polarization given by $TTT$, $TTE$, $TET$, $ETT$, $TEE$, $ETE$, $EET$ and $EEE$; the combinatorial factor $\Delta_{\ell_1  \ell_2 \ell_3}$ equals $1$ when all $\ell$'s are different, $6$ when $\ell_1 =  \ell_2 = \ell_3$, and $2$ otherwise. The covariance matrix {\bf Cov} is obtained in terms of $C_\ell^{TT}$, $C_\ell^{EE}$,  and $C_\ell^{TE}$  (see~\cite{BZ04,YKW07}) by applying Wick's theorem.

For non-Gaussianity of the local type, for which the functional form $F(k_1,k_2,k_3)$ is given by Eq.~(\ref{eq:Flocal}), we have
\be
\frac{\partial B^{~ijk}_{\ellt}}{\partial\fnl}  =  \sqrt{\frac{(2 \ell_1+1) (2 \ell_2+1)(2 \ell_3+1)}{4 \pi}} \left(\begin{array}{ccc} \ell_1 & \ell_2 & \ell_3 
\\ 0 & 0 & 0 \end{array}\right)~ 2 \int_0^\infty \!\!r^2 dr \left[ -\alpha^i_{\ell_1} \beta^j_{\ell_2} \beta^k_{\ell_3} 
+ 2 \;{\rm perm.}\right], 
\ee
where the functions $\alpha$ and $\beta$ are given by

\begin{eqnarray}
\alpha^i_\ell(r) & \equiv & \frac2\pi \int \!\!\! dk \; k^2 \, g^i_\ell(k) ~j_\ell(k r)\,, 
\\
\beta^i_\ell(r) & \equiv & \frac2\pi \int \!\!\! dk \; k^{-1} \, g^i_\ell(k) ~j_\ell(k r)\Delta_\Phi\, k^{n_s-1}\,, 
\end{eqnarray}
In the expression above we use the dimensionless power spectrum amplitude $\Delta_\Phi$, which is defined by $P_\Phi(k)=\Delta_\Phi k^{-3+(n_s-1)}$, 
where $n_s$ is the tilt of the primordial power spectrum. One can compute the transfer functions $g^T_{\ell}(k)$ and $g^E_{\ell}(k)$ using publicly available codes such as CMBfast~\cite{cmbfast} and CAMB~\cite{camb}

In a similar way, from Eq.~(\ref{eq:Fequil}), one can derive the following expressions for the bispectrum derivatives in the equilateral case,
\bea
\frac{\partial B^{~ijk}_{\ellt}}{\partial\fnl} & = & \sqrt{\frac{(2 \ell_1+1) (2 \ell_2+1)(2 \ell_3+1)}{4 \pi}} \left(\begin{array}{ccc} \ell_1 & \ell_2 & \ell_3 
\\ 0 & 0 & 0 \end{array}\right)  
\nonumber\\ 
& & \times ~6 \int\!\!r^2 dr \left[ -\alpha^i_{\ell_1} \beta^j_{\ell_2} \beta^k_{\ell_3} 
+ 2 \;{\rm perm.}
+ \beta^i_{\ell_1} \gamma^j_{\ell_2} \delta^k_{\ell_3} + 5 \;{\rm perm.}
- 2 \delta^i_{\ell_1} \delta^j_{\ell_2} \delta^k_{\ell_3} 
\right],  
\eea
where the functions $\delta,$ and $\gamma$ are given by

\begin{eqnarray}
\gamma^i_\ell(r) & \equiv & \frac2\pi \int \!\!\! dk \; k \, g^i_\ell(k) ~j_\ell(k r) \Delta_\Phi^{1/3}\,k^{(n_s-1)/3}\,,
\\
\delta^i_\ell(r) & \equiv & \frac2\pi \int \!\!\! dk  \, g^T_\ell(k) ~j_\ell(k r) \Delta_\Phi^{2/3}\,k^{2(n_s-1)/3}.
\end{eqnarray}
 Recently a new bispectrum template shape, an orthogonal shape has been introduced~\cite{Senatore:2009gt} which characterizes the size of the signal ($f^{ortho}_{NL}$) which peaks both for equilateral and flat-triangle configurations. The shape of non-Gaussianities associated with $f^{ortho}_{\rm NL}$ is orthogonal to the one associated to $f^{equil}_{NL}$. The bispectrum for orthogonal shape can be written as~\cite{Senatore:2009gt}
\bea
\frac{\partial B^{~ijk}_{\ellt}}{\partial\fnl^{{\text ortho}}} & = & \sqrt{\frac{(2 \ell_1+1) (2 \ell_2+1)(2 \ell_3+1)}{4 \pi}} \left(\begin{array}{ccc} \ell_1 & \ell_2 & \ell_3 
\\ 0 & 0 & 0 \end{array}\right)  
\nonumber\\ 
& & \times ~18 \int\!\!r^2 dr \left[ -\alpha^i_{\ell_1} \beta^j_{\ell_2} \beta^k_{\ell_3} 
+ 2 \;{\rm perm.}
+ \beta^i_{\ell_1} \gamma^j_{\ell_2} \delta^k_{\ell_3} + 5 \;{\rm perm.}
- \frac{2}{3} \delta^i_{\ell_1} \delta^j_{\ell_2} \delta^k_{\ell_3} 
\right] \,. 
\eea

\subsection{Estimator}
\label{sec:estimator}
An unbiased bispectrum-based minimum variance estimator for the nonlinearity parameter in the limit of full sky and homogeneous noise  can be written as~\cite{KSW05,Creminelli_wmap1,Yadav_etal08a}
\begin{eqnarray}
\hat f_{\rm NL} =\frac{1}{N}\sum_{\ell_i m_i} \left(\begin{array}{ccc} \ell_1 & \ell_2 & \ell_3 \\ m_1 & m_2 & m_3 \end{array} \right)  \frac{B_{\ell_1\ell_2\ell_3}}{C_{\ell_1}C_{\ell_2}C_{\ell_3}} \,\, a_{\ell_1m_1}a_{\ell_2m_2}a_{\ell_3m_3} 
\label{eq:estimator}
\end{eqnarray}
where $B_{\ell_1\ell_2\ell_3}$ is angle averaged theoretical CMB bispectrum for the model in consideration. The normalization $N$ can be calculated to require the estimator to be unbiased, $\langle \hat \fnl\rangle=\fnl$. If the bispectrum $B_{\ell_1\ell_2\ell_3}$ is calculated for $\fnl=1$ then the normalization takes the following form
\begin{eqnarray}
N=\sum_{\ell_i}\frac{(B_{\ell_1\ell_2\ell_3})^2}{C_{\ell_1}C_{\ell_2}C_{\ell_3}}
\end{eqnarray} 
The estimator for non-Gaussianity, Eq.~(\ref{eq:estimator}), can be simplified using Eq.~(\ref{phi_alm}) to yield
\begin{eqnarray}
{\hat f}_{\rm NL} & = & \frac1N \cdot \sum_{l_i m_i} \int 
d^2\hat{n} \; Y_{l_1 m_1}(\hat{n}) Y_{l_2 m_2}(\hat{n})Y_{l_3 m_3}(\hat{n}) 
\int \limits^{\infty}_0 r^2 dr \; j_{l_1}(k_1r) j_{l_2}(k_2r) j_{l_3}(k_3r) \; C_{l_1}^{-1} C_{l_2}^{-1} C_{l_3}^{-1} 
\nonumber \\ & & \int \frac{2 k^2_1 dk_1}{\pi} \frac{2 k^2_2 dk_2}{\pi} \frac{2 k^2_3 dk_3}{\pi} 
F(k_1,k_2,k_3) \Delta^T_{l_1}(k_1)
\Delta^T_{l_2}(k_2) \Delta^T_{l_3}(k_3) \; a_{l_1 m_1}a_{l_2 m_2}a_{l_3 m_3} \;,
\end{eqnarray}
where $F(k_1,k_2,k_3)$ is a shape of 3-point function as defined in Eq.~(\ref{eq:3pt}). Given the shape $F(k_1,k_2,k_3)$, one is interested in, it is conceptually straightforward to constrain the non-linearity parameter from the CMB data. Unfortunately the computation time for the estimate scales as $N_{pix}^{5/2}$, which is computationally challenging as even for the WMAP data the number of pixels is of order $N_{pix}\sim 10^6$. The scaling can be understood by noting that the each spherical harmonic transform scales as $N^{3/2}_{pix}$ and the estimator requires $\ell^2 (\propto N_{pix})$ number of spherical harmonic transforms. 

The computational cost decreases if the shape can be factorized as
\begin{eqnarray}
F(k_1,k_2,k_3)=f_1(k_1)f_2(k_2)f_3(k_3)\,,
\end{eqnarray}
with which the estimator simplifies to
\begin{eqnarray}
{\hat f}_{{\rm NL}} = \frac1N \cdot \int 
d^2\hat{n} \;
\int \limits^{\infty}_0 r^2 dr \; \prod^3_{i=1}\sum_{l_i m_i}\int \frac{2 k^2 dk}{\pi} j_{l_i}(k r)
f_i(k) \Delta^T_{l_i}(k) C_{l_i}^{-1} a_{l_i m_i} Y_{l_i m_i}(\hat{n})\
\end{eqnarray}
and computational cost now scales as $N^{3/2}_{pix}$. For Planck ($N_{pix} \sim  5 \times 10^7$) this translates into a  speed-up by factors of
millions, reducing the required computing time from thousands of years to
just hours and thus making $\fnl$ estimation feasible for future
surveys. The speed of the
estimator now allows sufficient number of Monte Carlo simulations
 to characterize its statistical properties in the presence of real
 world issues such as
instrumental effects, partial sky coverage, and foreground
contamination. Using the Monte Carlo simulations it has been shown that estimator is indeed optimal, where
optimality is defined by  saturation of the Cramer Rao bound, if noise
 is homogeneous. Note that even for the non-factorizable shapes, by using the flat sky approximation and interpolating between the modes, one can estimating $\fnl$ in a computationally efficient way~\cite{2009PhRvD..80d3510F}. 

The extension of the estimator of $\fnl$ from the
temperature data \cite{KSW05} to include both the temperature and polarization data of the CMB is discussed in Yadav et al.~\cite{BZ04, YW05,YKW07,Yadav_etal08a}. Summarizing briefly, we
construct a  cubic statistic as a combination of (appropriately
filtered) temperature and polarization maps which is specifically  sensitive to the
primordial  perturbations. This  is done  by reconstructing  a  map of
primordial perturbations,  and using that to define a fast estimator. We
also show  that this fast estimator is equivalent to the optimal estimator by demonstrating that the inverse of  the covariance matrix  for the optimal
estimator~\cite{BZ04} is the  same as the product of  inverses we get
in the  fast estimator. The estimator still takes  only $N^{3/2}_{pix}$ operations
in comparison to the full bispectrum calculation which takes $N^{5/2}_{pix}$
operations.

For a given shape the estimator for non-linearity parameter can be written as $\hat{f}_{\mathrm{NL}} = \frac{\hat{S}_{shape}}{N_{shape}}$, where for the equilateral, local and orthogonal shapes, the $S_{shape}$ can be written as
\begin{eqnarray}
\hat{S}_{equilateral}&=&\frac{3}{f_{sky}}\int     r^2dr     \int     d^2\hat{n}
[ B(\hat{n},r) B(\hat{n},r)A (\hat{n},r) +\frac{2}{3}D(\hat{n},r)^3-2 B (\hat{n},r)C (\hat{n},r)D (\hat{n},r) ]\\
\hat{S}_{local} &=&\frac{1}{f_{sky}}\int     r^2dr     \int     d^2\hat{n}
B(\hat{n},r) B(\hat{n},r)A (\hat{n},r) \\
\hat{S}_{orthogonal}&=&\frac{9}{f_{sky}}\int     r^2dr     \int     d^2\hat{n}
[ B(\hat{n},r) B(\hat{n},r)A (\hat{n},r) +\frac{8}{9}D(\hat{n},r)^3-2 B (\hat{n},r)C (\hat{n},r)D (\hat{n},r) ]
\label{s_prim}
\end{eqnarray} 
with \begin{eqnarray}
\label{B}
B(\hat{n},r)\equiv           \sum_{ip}\sum_{lm}(C^{-1})^{ip}a^{i}_{\ell
m}\beta^p_{\ell}(r)Y_{\ell m}(\hat{n}), \hspace{1cm} C(\hat{n},r)\equiv           \sum_{ip}\sum_{lm}(C^{-1})^{ip}a^{i}_{\ell
m}\beta^p_{\ell}(r)Y_{\ell m}(\hat{n}), 
\end{eqnarray}
\begin{eqnarray}
\label{A}
A(\hat{n},r)\equiv           \sum_{ip}\sum_{lm}(C^{-1})^{ip}a^{i}_{\ell
m}\alpha^p_{\ell}(r)Y_{\ell m}(\hat{n}), \hspace{1cm} D(\hat{n},r)\equiv           \sum_{ip}\sum_{lm}(C^{-1})^{ip}a^{i}_{\ell
m}\beta^p_{\ell}(r)Y_{\ell m}(\hat{n}), 
\end{eqnarray}
and $f_{sky}$ is a  fraction of sky. Index $i$ and $p$
can  either be  $T$ or  $E$. 
\begin{eqnarray}
N=  \sum_{i j  k  p q
r}\sum_{2 \le \ell_1\le  \ell_2\le  \ell_3} \frac{1}{\Delta_{\ell_1 \ell_2 \ell_3}} B^{p q  r,prim}_{\ell_1  \ell_2
\ell_3}(C^{-1})^{ip}_{\ell_1}(C^{-1})^{j         q}_{\ell_2}(C^{-1})^{k
r}_{\ell_3} B^{i j k,prim}_{\ell_1 \ell_2 \ell_3 },
\end{eqnarray}

Indices
$i, j, k, p, q$ and $r$ 
can  either be  $T$ or  $E$. Here, $\Delta_{\ell_1  \ell_2 \ell_3}$ is 1
when  $\ell_1 \neq \ell_2 
\neq \ell_3$, 6 when $\ell_1 =  \ell_2 = \ell_3$, and 2 otherwise, $B^{p q r,prim}_{\ell_1 \ell_2  \ell_3}$ is the
 theoretical bispectrum for $\fnl=1$~\cite{YKW07}.

It has been shown that the above estimators defined in Eq.~(\ref{s_prim}) are minimum variance amongst bispectrum-based estimators
for full sky coverage and homogeneous noise \cite{YKW07}. To be able to
deal with the realistic data, the estimator has to be able to deal with
the inhomogeneous noise and foreground masks. The estimator can be generalized to deal with partial sky
 coverage as well as inhomogeneous noise by
adding a linear term to $\hat{S}_{prim}$: $\hat{S}_{prim}\rightarrow 
\hat{S}_{prim}+\hat{S}_{prim}^{linear}$. 
For the temperature only case, this
has been done in~\cite{Creminelli_wmap1}. Following the same argument,
we find that 
the linear term for the combined analysis of CMB temperature and
polarization data is given by 
\begin{eqnarray}
\label{slin}
\hat{S}^{linear}_{prim}=\frac{-1}{f_{sky}}\int  r^2dr  \int d^2\hat{n}
\left   \{  2 B(\hat{n},r)\, \langle A_{sim}(\hat{n},r)B_{sim}(\hat{n},r)\rangle_{MC}  +     A(\hat{n},r)\, \langle B^2_{sim}(\hat{n},r) \rangle_{MC}
 \right \},
\end{eqnarray} 
where  $A_{sim}(\hat{n},r)$ and $B_{sim}(\hat{n},r)$ are the $A$ and $B$
maps generated from Monte Carlo simulations that contain signal and
noise,  and $\langle ..\rangle$ denotes the average over the Monte Carlo
simulations.

The generalized estimator is given by 
\begin{equation}
 \hat{f}_{\mathrm{NL}} =
\frac{\hat{S}_{prim}+\hat{S}^{linear}_{prim}}{N}\,.
\end{equation}
Note that
$\langle\hat{S}^{linear}_{prim}\rangle_{MC}=-\langle\hat{S}_{prim}\rangle_{MC}$,
and this relation also holds for the equilateral shape. Therefore, it is
straightforward 
to find the generalized estimator for the equilateral shape: first, find
the cubic estimator of the equilateral shape, $\hat{S}_{equil.}$,
and take the Monte Carlo average,
$\langle\hat{S}_{equil.}\rangle_{MC}$. 
Let us suppose that $\hat{S}_{equil.}$ contains terms in the form
of $ABC$, where $A$, $B$, and $C$ are some filtered maps.
Use the Wick's theorem to
re-write the average of a cubic product as 
$\langle ABC\rangle_{MC}=
\langle A\rangle_{MC}\langle BC\rangle_{MC}
+\langle B\rangle_{MC}\langle AC\rangle_{MC}
+\langle C\rangle_{MC}\langle AB\rangle_{MC}$. Finally, 
remove the MC average from single maps, and replace maps in the product
with the simulated maps 
$\langle A\rangle_{MC}\langle BC\rangle_{MC}
+\langle B\rangle_{MC}\langle AC\rangle_{MC}
+\langle C\rangle_{MC}\langle AB\rangle_{MC}
\rightarrow
A\langle B_{sim}C_{sim}\rangle_{MC}
+B\langle A_{sim}C_{sim}\rangle_{MC}
+C\langle A_{sim}B_{sim}\rangle_{MC}$. This operation gives the correct
expression for the linear term, both for the local form and the
equilateral form.

 The main  contribution to the
linear term  comes from  the inhomogeneous noise and sky cut.
For the
temperature only 
case, most of the contribution to the linear term comes from the
inhomogeneous noise, and the partial sky coverage does not contribute
much to the linear term. This is because the sky-cut 
induces a monopole contribution outside the mask. In the analysis one subtracts the monopole
from outside the mask before measuring $\hat{S}_{prim}$, which makes
the linear contribution from the mask small \cite{Creminelli_wmap1}. 
For a combined analysis of the temperature and polarization maps,
however, the linear term does  get a
significant contribution from a partial sky coverage. Subtraction of the monopole outside of the mask
is of no help for polarization, as the monopole does not exist in the
polarization maps by definition. (The lowest relevant multipole for
polarization is $l=2$.)

The estimator is still computationally efficient, taking only
$N^{3/2}_{pix}$ (times the $r$ sampling, which is of order 100) operations in comparison  to the full bispectrum calculation
which takes $N^{5/2}_{pix}$ operations. Here  $N_{pix}$ refers to the total number
of  pixels.  For Planck,  $N_{pix}  \sim  5 \times  10^7$,  and  so the  full
bispectrum analysis is not feasible while our analysis is.

\section{Constraints from the CMB Bispectrum} 
\label{sec:constraints}
\subsection{Current Status} 
 Currently the the Wilkinson Microwave Anisotropy Probe	(WMAP) satellite provides the ``best'' (largest number of signal dominated modes) CMB data for non-Gaussianity analysis. Over the course of WMAP operation the field of non-Gaussianity has made vast progress both in terms of theoretical predictions of non-Gaussianities from inflation and improvement in the bispectrum based estimators.
At the time of  WMAP's first data release in 2003 the estimator was sub-optimal in the presence of partial sky coverage and/or inhomogeneous noise. With the sub-optimal estimator one could not use the entirety of WMAP data and only the data up to $l_{max}~350$ were used to obtain the constraint $f^{local}_{\mathrm{NL}}=38\pm 96 (2\sigma)$~\cite{nong_wmap}. These limits were around 30 times better than the previous constraints of $\vert \fnl\vert < 1.5 \times 10^{3}$ from the Cosmic Background Explorer (COBE) satellite~\cite{fnl_cobe}. 

By the time of second WMAP release in 2007 the estimator was generalized by adding a linear to the KSW estimator which allows to deal with partial sky coverage and inhomogeneous noise. The idea of adding a linear term to reduce excess variance due to noise inhomogeneity was introduced in~\cite{Creminelli_wmap1}. Applied to a combination of the Q, V and W channels of the WMAP 3-year data up to $l_{max} \sim 400$ this estimator had yielded the tightest constraint at the time on $\fnl$ as: $ -36 < \fnl < 100 (2\sigma)$~\cite{creminelli_wmap2}. This estimator was further generalized to utilize both the temperature and E-polarization information in~\cite{Yadav_etal08a}, where it was pointed out that the linear term had been incorrectly implemented in Eq. 30 of~\cite{Creminelli_wmap1}. Using Monte-Carlo simulations it has been shown that this corrected estimator is nearly optimal and enables analysis of the entire WMAP data without suffering from a blow-up in the variance at high multipoles\footnote{We will refer to the estimator in Ref.~\cite{Creminelli_wmap1} as a near-optimal-v1, while the corrected estimator of Ref.~\cite{YW08} as near-optimal.}. The first analysis using this estimator shows an evidence of non-Gaussianity of local type at around $2.8\sigma$ in the WMAP 3-year data. Independent analysis shows the evidence of non-Gaussianity at lower significance, around $2.5\sigma$ (see Table~\ref{tab:fnl}).
 
\begin{figure*}[]
\includegraphics[height=17cm, angle=0]{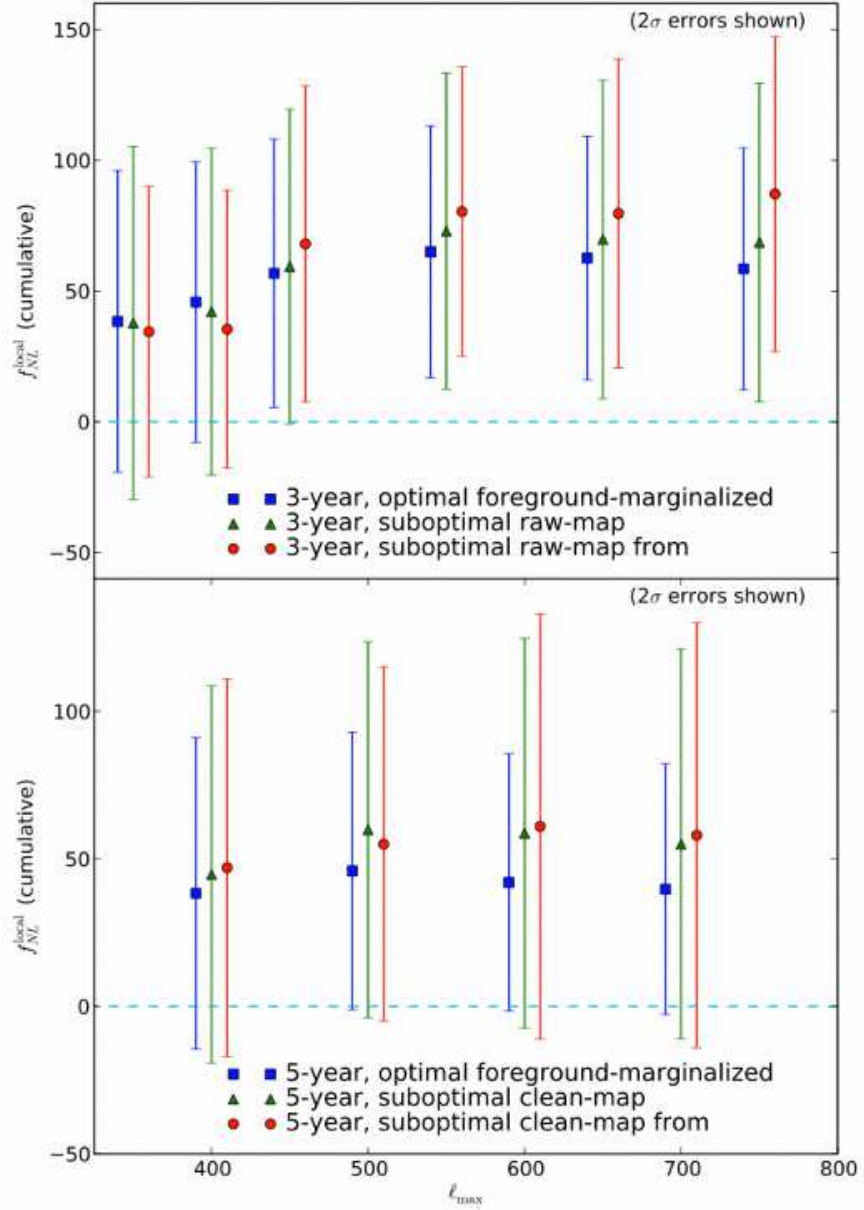}
\caption{{\it Top panel:} Constraints on local $\fnl$ using WMAP 3-year data as a function of maximum multipole $\ell_{max}$ used in the analysis. The red circles are the results obtained using near optimal estimator by Yadav $\&$ Wandelt~\cite{YW08}. The green triangles are using the the near-optimal estimator by Smith et al.~\cite{Smith:2009jr}. The blue square results are obtained using either the optimal estimator by Smith et al.~\cite{Smith:2009jr}. For all the three analysis Kp0 mask was used. {\it Bottom panel:} Comparison between 5-year results (optimal estimator, raw maps) reported in Komatsu et al.~\cite{wmap5_cosmol} and results obtained using the optimal or suboptimal estimator by Smith et al.~\cite{Smith:2009jr}.
  \label{fig:fnl}}
\end{figure*}

By the time of the third WMAP data release (with 5-year obsevational data) in 2008 the $\fnl$ estimation technique was improved further by implementing the covariance matrix including inhomogeneoous noise to make the estimator completely optimal~\cite{Smith:2009jr}. Using the optimal estimator and using the entirety of WMAP 3-year data there is an evidence for non-Gaussianity of local type at around $2.5\sigma$ level $\fnl\approx 58\pm 23 (1\sigma)$~\cite{Smith:2009jr}. However with WMAP 5-year data the significance goes down from $\sim 2.5\sigma$ to $\sim 1.8\sigma$~\cite{Smith:2009jr}. Table~\ref{tab:fnl} compares the constraints obtained by different groups using WMAP 3-year and WMAP 5-year data. Fig~\ref{fig:fnl} shows this comparison in more detail, showing the constraints also as a function of maximum multipole $l_{max}$ used in the analysis. Few comments are in place: 1) constraints on $\fnl$ from WMAP 3-year data as a function of $l_{max}$ show a trend where the mean value rises at around $l_{max}=450$, below which data is consistent with Gaussianity and above which there is deviation from Gaussianity at above $2\sigma$. The result becomes roughly independent of $\ell_{max}>550$ with evidence for non-Gaussianity at around $2.5\sigma$ level, 2) independent analysis and using different estimators  (optimal and near-optimal with linear term) sees this deviation from non-Gaussianity at around $2.5\sigma$ in WMAP 3-year data, 3) significance of non-Gaussianity goes down to around $2\sigma$ with WMAP 5-year data. The drop in the mean value between WMAP 3-year and 5-year data can be attributed to statistical shift. 

The best constraints on the equilateral and  orthogonal shape of non-Gaussianity using the WMAP 5-year data are $f^{\text{equil}}_{\mathrm{NL}}=155\pm140 (1\sigma)$ and $f^{\text{orthog.}}_{\mathrm{NL}}=149\pm110 (1\sigma)$ respectively~\cite{Senatore:2009gt}.

As we were completing this article, the WMAP 7-year data was released, with constraints $\fnl^{\text{local}}=32\pm 21 (1\sigma), \fnl^{\text{equil.}}=26\pm140 (1\sigma)$ and $\fnl^{\text{orthog}}=-202\pm 104 (1\sigma)$~\cite{Komatsu:2010fb}.

\subsection{Future Prospects} 
Now we discuss the future prospects of using the bispectrum estimators for constraining the non-linearity parameter $\fnl$ for local and equilateral shapes. We compute the Fisher bounds for three experimental setups, (1) cosmic variance limited experiment with perfect beam (ideal experiment hereafter), (2) Planck satellite with and noise sensitivity $\Delta_p=56 \mu$K-arcmin and beam FWHM $\sigma=7'$  (3) a futuristic CMBPol like satellite experiment with noise sensitivity $\Delta_p=1.4 \mu$K-arcmin and beam FWHM $\sigma=4'$ (CMBPol hereafter).  Beside $\fnl$ we fix all the other cosmological parameters to a standard
fiducial model with a flat $\Lambda CDM$ cosmology, with parameters
described by the best fit to WMAP 5-year results~\cite{wmap5_cosmol}, given by
$\Omega_b=0.045, \Omega_c=0.23, H_0=70.5, n_s=0.96, n_t=0.0,$ and
$\tau=0.08$. We calculate the theoretical CMB transfer functions and power spectrum from publicly available code CMBFAST~\cite{cmbfast}. We also neglect any non-Gaussianity which can be generated during recombination or there after. We discuss the importance and effect of these non-primordial non-Gaussianities in next section.

\begin{figure*}[]
\includegraphics[height=5.9cm, angle=-90]{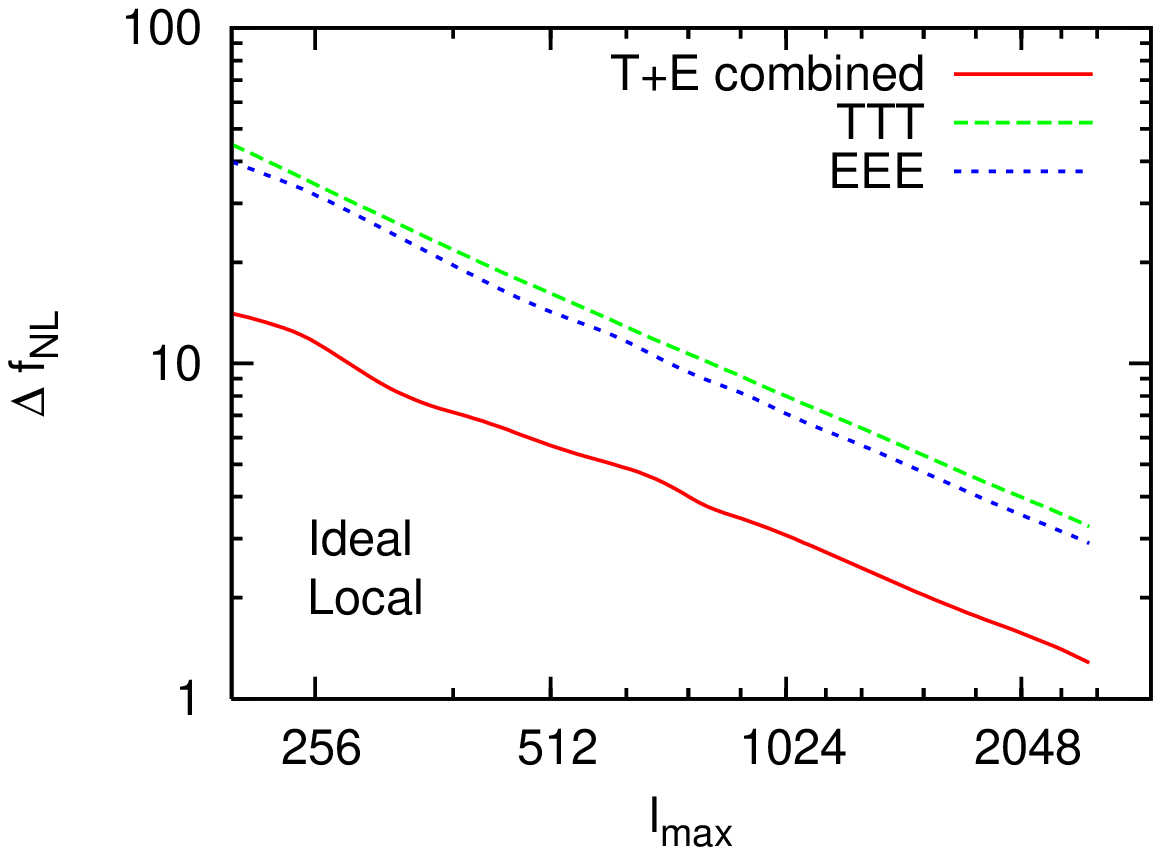}
\includegraphics[height=5.9cm, angle=-90]{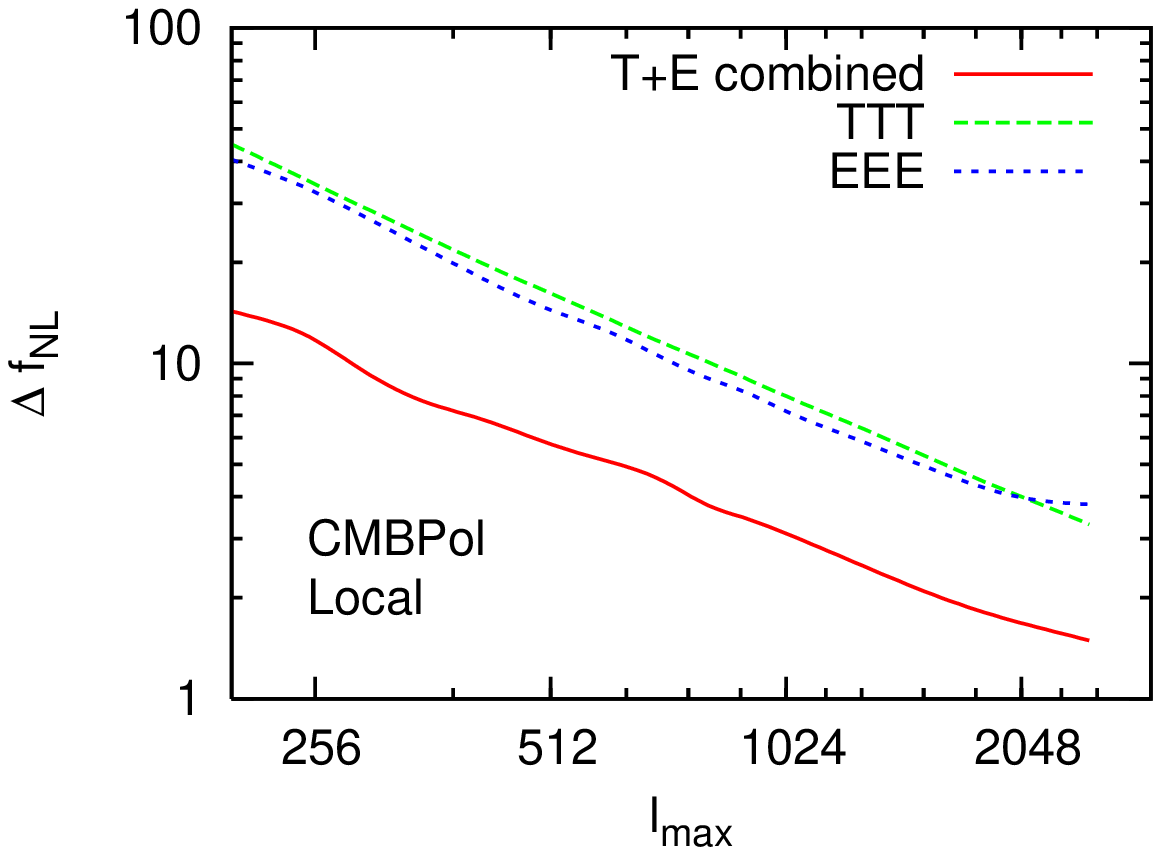}
\includegraphics[height=5.9cm, angle=-90]{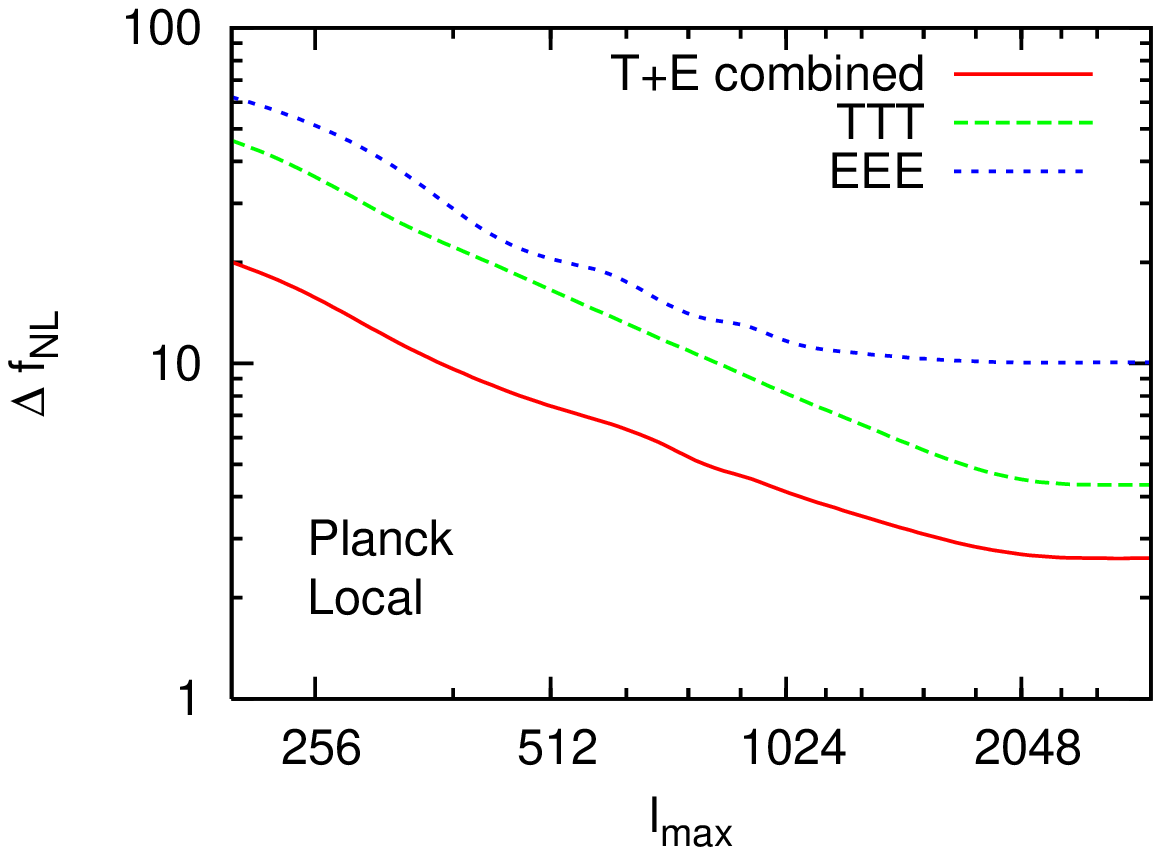}
\includegraphics[height=5.9cm, angle=-90]{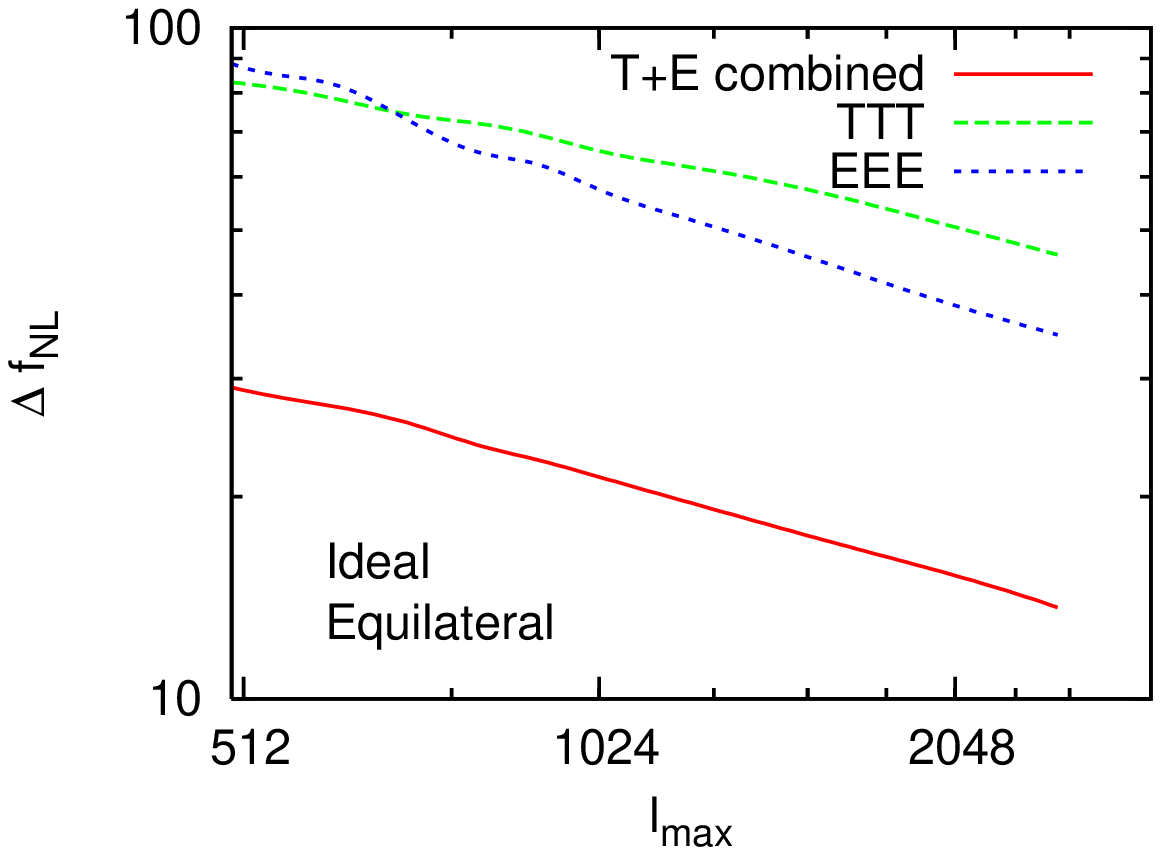}
\includegraphics[height=5.9cm, angle=-90]{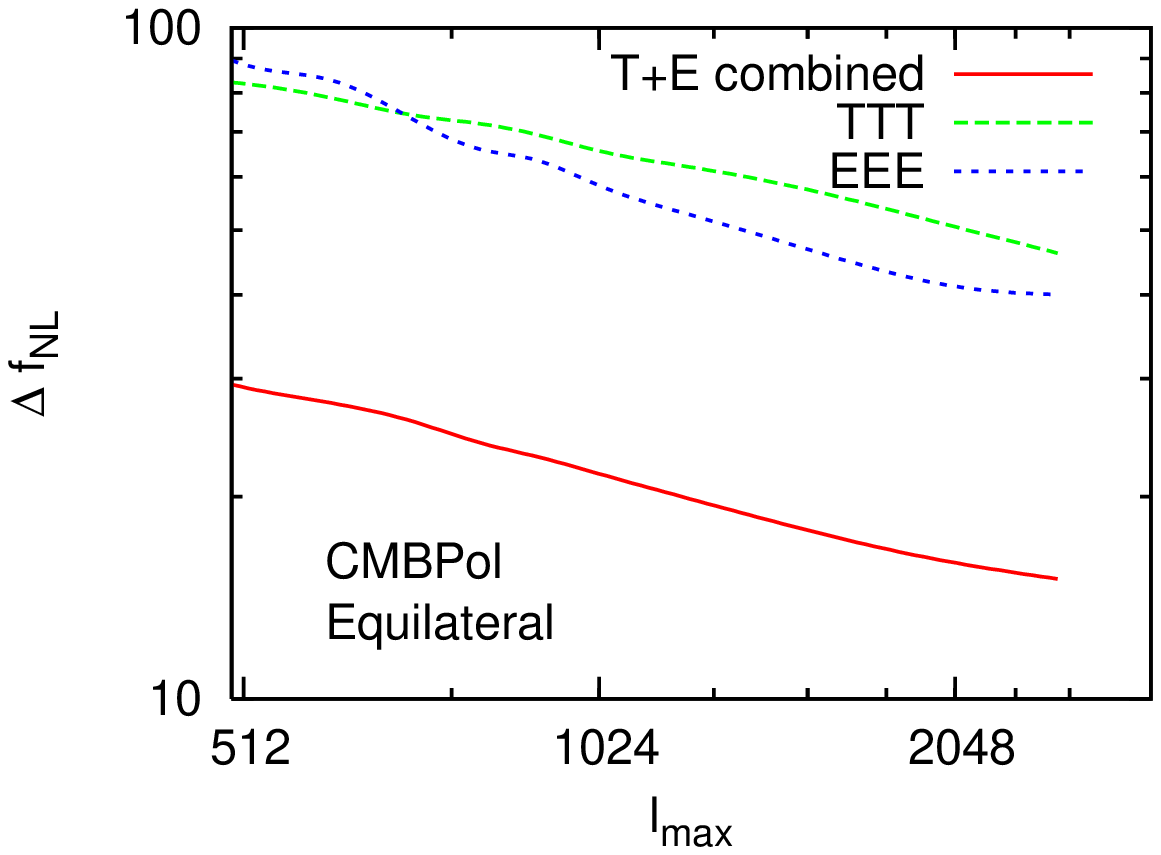}
\includegraphics[height=5.9cm, angle=-90]{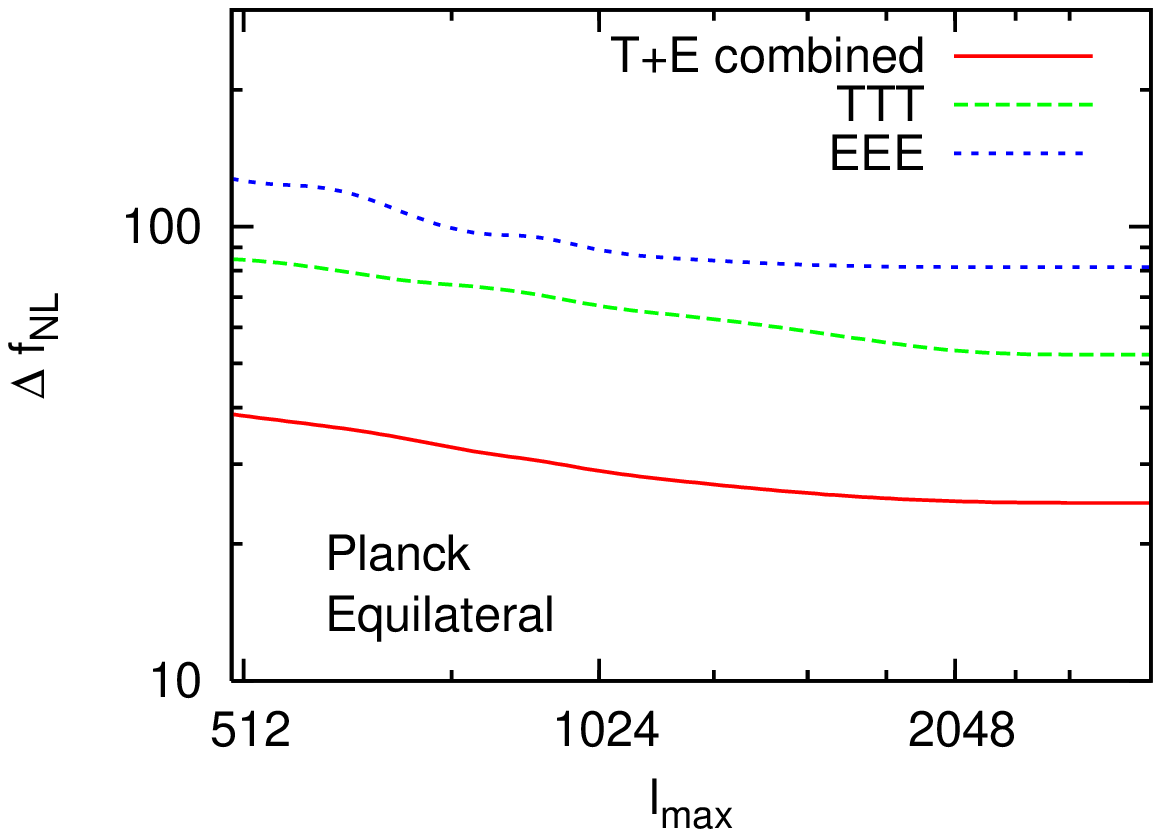}
\caption{Fisher predictions for minimum detectable  $\fnl$ (at 1 $\sigma$) as a function of maximum multipole $\ell_{max}$. Upper panels are for the local model while lower panels are for the equilateral model. Left
panels shows an ideal experiment, middle panels are for CMBPol like experiment with noise sensitivity $\Delta_p=1.4 \mu$K-arcmin and beam FWHM $\sigma=4'$ and right panels are for Planck like satellite with and noise sensitivity $\Delta_p=40 \mu$K-arcmin and beam FWHM $\sigma=5'$. In all the panels, the solid lines represent temperature and polarization combined analysis;  dashed  lines represent  temperature only analysis; dot-dashed lines represent polarization only analysis.
\label{fig:fnl_fisher}}
\end{figure*}

The scaling of signal-to-noise as a maximum multipole $l_{max}$ for the local~\cite{creminelli_estimators_nong,Liguori_Yadav_etal07} and equilateral model~\cite{Bartolo:2008sg} are 
\begin{eqnarray}
(S/N)_{local}\propto \ln\ell_{max}\,, \quad
(S/N)_{equil}\propto \sqrt{\ell_{max}}\,. 
\end{eqnarray}
In principle one could go to arbitrary high $l_{max}$ but in reality secondary signals will certainly overwhelm primary signal beyond $l_{max}>3000$, we restrict to the analysis to $l_{max}=3000$. 
 In Figure~\ref{fig:fnl_fisher} we show the $1\sigma$ Fisher bound as function of maximum multipole $\ell_{max}$, for local and equilateral type of non-Gaussianity. For both local and equilateral case we show the Fisher bound for the analysis using only the CMB temperature information (TTT), only the CMB polarization information (EEE), and the combined temperature and polarization analysis. Note that by having both the temperature and $E$-polarization information one can improve the sensitivity by combining the information. Apart from combining the T and E signal, one can also do cross-checks and diagnostics by independently analysing the data. Temperature and polarization will have different foregrounds and instrumental systematics.

A CMBPol like experiment will be able to achieve the sensitivity of $\Delta \fnl^{local} \simeq 2 (1\sigma)$ 
for non-Gaussianity of local type and  $\Delta \fnl^{equil.} \simeq 13 (1\sigma)$  for non-Gaussianity of equilateral type. For the local type of non-Gaussianity this amounts to an improvement of about a factor of 2 over the Planck satellite and about a factor of 12 over current best constraints. These estimates assume that foreground cleaning can be done perfectly, i.e. the effect of residual foregrounds has been neglected. Also the contribution from unresolved point sources and secondary anisotropies such as ISW-lensing and SZ-lensing has been ignored.
\begin{table}
\begin{center}
\caption{Hint of local non-Gaussianity at $2\sigma$ level}
\begin{tabular}{l| c|c|l }
\hline
\hline
data (mask, estimator) & $f^{local}_{\mathrm{NL}}\pm 1\sigma$ error & deviation from Gaussianity &\\
\hline
WMAP 3-year (Kp0, near-optimal) & $87\pm 31$ & $2.8\sigma$ & Yadav and Wandelt~\cite{YW08} \\
WMAP 3-year (KQ75, optimal) & $58\pm 23$ & $2.5\sigma$ & Smith et al.~\cite{Smith:2009jr} \\
WMAP 3-year (Kp0, near-optimal) & $69\pm 30$ & $2.3\sigma$ & Smith et al.~\cite{Smith:2009jr} \\
\hline
WMAP 5-year (KQ75, near-optimal)& $51\pm 30$ & $1.7\sigma$ & Komatsu et al.~\cite{wmap5_cosmol} \\
WMAP 5-year (KQ75, optimal) & $38\pm 21$ & $1.8\sigma$ & Smith et al.~\cite{Smith:2009jr} \\
\end{tabular}
\footnotetext{The difference between the results by Ref.~\cite{YW08} and~\cite{Smith:2009jr} for WMAP 3-year data using the Kp0 mask can be a result of different choices of weighting in the near-optimal estimator. The optimal estimator has a unique weighting scheme.}
\end{center}
\label{tab:fnl}
\end{table}

\begin{figure}[!t]
\begin{center}
{\includegraphics[width=0.38\textwidth]{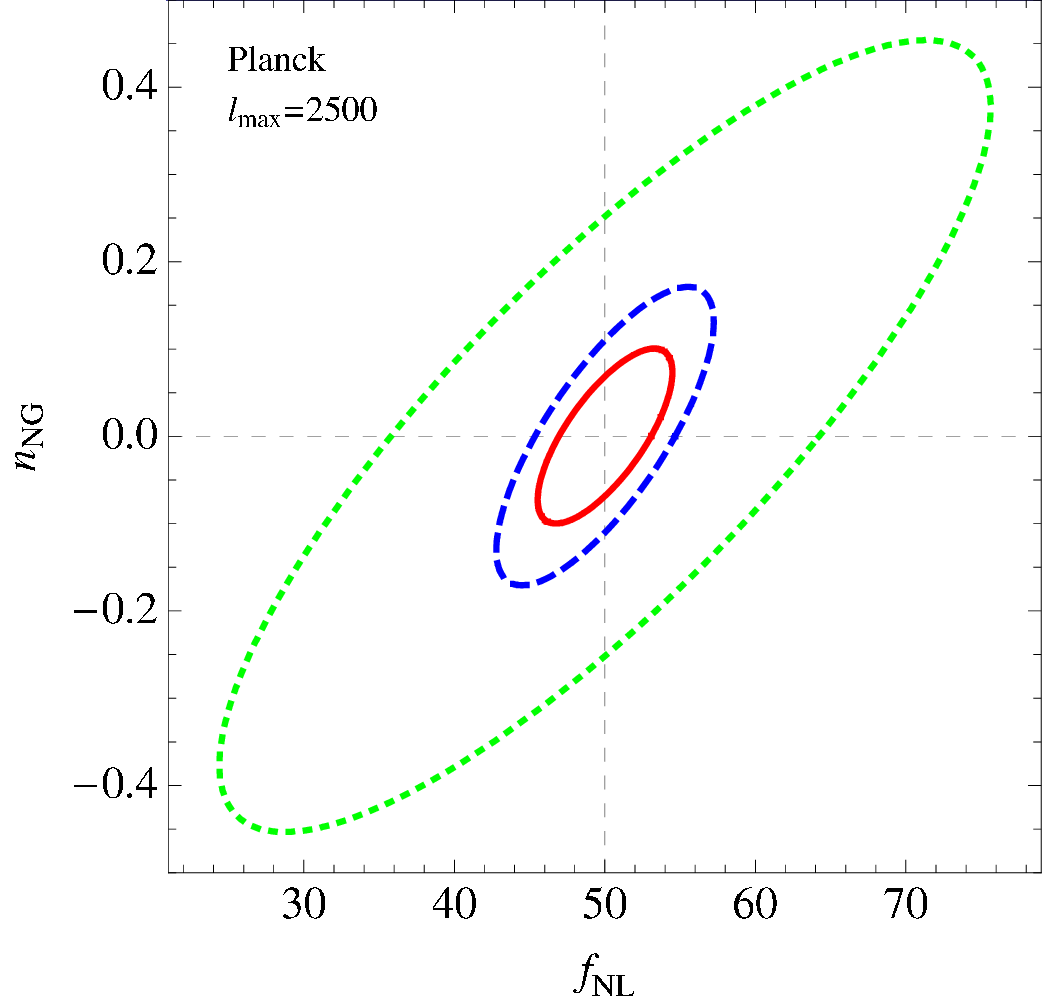}}
{\includegraphics[width=0.38\textwidth]{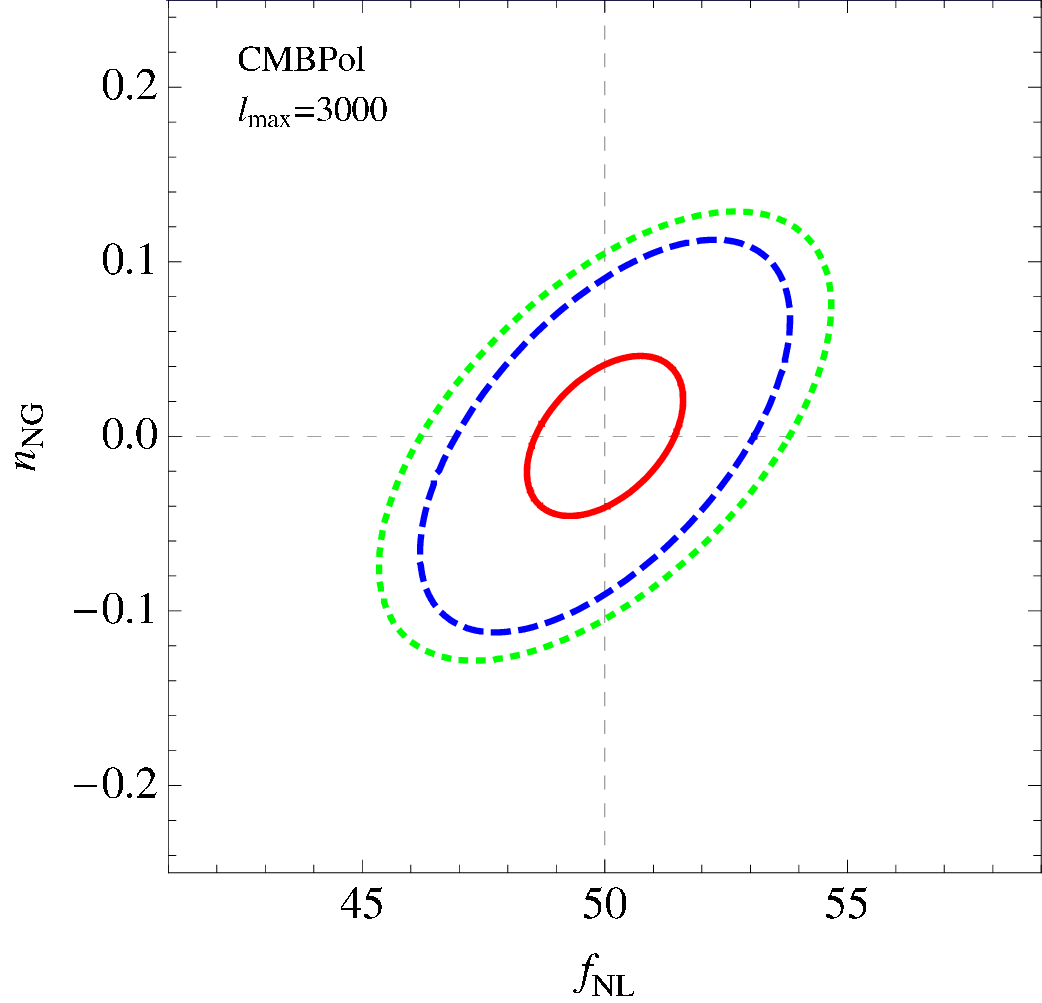}}                               
{\includegraphics[width=0.38\textwidth]{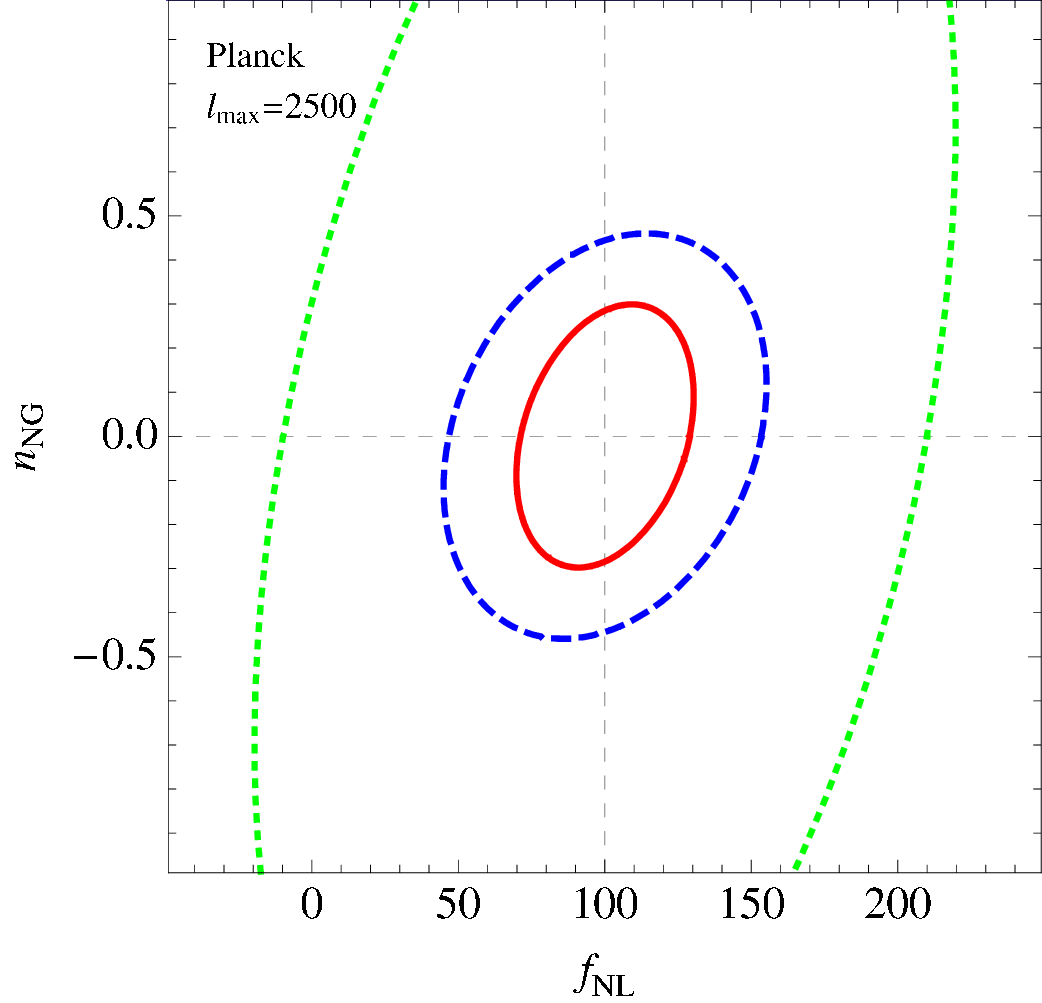}}
{\includegraphics[width=0.38\textwidth]{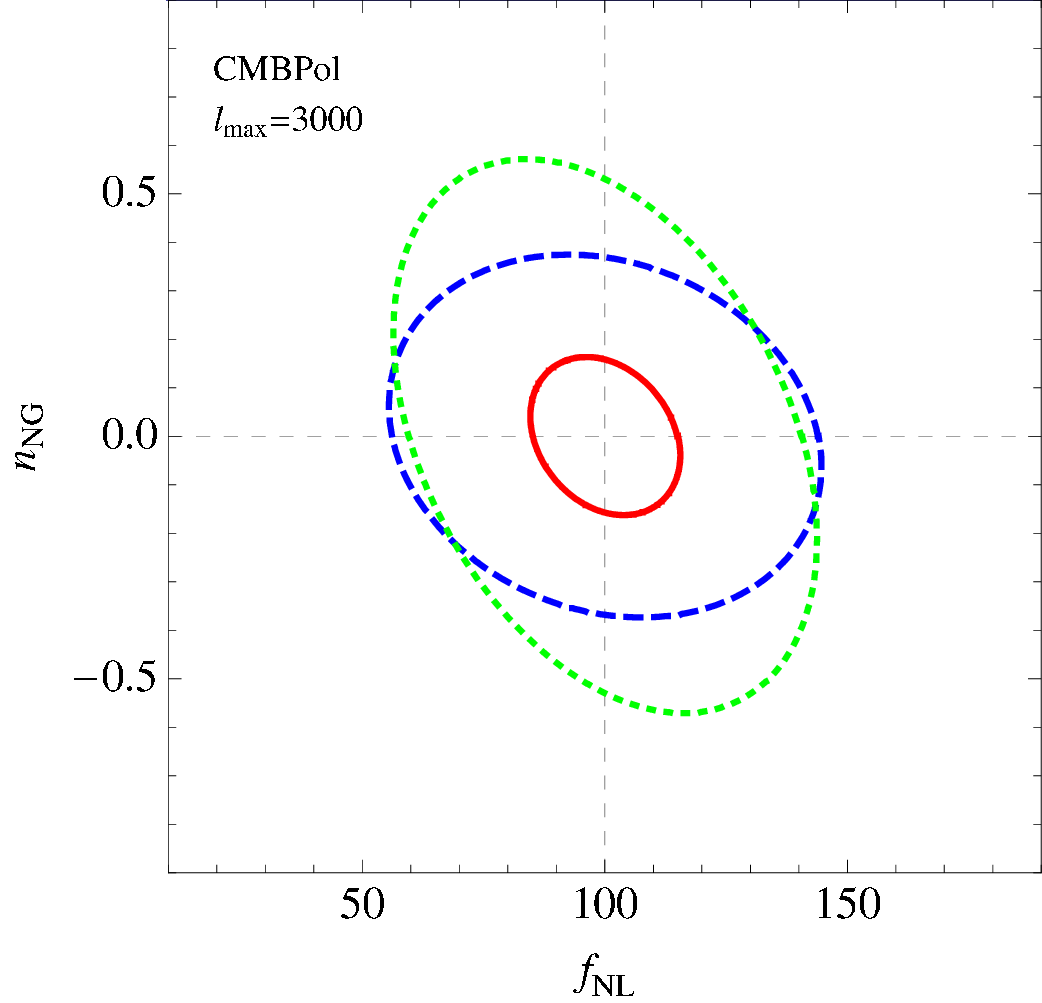}}
\caption{1-$\sigma$ constraints on $\fnl$ and $\nng$ for local ({\it upper panels}) and equilateral model ({\it lower panels}) assuming $k_p=0.04$ Mpc$^{-1}$ and fiducial values $\fnl=50$, $\nng=0$ for the local case and $\fnl=100, \nng=0$ for the equilateral case. Dashed lines correspond to the limits from the temperature information alone, dotted lines to polarization (EEE), while the continuous lines correspond to all bispectrum combinations. We consider Planck ({\it left panels}) and CMBPOl like ({\it right panels}) CMB experiment.}
\label{fig:running}
\end{center}
\end{figure}

{\it Running non-Gaussianity:} The primordial non-Gaussian parameter $\fnl$ has been shown to be scale-dependent in several models of inflation with a variable speed of sound, such as Dirac-Born-Infeld (DBI) models. Starting from a simple ansatz for a scale-dependent amplitude of the primordial curvature bispectrum for primordial non-Gaussianity, 
\be
\fnl \rightarrow \fnl \Bigg( \frac{K}{k_p}\Bigg)^{n_{NG}}
\ee
where $K\equiv (k_1k_2k_3)^{1/3}$ and $k_p$ is a pivot point. The primordial bispectrum is therefore determined in terms of two parameters: the amplitude $\fnl$ and the new parameter $n_{NG}$ quantifying its running.
One can generalize the Fisher matrix analysis of the bispectra of the temperature and polarization of the CMB radiation and derive the expected constraints on the parameter $n_{NG}$ that quantifies the running of $\fnl(k)$ for current and future CMB missions such as WMAP, Planck and CMBPol. We will consider some non-zero $\fnl$ as our fiducial value for the Fisher matrix evaluation. Clearly, in order to be able to constrain a scale-dependence of $\fnl$, its amplitude must be large enough to produce a detection. If $\fnl$ is too small to be detected ($\fnl< 2$ is a lowest theoretical limit even for the ideal experiment), we will obviously not be able to measure any of its features, either. In the following we will then always consider a fiducial value of $\fnl$ large enough to enable a detection. Figure~\ref{fig:running} shows the $1-\sigma$ joint constraints on $\fnl$ and $n_{NG}$. In the event of a significant detection of the non- Gaussian component, corresponding to $\fnl = 50$ for the local model and $\fnl = 100$ for the equilateral model of non-Gaussianity, is able to determine $n_{NG}$ with a $1-\sigma$ uncertainty of $\Delta n_{NG} \simeq 0.1$ and $\Delta n_{NG} \simeq 0.3$, respectively, for the Planck mission and a factor of two better for CMBPol. In addition to CMB one can include the information of the galaxy power spectrum, galaxy bispectrum, and cluster number counts as a probe of non- Gaussianity on small scales to further constrain the two parameters~\cite{Sefusatti:2009xu}.


\subsection{Contaminations} A detection of non-Gaussianity has profound implications on our understanding of
the early Universe. Hence it is important to know and quantify all the possible sources of non-Gaussianities in the CMB. Here we highlight some sources of non-Gaussianities due to second-order anisotropies after last scattering surface and during recombination. The fact that Gaussian initial conditions imply Gaussianity of the CMB is only true at linear order. We will also discuss the effects of instrumental effects and uncertainties in the cosmological parameters on the bispectrum estimate.

\subsubsection*{Secondary non-Gaussianities}
Current analysis of the CMB data ignore the contributions from the secondary non-Gaussianities. For WMAP resolution it may not be a bad approximation. Studies of the dominant secondary anisotropies conclude that they are
negligible for the analysis of the WMAP data for $l_{max}<800$ \cite{KS2001,SerraCooray08}.  However on smaller angular
scales several effects starts to kick in, for example, 1) the bispectrum contribution due to unresolved point source like thermal Sunyaev-Zeldovich clusters or standard radio sources, 2) three way correlations between primary CMB, lensed CMB and secondary anisotropies. We will refer to the bispectrum generated due to this three way correlation as $B^{secondary-\kappa}$, where some secondaries are, the integrated Sachs-Wolfe (ISW)  $B^{ISW-\kappa}$, Sunyaev-Zeldovich signal and Rees-Sciama ~\cite{Goldberg_spergel_1999, Spergel_Goldberg_1999, Cooray_Hu_2000, KS2001,nong_wmap,SerraCooray08}. 

For Future experiments such as Planck and CMBPOl the joint estimation of primordial and secondary bispectrum will be required. The observed bispectrum in general would take the following form
\begin{eqnarray}
\hat B^{obs}_{\ell_1\ell_2\ell_3}=  \fnl B^{prim}_{\ell_1\ell_2\ell_3}+b_{ps} B^{ps}_{\ell_1\ell_2\ell_3} + A_{SZ} B^{SZ-\kappa}_{\ell_1\ell_2\ell_3}+ A_{ISW} B^{ISW-\kappa}_{\ell_1\ell_2\ell_3}+...
\end{eqnarray}
The amplitude of bispectrum due to primary-lensing-secondary cross-correlation is proportional to the product of primary CMB power-spectrum and power spectrum of cross-correlation between secondary and lensing signal. 

The reduced bispectrum from the residual point sources (assuming Poisson distributed) is constant i.e. $b^{ps}_{\ell_1 \ell_2 \ell_3}=\text{constant.}$ The value of the constant will depend on the flux limit at which the point source can be detected and on assumed flux and frequency distribution of the sources.

Depending on the shape of primordial bispectrum in consideration, some secondary bispectra are more dangerous than others. For example, ISW-lensing $B^{ISW-\kappa}$ peaks at the ``local'' configurations, hence is more dangerous for local primordial shape than the equilateral primordial shape. For example for the Planck satellite if the secondary bispectrum is not incorporated in the analysis, the ISW-lensing contribution will bias the estimate for the local $\fnl$ by around $\Delta f^{local}_{\mathrm{NL}} \approx 10$~\cite{SmithZaldarriaga}. The bispectrum contribution from primary-lensing-Rees-Sciama signal also peaks at squeezed limit and contribute to effective local $f^{local}_{\mathrm{NL}}\approx 10$~\cite{Mangilli:2009dr}. For Planck sensitivity the point source will contamination the local non-Gaussianity by around  $\Delta f^{local}_{\mathrm{NL}}\sim 1$~\cite{Babich:2008uw}. A recent analysis of the full second order Boltzmann equation for photons~\cite{2010arXiv1003.0481P} claims that second order effects add a contamination  $\Delta\fnl\sim 5$.

The generalization of the Fisher matrix given by Eq.~(\ref{eq:fisher}) to include multiple bispectrum contribution is
\be
{\mathcal F}^{(XY)}_{ab} = 
\sum_{\left\{i\!jk,~pqr\right\}}\sum_{\ell_1\le\ell_2\le\ell_3}
 \frac{1}{\Delta_{\ellt}} 
 \frac{\partial B^{pqr, (X)}_{\ellt}}{\partial p_a}
 \left({\bf Cov}^{-1}\right)_{i\!jk,~pqr}
 \frac{\partial B^{\,i\!jk, (Y)}_{\ellt}}{\partial p_b}\,,
\ee
where the additional $X$ and $Y$ indices denote a component such as primordial, point-sources, ISW-lensing etc. For fixed cosmological parameters, the signal-to-noise $(S/N)_i$ for the component $i$ is 
\be
\Bigg( \frac{S}{N}\Bigg)_i=\frac{1}{\sqrt{{\mathcal F}^{(ii)}}}
\ee
 
\subsubsection*{Non-Gaussianities from recombination} Non-Gaussianities can be generated during recombination. One requires to solve second order Boltzmann and Einstein equations to evaluate the effect. The second order effect on CMB is an active field of study~\cite{Hu:1993tc,Dodelson:1993xz,Pyne:1995bs,Mollerach:1997up,Matarrese:1997ay,2004JHEP...04..006B,2004JCAP...01..003B,Creminelli:2004pv,Bartolo:2004ty,Tomita:2005et,Bartolo:2005fp,Bartolo:2005kv,Tomita:2007kc,Pitrou:2008ak,Bartolo:2006cu,Bartolo:2006fj,Pitrou:2008hy,Senatore:2008vi,2009JCAP...08..029B,2010ApJ...711.1310K}, see Ref.~\cite{2010arXiv1001.3957B} for a recent review. The non-Gaussianities produced during recombination comprise of various effects, for example see Ref.~\cite{Bartolo:2006fj}.

The dominant bispectrum due to perturbative recombination comes from perturbations in the electron density. The amplitude of perturbations of the free electron density $\delta_e$ is around a factor of 5 larger than the baryon density perturbations~\cite{Novosyadlyj:2006fw}. The bispectrum generated due to $\delta_e$ peaks around the ``local'' configuration with corresponding effective non-linearity amplitude $\fnl\sim \text{few}$~\cite{Khatri:2008kb,Senatore:2008wk,2010PhRvD..81j3518K}.   

The bispectrum contribution due to second order terms which are the products of the first order perturbations is calculated in Ref.~\cite{Nitta:2009jp}. The bispectrum contribution from these terms which also peak for the squeezed triangles is small and can be neglected in the analysis. For example the signal-to-noise is about  0.4 at $l_{max}=2000$ for a full-sky, cosmic variance limited experiment.

Another contribution to bispectrum which peaks for the equilateral configurations comes from the non-linear evolution of the second order gravitational potential. Because of this effect the minimum detectable non-Gaussianity parameter $f^{equil.}_{\mathrm{NL}}$ changes by $\Delta f^{equil.}_{\mathrm{NL}}={\cal O}(10)$ for Planck like experiment~\cite{Bartolo:2008sg}. The bispectrum peaks for the equilateral shape because the the growth of potential happens on scales smaller than the horizon size.

 On large scales, in the absence of primordial non-Gaussianities and assuming matter domination (so that the early and late ISW can be neglected), it has been shown in Ref.~\cite{2009JCAP...08..029B} that for the squeezed limit the effective $\fnl$ generated by second order gravitational effects on the CMB is $\fnl= - 1/6 - \cos(2 \theta)$ (also see~\cite{2004JHEP...04..006B,2004JCAP...01..003B, Creminelli:2004pv}). Here $\theta$ is the angle between the short and the long modes. The angle dependent contribution comes from lensing.  

\subsubsection*{Effect of Cosmological parameter uncertainties} Impact of uncertainties on the cosmological parameters effect the error bar on $\fnl$. The effect of cosmological parameters have been discussed in Ref.~\cite{Creminelli_wmap1,creminelli_wmap2,YW08,Liguori:2008vf}. The cosmological parameters are determined using the 2-point statistics of the CMB and therefor we expect the largest effect of $\fnl$ would come from those parameters which leave the CMB power spectrum unchanged while change the bispectrum. The expectation value of the estimator
\be
\langle \hat f_{NL}\rangle= \frac{1}{N}\sum_{\ell_1,\ell_2,\ell_3} \frac{B_{\ell_1\ell_2\ell_3} \hat B_{\ell_1\ell_2\ell_3}}{C_{\ell_1}C_{\ell_2} C_{\ell_3}}\,,
\ee
 changes with the change in cosmological parameters. Here $\hat B_{\ell_1 \ell_2 \ell_3}$ is the true CMB bispectrum. When changing the parameters the normalization $N$ should be changed to make the estimator unbiased. In general for a set of cosmological paramerets $\{p_i\}$, the error in $\fnl$ is given by~\cite{Liguori:2008vf}
\begin{eqnarray}
\delta  \hat f_{NL}  = \sqrt{\sum_{ij}\left. \frac{\partial \fnl}{\partial
   p_i}\right|_{p_i = \bar{p}_i}
                                \left.\frac{\partial \fnl}{\partial
   p_j}\right|_{p_j=\bar{p}_j} {\rm Cov}(p_i,p_j)} \,.
\end{eqnarray} 
Here the average parameter values $\bar{p}_i$ and their covariance matrix ${\rm Cov}(p_i,p_j)$ can be determined using CMB-likelihood analysis tools.

If the parameters are allowed to vary in the analysis then for WMAP this increases the  $1\sigma$ uncertainty in $\fnl$ by $\delta f^{local}_{\mathrm{NL}}/\fnl\approx 16\%$ for the local shape and $\delta f^{equil}_{\mathrm{NL}}/\fnl\approx 14\%$ for the equilateral shape. For Planck experiment the increases in $1\sigma$ uncertainity is $\delta f^{local}_{\mathrm{NL}}/\fnl\approx 5\%$ for local shape and $\delta f^{equil.}_{\mathrm{NL}}/\fnl\approx 4\%$ for the equilateral shape. Most of the contribution to the error comes from three cosmological parameters, the amplitude of scalar perturbations $\Delta_\Phi$, the tilt of the power spectrum of the saclar perturbations $n_S$, and re-ionization optical depth $\tau$.

For modes inside the horizon during reionization, the reionization optical depth $\tau$ appears as a multiplicative factor $e^{-\tau}$ in front of transfer function $g^{i}_{\ell}$. For local model one of the mode is outside so the effect on bispectrum $\tilde b^{local}_{\ell_1 \ell_2 \ell_3}=\exp(-2\tau)  b^{local}_{\ell_1 \ell_2 \ell_3}$ and for equilateral model all the modes are inside the horizon so $\tilde b^{equil.}_{\ell_1 \ell_2 \ell_3}=\exp(-3\tau)  b^{equil.}_{\ell_1 \ell_2 \ell_3}$. This reduces to $\delta f^{local}_{\mathrm{NL}}\simeq -2 \fnl \tau$ for local model and  $\delta f^{equil.}_{\mathrm{NL}}\simeq -3 \fnl \tau$ for equilateral model.

 The effect of amplitude of perturbations can be seen by noting that the level of non-Gaussianity is given by $\fnl \cdot \Delta^{1/2}_{\Phi}$. Hence the decrease (increase) in the amplitude of perturbations relax (tighten) the constraints on $\fnl$. The effect of red tilt ($n_s < 1$) can be thought of as a reduction in power on at scales shorter than first peak and enhancement of power on larger scales. The effect of blue tilt is just opposite of red tilt. For local shape the limit on $\fnl$ becomes tighter proportional to $\Delta^{1/2}_{long}$~\cite{creminelli_wmap2}. 
Note that Ref.~\cite{Liguori:2008vf} show that the effect of cosmological parameters is negligible if the parameters are allowed to vary in the analysis and then marginalize over. 

\subsubsection*{Instrumental effects and distortions along the line of sight}
Here we point out that any cosmological or instrumental effect that can be modelled as a line of a sight CMB distortions of the primary CMB do not generate new bispectrum contribution. Although they can modify the the primordial bispectrum. A general model of line of sight distortions of the primary CMB are described in Ref.~\cite{HHZ,SYZ09,YSZ09} where the changes in the Stokes parameter of the CMB due to distortions along the line-of-sight can be written as
\begin{eqnarray}
\delta [Q \pm i U](\bn) &=&
            [\calb \pm i 2 \rot](\bn)  [\tilde Q \pm i \tilde U](\bn) + [f_1 \pm i f_2](\bn)   [\tilde Q \mp i \tilde U](\bn) + [\gamma_1 \pm i \gamma_2](\bn) \tilde T(\bn) \nonumber \\
 &&+\sigma {\bf p}(\bn) \cdot \nabla [\tilde Q \pm i \tilde U](\bn;\sigma) + \sigma [d_1 \pm i d_2](\bn) [\partial_1 \pm i\partial_2] \tilde T(\bn;\sigma)
 + \sigma^2 q(\bn) [\partial_1 \pm i \partial_2]^2 \tilde T(\bn;\sigma)+\ldots \,.
\label{eqn:model_distortions}
\end{eqnarray}
The first line  captures the distortions in a  single perfectly known direction $\bn$. The distortions in second line capture mixing of the polarization fields in a local region of length scale $\sigma$ around $\bn$. We Taylor expand the CMB fields $Q, U,$ and $ T$ around the point $\bn$ and consider the leading order terms. Here $\tilde Q, \tilde U,$ and $\tilde T$ stands for primordial (un-distorted) CMB fields. Since $(Q \pm iU)(\bn)$ is spin $\pm 2$ field, $a(\bn)$ is a scalar field that describes modulation in the amplitude of the fields in a given direction $\bn$; $\rot(\bn)$ is also a scalar field that describes the rotation of the plane of polarization, $(f_1\pm if_2)$ are spin $\pm 4$ fields that describe the coupling between two spin states (spin-flip),
 and $(\gamma_1\pm i \gamma_2)(\bn)$ are spin $\pm2$ fields that describe leakage from the temperature to polarization (monopole leakage hereon). Distortions in the second line of Eqn.~(\ref{eqn:model_distortions}), $(p_1\pm p_2), (d_1 \pm d_2)$, and $q$ are measured in the units of the length scale $\sigma$.
The field $(p_1\pm ip_2)(\bn)$ is a spin $\pm 1$ field and describes the change in the photon direction; we will refer to it as a deflection field. Finally $(d_1\pm d_2)(\bn)$ and $q(\bn)$ describe leakage from temperature to polarization, $(d_1\pm d_2)(\bn)$ is spin $\pm1$ field and we will refer to it as dipole leakage; $q(\bn)$ is a scalar field that we will call quadrupole leakage.

These distortions can be produced by various cosmological processes such as weak gravitational lensing of the CMB, screening effects from patchy reionization, rotation of the plane of polarization due to magnetic fields or parity violating physics and various instrumental systematics such as gain fluctuations, pixel rotation, differential gain, pointing, differential ellipticity are also captured via line of sight distortions. All these distortions modify the primordial bispectrum as

\begin{eqnarray}
&&\tilde B_{({\bf \ell}_1,{\bf \ell}_2,{\bf \ell}_3)} = B_{({\bf \ell}_1,{\bf \ell}_2,\bf{\ell}_3)} + \int \frac{d^2 {\ell'}}{(2\pi)^2} C^{\cal D \cal D}_{l'} \Big[B_{(\bf \ell_1,\bf \ell_2-\bf \ell',\bf \ell_3 +\bf \ell')}W^{\cal D}({\bf \ell}_1-{\bf \ell}')W^{\cal D}({\bf \ell}_2-{\bf \ell}'') + \text{perm.} \Big] \,,
 \label{eq:Ns}
\end{eqnarray}
where $W$ is a window which depends on the type of distortion in consideration and tells how the primordial CMB bispectrum modes are coupled to the distortion field power spectrum $C^{\cal D \cal D}_\ell$. The effect of the distortions on the bispectrum is to smooth out the acoustic features. These effects for the case of lensing have been shown to be small and can be neglected~\cite{Hanson:2009kg,Cooray:2008xz}. 

In Ref.~\cite{Donzelli:2009ya} the impact of the $1/f$ noise and asymmetric beam on local $f^{local}_{\mathrm{NL}}$ has been found insignificant in the context of a Planck-like experiment.

\section{Other probes of non-Gaussianity in the CMB}
\label{sec:otherprobes}
Although using the full bispectrum is the most sensitive  cubic statistic other statistical methods may be sensitive to different aspects  of non-Gaussianity and,  more importantly, different methods have different systematic effects. Therefore it is important to study various probes. In this section we will discuss some of the methods which have been recently used or developed to test for primordial non-Gaussianities in the CMB.

\subsubsection*{Trispectrum} The four-point function in harmonic space is called trispectrum, which can be written as   
\begin{eqnarray}
\langle a_{l_1 m_1}a_{l_2 m_2}a_{l_3 m_3}a_{l_4 m_4} \rangle
=\sum_{LM}(-1)^M
\left(\begin{array}{ccc} \ell_1 & \ell_2 & L
\\ m_1 & m_2 & m_3 \end{array}\right)  \left(\begin{array}{ccc} \ell_3 & \ell_4 & L 
\\ m_1 & m_2 & m_3 \end{array}\right) 
T^{l_1 l_2}_{l_3 l_4}(L),
\end{eqnarray}
where $T^{l_1 l_2}_{l_3 l_4}(L)$ is the angular averaged trispectrum, 
$L$ is the length of a diagonal that forms triangles 
with $l_1$ and $l_2$ and with $l_3$ and $l_4$,
and the matrix is the Wigner 3-$j$ symbol. The trispectrum contains unconnected part, $T_G$,
\begin{eqnarray}
T_G{}^{l_1 l_2}_{l_3 l_4}(L)
&=&(-1)^{l_1+l_3}\sqrt{(2l_1+1)(2l_3+1)}C_{l_1}C_{l_3}
   \delta_{l_1 l_2}\delta_{l_3 l_4}\delta_{L0} \nonumber \\
&&+(2L+1)C_{l_1}C_{l_2}
   \left[ (-1)^{l_1+l_2+L}
          \delta_{l_1 l_3}\delta_{l_2 l_4}
         +\delta_{l_1 l_4}\delta_{l_2 l_3} \right].
\label{TRIG}
\end{eqnarray}
which comes from the Gaussian part of the perturbations, and the connected part $T_c$ which contains non-Gaussian signatures.
Using permutation symmetry,
one may write the connected part of the trispectrum as
\begin{eqnarray}
T_c{}^{l_1 l_2}_{l_3 l_4}(L)
=P^{l_1 l_2}_{l_3 l_4}(L)
+(2L+1)\sum_{L'}
   \left[ (-1)^{l_2+l_3} \left\{\begin{array}{ccc} \ell_1 & \ell_2 & L 
\\ \ell_4 & \ell_3 & L' \end{array}\right\} 
   P^{l_1 l_3}_{l_2 l_4}(L')
+(-1)^{L+L'} \left\{\begin{array}{ccc} \ell_1 & \ell_2 & L 
\\ \ell_3 & \ell_4 & L' \end{array}\right\} 
   P^{l_1 l_4}_{l_3 l_2}(L') \right ],
\label{TRIC}
\end{eqnarray}
where
\begin{eqnarray}
P^{l_1 l_2}_{l_3 l_4}(L)
=t^{l_1 l_2}_{l_3 l_4}(L)
+(-1)^{2L+l_1+l_2+l_3+l_4}t^{l_2 l_1}_{l_4 l_3}(L)
+(-1)^{L+l_3+l_4}t^{l_1 l_2}_{l_4 l_3}(L)
+(-1)^{L+l_1+l_2}t^{l_2 l_1}_{l_3 l_4}(L).
\label{TRIP}
\end{eqnarray}
Here, the matrix is the Wigner 6-$j$ symbol,
and $t^{l_1 l_2}_{l_3 l_4}(L)$ is called the reduced trispectrum,
which contains all the physical information about non-Gaussianities. For non-Gaussianity of local-type for which
\be
\Phi({\bf x}) = \Phi_L({\bf x}) + \fnl \big[ \Phi^2_L- \langle \Phi^2_L \rangle \big] + g_{\mathrm{NL}} \Phi_L^3\,, 
\ee
both $\fnl$ and $g_{\mathrm{NL}}$ contribute to the trispectrum, but only $\fnl$ contributes to the bispectrum. Tripectrum based estimators for measuring $\fnl$ and $g_{\mathrm{NL}}$ have been developed~\cite{Okamoto:2002ik,Kogo:2006kh,Creminelli:2006gc,Munshi:2009wy,2010arXiv1004.2915R}. For local template, the bispectrum nearly contains all the information on $\fnl$~\cite{Creminelli:2006gc}, however if the non-Gaussianity is seen in bispectrum, trispectrum can serve as a important cross-check. Generically for single field slow-roll models the trispectrum is small and un-observable~\cite{Seery:2006vu} however for more general single field models whenever the equilateral bispectrum is large,  the trispectrum is large as well~\cite{2006PhRvD..74l1301H,2009JCAP...08..008C,2009PhRvD..80d3527A}. For example for equilateral non-Gaussianity Ref.~\cite{Engel:2008fu} study how to tune the model parameters to get large trispectrum and small bispectrum. For multi-field inflation one can construct models that predicts small $\fnl$ but large $g_{\mathrm{NL}}$, for example Ref.~\cite{Byrnes:2006vq} discusses the local form from a multi-field inflation, and briefly mentioned the condition in their class of models to get the large trispectrum and small bispectrum. Joint constraints on both $\fnl$ and $g_{\mathrm{NL}}$ have the potential to add to the specificity of the search for primordial non-Gaussianity. For a given model these two numbers will often be predicted in terms of a single model parameter, such as a coupling constant, see e.g.~\cite{2009PhRvD..80j3520L} for the case of ekyprotic models. Using WMAP 5-year data, the constraints on $g_{\mathrm{NL}}$ using the trispectrum are $-7.4<g_{\mathrm{NL}}/10^5<8.2$ at $2\sigma$~\cite{Smidt:2010sv}.

\subsubsection*{Minkowski Functionals} Minkowski Functionals (MFs) describe morphological properties (such as area, circumference, Euler characteristic) of fluctuating fields~\cite{Mecke_etal1994,Schmalzing:1997aj,Schmalzing:1997uc,Winitzki:1997jj}. For a $d$-dimensional fluctuating field, $f$, the $k$-th Minkowski Functionals of weakly
non-Gaussian fields in, $V_k^{(d)}(\nu)$  for a given threshold $\nu=f/\sigma_0$ can be written as\cite{Matsubara:1994wn,Matsubara2003}
\begin{eqnarray}
\label{eq:MFs_perturb}
\nonumber
V_k^{(d)}(\nu) 
& = & \frac1{(2\pi)^{(k+1)/2}}\frac{\omega_d}{\omega_{d-k}\omega_k}
\left(\frac{\sigma_1}{\sqrt{d}\sigma_0}\right)^ke^{-\nu^2/2}\Bigg\{H_{k-1}(\nu)+
\Big[\frac16S^{(0)}H_{k+2}(\nu)
+\frac{k}3S^{(1)}H_k(\nu)\\
& &+\frac{k(k-1)}6S^{(2)}H_{k-2}(\nu)\Big]\sigma_0+
{\cal O}(\sigma_0^2)\Bigg\},
\end{eqnarray}
 where $\sigma_0\equiv \langle f^2\rangle^{1/2}$
is the variance of the fluctuating field, $H_n(\nu)$ are the Hermite polynomials, $\omega_k\equiv
\pi^{k/2}/{\Gamma(k/2+1)}$, and finally $S^{(i)}$ are the
``skewness parameters'' defined as 
\begin{eqnarray}
S^{(0)}&\equiv& \frac{\langle f^3\rangle}{\sigma_0^4},\\
S^{(1)}&\equiv& -\frac34\frac{\langle
 f^2(\nabla^2f)\rangle}{\sigma_0^2\sigma_1^2},\\ 
S^{(2)}&\equiv& -\frac{3d}{2(d-1)}\frac{\langle (\nabla f)\cdot(\nabla
 f)(\nabla^2f)\rangle}{\sigma_1^4}, 
\end{eqnarray}
which characterize the skewness of fluctuating fields and  their
derivatives. Here $\sigma_i$ characterizes the variance of the fluctuating field and is given by
\begin{eqnarray}
\sigma_i^2=\frac{1}{4\pi} \sum_l (2\ell+1)[\ell(\ell+1)]^i C_\ell^2
\end{eqnarray}
For CMB, for which $d=2$ and $f=\Delta T/T$, the skewness parameters are~\cite{Hikage_Komatsu_etal06}
\begin{eqnarray}
S^{(0)}
&=&
\frac1{4\pi\sigma_0^4}\sum_{l_im_i}
B_{l_1l_2l_3}^{m_1m_2m_3}{\cal G}_{l_1l_2l_3}^{m_1m_2m_3}
W_{l_1}W_{l_2}W_{l_3},\\
S^{(1)}
&=&
\nonumber
\frac3{16\pi\sigma_0^2\sigma_1^2}\sum_{l_im_i}
\frac{l_1(l_1+1)+l_2(l_2+1)+l_3(l_3+1)}3 
B_{l_1l_2l_3}^{m_1m_2m_3}{\cal G}_{l_1l_2l_3}^{m_1m_2m_3}
W_{l_1}W_{l_2}W_{l_3}, \\
S^{(2)}
&=&
\nonumber
\frac3{8\pi\sigma_1^4}\sum_{l_im_i}
\left\{
\frac{\left[l_1(l_1+1)+l_2(l_2+1)-l_3(l_3+1)\right]}3 \right.
\left.\times l_3(l_3+1)+({\rm cyc.})\right\}
B_{l_1l_2l_3}^{m_1m_2m_3}{\cal G}_{l_1l_2l_3}^{m_1m_2m_3} \times W_{l_1}W_{l_2}W_{l_3},
\end{eqnarray}
where $B_{l_1 l_2 l_3}$ is the CMB bispectrum, $W_l$ represents a smoothing kernel which  depends on the experiment beam and ${\cal G}^{m_1 m_2 m_3}_{l_1, l_2, l_3}$ is the usual Gaunt function.

Since MFs can be determined as weighted sum of the bispectrum, they contain less information than the bispectrum. MFs can still be useful because they perhaps suffer from different systematics, though they are less specific to primordial non-Gaussianity since they measure a smaller number of independent bispectrum modes. Also, the bispectrum is defined in Fourier (or harmonic) space  while the MFs are defined in real space. Limits on non-Gaussianity of local-type from the MFs of the WMAP 5-year temperature data are $-70 < \fnl < 91 (2\sigma)$~\cite{Hikage:2008gy}. The MFs from the Planck temperature data should be sensitive to $\fnl\sim 20 $ at $1\sigma$ level~\cite{Hikage:2006fe} in contrast to bispectrum which is sensitive to $\fnl\sim 5$ at $1\sigma$ level. Note that polarization data further improves the sensitivity.

\subsubsection*{Wavelets}
Several studies  have used wavelet representations of the WMAP maps to search for a non-Gaussian signal~\cite{1999A&A...347..409A,1999MNRAS.309..125H,2004ApJ...609...22V,2004A&A...416....9S,2005JASP...15.2470J,2006NewAR..50..880V,2007JFAA...13..495M}. In most of these studies, wavelets were used as a tool for blind searches of non-Gaussian anomalies  in a basis with resolution in both scale and location. However, in some more recent studies, wavelets were tuned to look for non-Gaussianity of a particular type. In the context of searches for primordial non-Gaussianity of local type, wavelet based estimators for $\fnl$ have been built by extracting a signature of local non-Gaussianity that is cubic in the wavelet coefficients from simulations of non-Gaussian skies and searching for this signature in data. This ability to calibrate on a set of simulations makes the wavelet approach very flexible. While not optimal in a least-squared sense, using a wavelet representation can be thought of as a generalized cubic statistic with a different weighting scheme to the optimal bispectrum estimator. Using such estimators  therefore provides a useful exploration of nearly optimal cubic estimators similar to the full bispectrum estimator. Any believable detection of non-Gaussianity should be robust to such changes in the analysis. Similarly, contaminating non-Gaussianity from astrophysical and instrumental systematics will propagate through the analysis in a different way to the bispectrum-based analysis.

There are several constraints on local $\fnl$ using wavelet based estimators. For example, using the COBE data the constraints are $\vert \fnl \vert < 2200 (1 \sigma)$~\cite{2003MNRAS.339.1189C}. Using an estimator based on the skewness of the wavelet coefficients, Mukherjee and Wang constrain the $\fnl$ value for WMAP 1-yr data obtaining $\fnl = 50 \pm 160 (2 \sigma)$~\cite{hotcold}. Using an extension of the previous estimator by combining wavelet coefficients at different contiguous scales, Curto et al. obtain $-8 < \fnl < 111 (2 \sigma)$~\cite{2009MNRAS.393..615C}. Recently, using a generalized third order estimator based on the wavelet coefficients, Curto et al. obtain $-18 < \fnl < 80 (2 \sigma)$~\cite{2009ApJ...706..399C}. 

\subsubsection*{Needlet Bispectrum} Needlets are a family of spherical wavelets which are localized and asymptotically uncorrelated~\cite{MR2237162,MR2253732}. The needlet based statistics as been considered for testing Gaussianity and isotropy (for example see Ref.~\cite{baldi-2008-99,baldi-2009-37,baldi-2009-15,2008EJSta...2..332L, 2009MNRAS.396.1682P,2010MNRAS.402L..34P,2010MNRAS.tmp..504C}. Using the bispectrum of needlet coefficient, the constraints on non-Gaussianity of local-type using WMAP 5-year data yields $\fnl=73\pm 62 (2\sigma)$ ~\cite{Rudjord:2009mh,2010ApJ...708.1321R}. As is clear, the needlet based bispectrum is not as sensitive as the CMB bispectrum discussed in Sec.~\ref{sec:cmbbispectrum}, however again in the event of detection the needlet based methods can be calibrated on simulations and represent a different weighting scheme for handling the sky mask and anisotropic noise. Finally, needlets and wavelets allow for the possibility to analyze spatially localized regions in the sky.

\subsubsection*{Probing non-Gaussianity using Bayesian statistics}
A somewhat different approach to searching for non-Gaussianity is provided by the Bayesian approach. Here, the starting point is an explicit physical or statistical model for the data and the goal is to evaluate the posterior density of the parameters of the model and/or the relative probability of the Gaussianity and non-Gaussianity. 

On large scales, in the Sachs-Wolfe regime, one can simplify the Bayesian approach by modeling directly the temperature anisotropy.  Rocha et al.~(2001)~\cite{2001PhRvD..64f3512R} discuss a Bayesian exploration of a model where each spherical harmonic coefficient is drawn from a non-Gaussian distribution.  In this regime, the simple form  of the non-Gaussian potential for the local model \eqn{phi_bispec} also translates into a simple model for the temperature anisotropy. Ref. ~\cite{2008arXiv0812.1756V} develop several results for it, including an analytical expression of the evidence ratio of the Gaussian and non-Gaussian models.  At the level of current data, this approximation is too restrictive, since most of the information about $\fnl$ is contained near the resolution limit of the experiment, where most of the measured perturbation modes are concentrated.
 
A full implementation of a physical non-Gaussian model must include the effect of Boltzmann transport. In the context of local non-Gaussianity, the model equation \eqn{phi_bispec} suggests that a full Bayesian treatment may be feasible. At the time of writing, no fully Bayesian analysis for local $\fnl$ has been published. The effort has focused on developing approximations to the full Bayesian problem. 

Using a perturbative analysis, Ref.~\cite{2009PhRvD..80j5005E} relate the frequentist bispectrum estimator to moments of the Bayesian posterior distribution. Ref.~\cite{2010arXiv1002.1713E} described approximations to the full Bayesian treatment that simplify the analysis for high signal-to-noise maps and compare these to the full Bayesian treatment for a simple 1-D toy model of non-Gaussian sky where this analysis is feasible.

\begin{table}
\begin{center}
\caption{Summary of constraints on local non-Gaussianity}
\begin{tabular}{cc| l|c|l }
\hline
\hline
Year &data & Method &$f^{local}_{\mathrm{NL}}\pm 2\sigma $ error &\\ \hline
2002 &COBE & Bispectrum sub-optimal& $\vert \fnl \vert < 1500$ & Komatsu et al.~\cite{2002ApJ...566...19K} \\
2003 & MAXIMA & Bispectrum sub-optimal&$\vert \fnl \vert < 1900$ & Santos et al.~\cite{Santos:2002df} \\
2003 & WMAP 1-year & Bispectrum sub-optimal & $39.5\pm97.5$ & E. Komatsu et al.~\cite{nong_wmap}\\
2004 & VSA & Bispectrum sub-optimal&$\fnl < 5400 $ & Smith et al.~\cite{2004MNRAS.352..887S}\\
2005 & WMAP 1-year & Bispectrum sub-optimal-v1 & $ 47 \pm 74$ & Creminelli et al.~\cite{Creminelli_wmap1} \\
2006 & WMAP 3-year & Bispectrum sub-optimal & $30\pm 84$ & Spergel et al.~\cite{wmap_2nd_spergel}\\
2006 & WMAP 3-year & Bispectrum sub-optimal-v1  & $32 \pm 68$ & Creminelli et al.~\cite{creminelli_wmap2}\\
2007 &WMAP 3-year & Bispectrum near-optimal &$87\pm 62$ & Yadav and Wandelt~\cite{YW08} \\
2007 & Boomerang & Minkowski Functionals & $ 110 \pm 910$& De Troia et al.~\cite{2007NewAR..51..250D}\\
2008 &WMAP 3-year & Minkowski Functionals & $ 10.5 \pm 80.5$ & C. Hikage et al.~\cite{Hikage:2008gy} \\
2008 &WMAP 5-year & Bispectrum near-optimal &$51\pm 60$ & Komatsu et al.~\cite{wmap5_cosmol} \\
2008 & ARCHEOPS & Minkowski Functionals & $70^{1075}_{-950}$&Curto et al. 2008~\cite{Curto:2008gg}\\
2009 &WMAP 3-year & Bispectrum optimal &$58\pm 46 $ & Smith et al.~\cite{Smith:2009jr} \\
2009 &WMAP 5-year & Bispectrum optimal &$38\pm 42 $ & Smith et al.~\cite{Smith:2009jr} \\
2009 & WMAP 5-year &Spherical Mexican hat wavelet& $ 31\pm 49$ & Curto, A et. al.~\cite{2009ApJ...706..399C} \\
2009& BOOMERanG & Minkowski Functionals & $ -315 \pm 705 $ &P. Natoli et al.~\cite{Natoli:2009wk} \\
2009 & WMAP 5-year &Skewness power spectrum& $11 \pm 47.4 $ & Smidt, Joseph et al.~\cite{Smidt:2009ir} \\ 
2010 & WMAP 7-year &Bispectrum optimal& $32 \pm 42 $ & Komatsu et al.~\cite{Komatsu:2010fb} \\ 
\end{tabular}
\end{center}
\label{tab:fnl}
\end{table}

\section{Summary}
\label{sec:summary}
The physics of the early universe responsible for generating the seed perturbations in the CMB is not understood. Inflation which is perhaps the the most promising paradigm for generating seed perturbations allow for vast number of inflationary models that are compatible with data based on 2-point statistics like CMB power spectrum. Moreover, the alternatives to inflation such as cyclic models are also compatible with the data. Characterizing the non-Gaussianity in the primordial perturbations has emerged as probe for discriminating between different models of the early universe. Models based on slowly rolling single field produce undetectable amount of non-Gaussianity. Single field models without the slow roll can generate large (detectable with future experiments) non-Gaussianities but 1) can not produce large non-Gaussianity of local type unless inflation started with with excited vacuum state 2) if non-Gaussianity is produced it would naturally be as bispectrum while higher order such as trispectrum can be generated, it requires fine tuning.

The bispectrum of the CMB is one of the most promising tool for connecting the non-Gaussianities in the cosmic microwave background and the models of inflation. Bispectrum-based estimator which saturates Cramer-Rao bound has been developed and well characterised using non-Gaussian Monte-Carlos. Other statistics although not as sensitive to non-Gaussianity as an optimally weighted  bispectrum estimator, do provide independent checks and have different systematics. While Bayesian analysis has been applied in the context of non-Gaussianity analysis, this still appears to be an open area for fruitful investigation.  

Given the importance of detecting primordial non-Gussianity, it is crucial to characterise any non-primordial sources of non-Gaussianities. We describe several sources of non-Gaussianities such as from second order anisotropies after last scattering surface and during recombination. 

With Planck launched and taking data, we look forward to the next few years as an exciting time in the exploration of primordial non-Gaussianity in the cosmic microwave background.

\acknowledgments{A.P.S.Y. gratefully acknowledges support from the IBM Einstein Fellowship.


\begin{thebibliography}{224}
\expandafter\ifx\csname natexlab\endcsname\relax\def\natexlab#1{#1}\fi
\expandafter\ifx\csname bibnamefont\endcsname\relax
  \def\bibnamefont#1{#1}\fi
\expandafter\ifx\csname bibfnamefont\endcsname\relax
  \def\bibfnamefont#1{#1}\fi
\expandafter\ifx\csname citenamefont\endcsname\relax
  \def\citenamefont#1{#1}\fi
\expandafter\ifx\csname url\endcsname\relax
  \def\url#1{\texttt{#1}}\fi
\expandafter\ifx\csname urlprefix\endcsname\relax\def\urlprefix{URL }\fi
\providecommand{\bibinfo}[2]{#2}
\providecommand{\eprint}[2][]{\url{#2}}

\bibitem[{\citenamefont{Guth}(1981)}]{Guth81}
\bibinfo{author}{\bibfnamefont{A.~H.} \bibnamefont{Guth}},
  \bibinfo{journal}{Phys. Rev. D} \textbf{\bibinfo{volume}{23}},
  \bibinfo{pages}{347} (\bibinfo{year}{1981}).

\bibitem[{\citenamefont{Sato}(1981)}]{Sato81}
\bibinfo{author}{\bibfnamefont{K.}~\bibnamefont{Sato}}, \bibinfo{journal}{Phys.
  Lett.} \textbf{\bibinfo{volume}{{99B}}}, \bibinfo{pages}{{66}}
  (\bibinfo{year}{1981}).

\bibitem[{\citenamefont{Linde}(1982)}]{Linde81}
\bibinfo{author}{\bibfnamefont{A.~D.} \bibnamefont{Linde}},
  \bibinfo{journal}{Phys. Lett.} \textbf{\bibinfo{volume}{B108}},
  \bibinfo{pages}{389} (\bibinfo{year}{1982}).

\bibitem[{\citenamefont{Albrecht and Steinhardt}(1982)}]{Albrecht_Steinhardt82}
\bibinfo{author}{\bibfnamefont{A.}~\bibnamefont{Albrecht}} \bibnamefont{and}
  \bibinfo{author}{\bibfnamefont{P.~J.} \bibnamefont{Steinhardt}},
  \bibinfo{journal}{Phys. Rev. Lett.} \textbf{\bibinfo{volume}{48}},
  \bibinfo{pages}{1220} (\bibinfo{year}{1982}).

\bibitem[{\citenamefont{Guth and Pi}(1982)}]{Guth_Pi82}
\bibinfo{author}{\bibfnamefont{A.~H.} \bibnamefont{Guth}} \bibnamefont{and}
  \bibinfo{author}{\bibfnamefont{S.~Y.} \bibnamefont{Pi}},
  \bibinfo{journal}{Phys. Rev. Lett.} \textbf{\bibinfo{volume}{49}},
  \bibinfo{pages}{1110} (\bibinfo{year}{1982}).

\bibitem[{\citenamefont{Starobinsky}(1982)}]{Starobinsky82}
\bibinfo{author}{\bibfnamefont{A.~A.} \bibnamefont{Starobinsky}},
  \bibinfo{journal}{Phys. Lett. B.} \textbf{\bibinfo{volume}{117}},
  \bibinfo{pages}{175} (\bibinfo{year}{1982}).

\bibitem[{\citenamefont{Hawking}(1982)}]{Hawking82}
\bibinfo{author}{\bibfnamefont{S.~W.} \bibnamefont{Hawking}},
  \bibinfo{journal}{Phys. Lett. B.} \textbf{\bibinfo{volume}{115}},
  \bibinfo{pages}{295} (\bibinfo{year}{1982}).

\bibitem[{\citenamefont{Bardeen et~al.}(1983)\citenamefont{Bardeen, Steinhardt,
  and Turner}}]{Bardeen_etal83}
\bibinfo{author}{\bibfnamefont{J.~M.} \bibnamefont{Bardeen}},
  \bibinfo{author}{\bibfnamefont{P.~J.} \bibnamefont{Steinhardt}},
  \bibnamefont{and} \bibinfo{author}{\bibfnamefont{M.~S.}
  \bibnamefont{Turner}}, \bibinfo{journal}{Phys. Rev. D.}
  \textbf{\bibinfo{volume}{28}}, \bibinfo{pages}{679} (\bibinfo{year}{1983}).

\bibitem[{\citenamefont{Mukhanov
  et~al.}(1992{\natexlab{a}})\citenamefont{Mukhanov, Feldman, and
  Brandenberger}}]{Mukhanov_et92}
\bibinfo{author}{\bibfnamefont{V.~F.} \bibnamefont{Mukhanov}},
  \bibinfo{author}{\bibfnamefont{H.~A.} \bibnamefont{Feldman}},
  \bibnamefont{and} \bibinfo{author}{\bibfnamefont{R.~H.}
  \bibnamefont{Brandenberger}}, \bibinfo{journal}{Phys. Rep.}
  \textbf{\bibinfo{volume}{215}}, \bibinfo{pages}{203}
  (\bibinfo{year}{1992}{\natexlab{a}}).

\bibitem[{\citenamefont{{Salopek} and {Bond}}(1990)}]{Salopek_Bond90}
\bibinfo{author}{\bibfnamefont{D.~S.} \bibnamefont{{Salopek}}}
  \bibnamefont{and} \bibinfo{author}{\bibfnamefont{J.~R.}
  \bibnamefont{{Bond}}}, \bibinfo{journal}{Phys. Rev. D.}
  \textbf{\bibinfo{volume}{42}}, \bibinfo{pages}{3936} (\bibinfo{year}{1990}).

\bibitem[{\citenamefont{{Salopek} and {Bond}}(1991)}]{Salopek_Bond91}
\bibinfo{author}{\bibfnamefont{D.~S.} \bibnamefont{{Salopek}}}
  \bibnamefont{and} \bibinfo{author}{\bibfnamefont{J.~R.}
  \bibnamefont{{Bond}}}, \bibinfo{journal}{Phys. Rev. D.}
  \textbf{\bibinfo{volume}{43}}, \bibinfo{pages}{1005} (\bibinfo{year}{1991}).

\bibitem[{\citenamefont{Falk et~al.}(1993)\citenamefont{Falk, Rangarajan, and
  Srednicki}}]{Falk_et93}
\bibinfo{author}{\bibfnamefont{T.}~\bibnamefont{Falk}},
  \bibinfo{author}{\bibfnamefont{R.}~\bibnamefont{Rangarajan}},
  \bibnamefont{and}
  \bibinfo{author}{\bibfnamefont{M.}~\bibnamefont{Srednicki}},
  \bibinfo{journal}{ApJ.} \textbf{\bibinfo{volume}{403}}, \bibinfo{pages}{L1}
  (\bibinfo{year}{1993}).

\bibitem[{\citenamefont{Gangui et~al.}(1994)\citenamefont{Gangui, Lucchin,
  Matarrese, and Mollerach}}]{Gangui_etal94}
\bibinfo{author}{\bibfnamefont{A.}~\bibnamefont{Gangui}},
  \bibinfo{author}{\bibfnamefont{F.}~\bibnamefont{Lucchin}},
  \bibinfo{author}{\bibfnamefont{S.}~\bibnamefont{Matarrese}},
  \bibnamefont{and}
  \bibinfo{author}{\bibfnamefont{S.}~\bibnamefont{Mollerach}},
  \bibinfo{journal}{ApJ.} \textbf{\bibinfo{volume}{430}}, \bibinfo{pages}{447}
  (\bibinfo{year}{1994}).

\bibitem[{\citenamefont{Acquaviva et~al.}(2003)\citenamefont{Acquaviva,
  Bartolo, Matarrese, and Riotto}}]{Acquaviva02}
\bibinfo{author}{\bibfnamefont{V.}~\bibnamefont{Acquaviva}},
  \bibinfo{author}{\bibfnamefont{N.}~\bibnamefont{Bartolo}},
  \bibinfo{author}{\bibfnamefont{S.}~\bibnamefont{Matarrese}},
  \bibnamefont{and} \bibinfo{author}{\bibfnamefont{A.}~\bibnamefont{Riotto}},
  \bibinfo{journal}{Nucl. Phys. B} \textbf{\bibinfo{volume}{667}},
  \bibinfo{pages}{119} (\bibinfo{year}{2003}).

\bibitem[{\citenamefont{Maldacena}(2003{\natexlab{a}})}]{Maldacena03}
\bibinfo{author}{\bibfnamefont{J.}~\bibnamefont{Maldacena}},
  \bibinfo{journal}{J. High Energy Phys.} \textbf{\bibinfo{volume}{05}},
  \bibinfo{pages}{013} (\bibinfo{year}{2003}{\natexlab{a}}).

\bibitem[{\citenamefont{{Bartolo}
  et~al.}(2004{\natexlab{a}})\citenamefont{{Bartolo}, {Komatsu}, {Matarrese},
  and {Riotto}}}]{BKMR_04}
\bibinfo{author}{\bibfnamefont{N.}~\bibnamefont{{Bartolo}}},
  \bibinfo{author}{\bibfnamefont{E.}~\bibnamefont{{Komatsu}}},
  \bibinfo{author}{\bibfnamefont{S.}~\bibnamefont{{Matarrese}}},
  \bibnamefont{and} \bibinfo{author}{\bibfnamefont{A.}~\bibnamefont{{Riotto}}},
  \bibinfo{journal}{Physics Reports} \textbf{\bibinfo{volume}{402}},
  \bibinfo{pages}{103} (\bibinfo{year}{2004}{\natexlab{a}}).

\bibitem[{\citenamefont{Komatsu et~al.}(2005)\citenamefont{Komatsu, Spergel,
  and Wandelt}}]{KSW05}
\bibinfo{author}{\bibfnamefont{E.~N.} \bibnamefont{Komatsu}},
  \bibinfo{author}{\bibfnamefont{D.~N.} \bibnamefont{Spergel}},
  \bibnamefont{and} \bibinfo{author}{\bibfnamefont{B.~D.}
  \bibnamefont{Wandelt}}, \bibinfo{journal}{ApJ.}
  \textbf{\bibinfo{volume}{634}}, \bibinfo{pages}{14} (\bibinfo{year}{2005}).

\bibitem[{\citenamefont{{Cabella} et~al.}(2006)\citenamefont{{Cabella},
  {Hansen}, {Liguori}, {Marinucci}, {Matarrese}, {Moscardini}, and
  {Vittorio}}}]{IntegratedBispectrum}
\bibinfo{author}{\bibfnamefont{P.}~\bibnamefont{{Cabella}}},
  \bibinfo{author}{\bibfnamefont{F.~K.} \bibnamefont{{Hansen}}},
  \bibinfo{author}{\bibfnamefont{M.}~\bibnamefont{{Liguori}}},
  \bibinfo{author}{\bibfnamefont{D.}~\bibnamefont{{Marinucci}}},
  \bibinfo{author}{\bibfnamefont{S.}~\bibnamefont{{Matarrese}}},
  \bibinfo{author}{\bibfnamefont{L.}~\bibnamefont{{Moscardini}}},
  \bibnamefont{and}
  \bibinfo{author}{\bibfnamefont{N.}~\bibnamefont{{Vittorio}}},
  \bibinfo{journal}{Mon. Not. R. Astron. Soc.} \textbf{\bibinfo{volume}{369}},
  \bibinfo{pages}{819} (\bibinfo{year}{2006}), \eprint{astro-ph/0512112}.

\bibitem[{\citenamefont{{Smith} and
  {Zaldarriaga}}(2006{\natexlab{a}})}]{Smith_Zaldarriaga06}
\bibinfo{author}{\bibfnamefont{K.~M.} \bibnamefont{{Smith}}} \bibnamefont{and}
  \bibinfo{author}{\bibfnamefont{M.}~\bibnamefont{{Zaldarriaga}}},
  \bibinfo{journal}{ArXiv Astrophysics e-prints}
  (\bibinfo{year}{2006}{\natexlab{a}}), \eprint{astro-ph/0612571}.

\bibitem[{\citenamefont{{Yadav} et~al.}(2007)\citenamefont{{Yadav}, {Komatsu},
  and {Wandelt}}}]{YKW07}
\bibinfo{author}{\bibfnamefont{A.~P.~S.} \bibnamefont{{Yadav}}},
  \bibinfo{author}{\bibfnamefont{E.}~\bibnamefont{{Komatsu}}},
  \bibnamefont{and} \bibinfo{author}{\bibfnamefont{B.~D.}
  \bibnamefont{{Wandelt}}}, \bibinfo{journal}{Astrophys. J.}
  \textbf{\bibinfo{volume}{664}}, \bibinfo{pages}{680} (\bibinfo{year}{2007}),
  \eprint{arXiv:astro-ph/0701921}.

\bibitem[{\citenamefont{{Yadav} et~al.}(2008)\citenamefont{{Yadav}, {Komatsu},
  {Wandelt}, {Liguori}, {Hansen}, and {Matarrese}}}]{Yadav_etal08a}
\bibinfo{author}{\bibfnamefont{A.}~\bibnamefont{{Yadav}}},
  \bibinfo{author}{\bibfnamefont{E.}~\bibnamefont{{Komatsu}}},
  \bibinfo{author}{\bibfnamefont{B.}~\bibnamefont{{Wandelt}}},
  \bibinfo{author}{\bibfnamefont{M.}~\bibnamefont{{Liguori}}},
  \bibinfo{author}{\bibfnamefont{F.~K.} \bibnamefont{{Hansen}}},
  \bibnamefont{and}
  \bibinfo{author}{\bibfnamefont{S.}~\bibnamefont{{Matarrese}}},
  \bibinfo{journal}{Astrophys. J.} \textbf{\bibinfo{volume}{788}},
  \bibinfo{pages}{578} (\bibinfo{year}{2008}),
  \eprint{arXiv:astro-ph/0711.4933}.

\bibitem[{\citenamefont{Komatsu et~al.}(2002)\citenamefont{Komatsu, Wandelt,
  Spergel, Banday, and Gorski}}]{nong_bdw}
\bibinfo{author}{\bibfnamefont{E.~N.} \bibnamefont{Komatsu}},
  \bibinfo{author}{\bibfnamefont{B.~D.} \bibnamefont{Wandelt}},
  \bibinfo{author}{\bibfnamefont{D.~N.} \bibnamefont{Spergel}},
  \bibinfo{author}{\bibfnamefont{A.~J.} \bibnamefont{Banday}},
  \bibnamefont{and} \bibinfo{author}{\bibfnamefont{K.~M.}
  \bibnamefont{Gorski}}, \bibinfo{journal}{Astrophys. J.}
  \textbf{\bibinfo{volume}{566}}, \bibinfo{pages}{19} (\bibinfo{year}{2002}).

\bibitem[{\citenamefont{Komatsu et~al.}(2003)\citenamefont{Komatsu, Kogut,
  Nolta, Bennett, Halpern, Hinshaw, Jarosik, Limon, Meyer, Page
  et~al.}}]{nong_wmap}
\bibinfo{author}{\bibfnamefont{E.}~\bibnamefont{Komatsu}},
  \bibinfo{author}{\bibfnamefont{A.}~\bibnamefont{Kogut}},
  \bibinfo{author}{\bibfnamefont{M.~N.} \bibnamefont{Nolta}},
  \bibinfo{author}{\bibfnamefont{C.~L.} \bibnamefont{Bennett}},
  \bibinfo{author}{\bibfnamefont{M.}~\bibnamefont{Halpern}},
  \bibinfo{author}{\bibfnamefont{G.}~\bibnamefont{Hinshaw}},
  \bibinfo{author}{\bibfnamefont{N.}~\bibnamefont{Jarosik}},
  \bibinfo{author}{\bibfnamefont{M.}~\bibnamefont{Limon}},
  \bibinfo{author}{\bibfnamefont{S.~S.} \bibnamefont{Meyer}},
  \bibinfo{author}{\bibfnamefont{L.}~\bibnamefont{Page}}, \bibnamefont{et~al.},
  \bibinfo{journal}{ApJ.} \textbf{\bibinfo{volume}{148}}, \bibinfo{pages}{119}
  (\bibinfo{year}{2003}).

\bibitem[{\citenamefont{{Spergel} et~al.}(2007)\citenamefont{{Spergel}, {Bean},
  {Dore'}, {Nolta}, {Bennett}, {Hinshaw}, {Jarosik}, {Komatsu}, {Page},
  {Peiris} et~al.}}]{wmap_2nd_spergel}
\bibinfo{author}{\bibfnamefont{D.~N.} \bibnamefont{{Spergel}}},
  \bibinfo{author}{\bibfnamefont{R.}~\bibnamefont{{Bean}}},
  \bibinfo{author}{\bibfnamefont{O.}~\bibnamefont{{Dore'}}},
  \bibinfo{author}{\bibfnamefont{M.~R.} \bibnamefont{{Nolta}}},
  \bibinfo{author}{\bibfnamefont{C.~L.} \bibnamefont{{Bennett}}},
  \bibinfo{author}{\bibfnamefont{G.}~\bibnamefont{{Hinshaw}}},
  \bibinfo{author}{\bibfnamefont{N.}~\bibnamefont{{Jarosik}}},
  \bibinfo{author}{\bibfnamefont{E.}~\bibnamefont{{Komatsu}}},
  \bibinfo{author}{\bibfnamefont{L.}~\bibnamefont{{Page}}},
  \bibinfo{author}{\bibfnamefont{H.~V.} \bibnamefont{{Peiris}}},
  \bibnamefont{et~al.}, \bibinfo{journal}{ApJS} \textbf{\bibinfo{volume}{170}},
  \bibinfo{pages}{377} (\bibinfo{year}{2007}).

\bibitem[{\citenamefont{{Creminelli} et~al.}(2006)\citenamefont{{Creminelli},
  {Nicolis}, {Senatore}, {Tegmark}, and {Zaldarriaga}}}]{Creminelli_wmap1}
\bibinfo{author}{\bibfnamefont{P.}~\bibnamefont{{Creminelli}}},
  \bibinfo{author}{\bibfnamefont{A.}~\bibnamefont{{Nicolis}}},
  \bibinfo{author}{\bibfnamefont{L.}~\bibnamefont{{Senatore}}},
  \bibinfo{author}{\bibfnamefont{M.}~\bibnamefont{{Tegmark}}},
  \bibnamefont{and}
  \bibinfo{author}{\bibfnamefont{M.}~\bibnamefont{{Zaldarriaga}}},
  \bibinfo{journal}{Journal of Cosmology and Astro-Particle Physics}
  \textbf{\bibinfo{volume}{5}}, \bibinfo{pages}{4} (\bibinfo{year}{2006}),
  \eprint{astro-ph/0509029}.

\bibitem[{\citenamefont{{Creminelli}
  et~al.}(2007{\natexlab{a}})\citenamefont{{Creminelli}, {Senatore},
  {Zaldarriaga}, and {Tegmark}}}]{creminelli_wmap2}
\bibinfo{author}{\bibfnamefont{P.}~\bibnamefont{{Creminelli}}},
  \bibinfo{author}{\bibfnamefont{L.}~\bibnamefont{{Senatore}}},
  \bibinfo{author}{\bibfnamefont{M.}~\bibnamefont{{Zaldarriaga}}},
  \bibnamefont{and}
  \bibinfo{author}{\bibfnamefont{M.}~\bibnamefont{{Tegmark}}},
  \bibinfo{journal}{Journal of Cosmology and Astro-Particle Physics}
  \textbf{\bibinfo{volume}{3}}, \bibinfo{pages}{5}
  (\bibinfo{year}{2007}{\natexlab{a}}), \eprint{arXiv:astro-ph/0610600}.

\bibitem[{\citenamefont{{Chen} and {Szapudi}}(2006)}]{szapudi06}
\bibinfo{author}{\bibfnamefont{G.}~\bibnamefont{{Chen}}} \bibnamefont{and}
  \bibinfo{author}{\bibfnamefont{I.}~\bibnamefont{{Szapudi}}},
  \bibinfo{journal}{Astrophys. J. Lett.} \textbf{\bibinfo{volume}{647}},
  \bibinfo{pages}{L87} (\bibinfo{year}{2006}), \eprint{astro-ph/0606394}.

\bibitem[{\citenamefont{{Yadav} and {Wandelt}}(2008)}]{YW08}
\bibinfo{author}{\bibfnamefont{A.~P.~S.} \bibnamefont{{Yadav}}}
  \bibnamefont{and} \bibinfo{author}{\bibfnamefont{B.~D.}
  \bibnamefont{{Wandelt}}}, \bibinfo{journal}{Physical Review Letters}
  \textbf{\bibinfo{volume}{100}} (\bibinfo{year}{2008}),
  \eprint{arXiv:astro-ph/0712.1148}.

\bibitem[{\citenamefont{Mukherjee and Wang}(2004)}]{hotcold}
\bibinfo{author}{\bibfnamefont{P.}~\bibnamefont{Mukherjee}} \bibnamefont{and}
  \bibinfo{author}{\bibfnamefont{Y.}~\bibnamefont{Wang}},
  \bibinfo{journal}{ApJ.} \textbf{\bibinfo{volume}{613}}, \bibinfo{pages}{51}
  (\bibinfo{year}{2004}).

\bibitem[{\citenamefont{Larson and Wandelt}(2004)}]{larsonwandelt}
\bibinfo{author}{\bibfnamefont{D.~L.} \bibnamefont{Larson}} \bibnamefont{and}
  \bibinfo{author}{\bibfnamefont{B.~D.} \bibnamefont{Wandelt}},
  \bibinfo{journal}{Astrophys. J.} \textbf{\bibinfo{volume}{613}},
  \bibinfo{pages}{L85} (\bibinfo{year}{2004}).

\bibitem[{\citenamefont{Vielva et~al.}(2004)\citenamefont{Vielva, Gonzalez,
  Barreiro, Sanz, and Cayon}}]{nong_cmb1}
\bibinfo{author}{\bibfnamefont{P.}~\bibnamefont{Vielva}},
  \bibinfo{author}{\bibfnamefont{E.~M.} \bibnamefont{Gonzalez}},
  \bibinfo{author}{\bibfnamefont{R.~B.} \bibnamefont{Barreiro}},
  \bibinfo{author}{\bibfnamefont{J.~L.} \bibnamefont{Sanz}}, \bibnamefont{and}
  \bibinfo{author}{\bibfnamefont{L.}~\bibnamefont{Cayon}},
  \bibinfo{journal}{Astrophys. J} \textbf{\bibinfo{volume}{609}},
  \bibinfo{pages}{22} (\bibinfo{year}{2004}).

\bibitem[{\citenamefont{Chiang et~al.}(2003)\citenamefont{Chiang, P.~D,
  Verkhodanov, and Way}}]{nong_cmb2}
\bibinfo{author}{\bibfnamefont{L.~Y.} \bibnamefont{Chiang}},
  \bibinfo{author}{\bibfnamefont{N.}~\bibnamefont{P.~D}},
  \bibinfo{author}{\bibfnamefont{O.~V.} \bibnamefont{Verkhodanov}},
  \bibnamefont{and} \bibinfo{author}{\bibfnamefont{M.~J.} \bibnamefont{Way}},
  \bibinfo{journal}{Astrophys. J} \textbf{\bibinfo{volume}{590}},
  \bibinfo{pages}{L65} (\bibinfo{year}{2003}).

\bibitem[{\citenamefont{Chiang et~al.}(2004)\citenamefont{Chiang, Naselsky, and
  Coles}}]{nong_cmb3}
\bibinfo{author}{\bibfnamefont{L.~Y.} \bibnamefont{Chiang}},
  \bibinfo{author}{\bibfnamefont{P.~D.} \bibnamefont{Naselsky}},
  \bibnamefont{and} \bibinfo{author}{\bibfnamefont{P.}~\bibnamefont{Coles}},
  \bibinfo{journal}{Astrophys. J} \textbf{\bibinfo{volume}{602}},
  \bibinfo{pages}{1} (\bibinfo{year}{2004}).

\bibitem[{\citenamefont{Babich and Zaldarriaga}(2004)}]{BZ04}
\bibinfo{author}{\bibfnamefont{D.}~\bibnamefont{Babich}} \bibnamefont{and}
  \bibinfo{author}{\bibfnamefont{M.}~\bibnamefont{Zaldarriaga}},
  \bibinfo{journal}{Phys. Rev. D.} \textbf{\bibinfo{volume}{70}},
  \bibinfo{pages}{083005} (\bibinfo{year}{2004}), \eprint{astro-ph/040845}.

\bibitem[{\citenamefont{{Yadav} and {Wandelt}}(2005)}]{YW05}
\bibinfo{author}{\bibfnamefont{A.~P.} \bibnamefont{{Yadav}}} \bibnamefont{and}
  \bibinfo{author}{\bibfnamefont{B.~D.} \bibnamefont{{Wandelt}}},
  \bibinfo{journal}{Phys. Rev. D.} \textbf{\bibinfo{volume}{71}},
  \bibinfo{pages}{123004} (\bibinfo{year}{2005}),
  \eprint{arXiv:astro-ph/0505386}.

\bibitem[{\citenamefont{Kovac et~al.}(2002)\citenamefont{Kovac, Leitch, Pryke,
  Carlstrom, Halverson, and Holzapfel}}]{dasi_pol_02}
\bibinfo{author}{\bibfnamefont{J.~M.} \bibnamefont{Kovac}},
  \bibinfo{author}{\bibfnamefont{E.~M.} \bibnamefont{Leitch}},
  \bibinfo{author}{\bibfnamefont{C.}~\bibnamefont{Pryke}},
  \bibinfo{author}{\bibfnamefont{J.~E.} \bibnamefont{Carlstrom}},
  \bibinfo{author}{\bibfnamefont{N.~W.} \bibnamefont{Halverson}},
  \bibnamefont{and} \bibinfo{author}{\bibfnamefont{W.~L.}
  \bibnamefont{Holzapfel}}, \bibinfo{journal}{Nature}
  \textbf{\bibinfo{volume}{420}}, \bibinfo{pages}{772} (\bibinfo{year}{2002}).

\bibitem[{\citenamefont{{Kogut} et~al.}(2003)\citenamefont{{Kogut}, {Spergel},
  {Barnes}, {Bennett}, {Halpern}, {Hinshaw}, {Jarosik}, {Limon}, {Meyer},
  {Page} et~al.}}]{wmap_1st_pol}
\bibinfo{author}{\bibfnamefont{A.}~\bibnamefont{{Kogut}}},
  \bibinfo{author}{\bibfnamefont{D.~N.} \bibnamefont{{Spergel}}},
  \bibinfo{author}{\bibfnamefont{C.}~\bibnamefont{{Barnes}}},
  \bibinfo{author}{\bibfnamefont{C.~L.} \bibnamefont{{Bennett}}},
  \bibinfo{author}{\bibfnamefont{M.}~\bibnamefont{{Halpern}}},
  \bibinfo{author}{\bibfnamefont{G.}~\bibnamefont{{Hinshaw}}},
  \bibinfo{author}{\bibfnamefont{N.}~\bibnamefont{{Jarosik}}},
  \bibinfo{author}{\bibfnamefont{M.}~\bibnamefont{{Limon}}},
  \bibinfo{author}{\bibfnamefont{S.~S.} \bibnamefont{{Meyer}}},
  \bibinfo{author}{\bibfnamefont{L.}~\bibnamefont{{Page}}},
  \bibnamefont{et~al.}, \bibinfo{journal}{Astrophys. J. Suppl. S.}
  \textbf{\bibinfo{volume}{148}}, \bibinfo{pages}{161} (\bibinfo{year}{2003}),
  \eprint{astro-ph/0302213}.

\bibitem[{\citenamefont{{Page} et~al.}(2007)\citenamefont{{Page}, {Hinshaw},
  {Komatsu}, {Nolta}, {Spergel}, {Bennett}, {Barnes}, {Bean}, {Dore'},
  {Halpern} et~al.}}]{wmap_2nd_pol}
\bibinfo{author}{\bibfnamefont{L.}~\bibnamefont{{Page}}},
  \bibinfo{author}{\bibfnamefont{G.}~\bibnamefont{{Hinshaw}}},
  \bibinfo{author}{\bibfnamefont{E.}~\bibnamefont{{Komatsu}}},
  \bibinfo{author}{\bibfnamefont{M.~R.} \bibnamefont{{Nolta}}},
  \bibinfo{author}{\bibfnamefont{D.~N.} \bibnamefont{{Spergel}}},
  \bibinfo{author}{\bibfnamefont{C.~L.} \bibnamefont{{Bennett}}},
  \bibinfo{author}{\bibfnamefont{C.}~\bibnamefont{{Barnes}}},
  \bibinfo{author}{\bibfnamefont{R.}~\bibnamefont{{Bean}}},
  \bibinfo{author}{\bibfnamefont{O.}~\bibnamefont{{Dore'}}},
  \bibinfo{author}{\bibfnamefont{M.}~\bibnamefont{{Halpern}}},
  \bibnamefont{et~al.}, \bibinfo{journal}{APJS} \textbf{\bibinfo{volume}{170}},
  \bibinfo{pages}{335} (\bibinfo{year}{2007}), \eprint{astro-ph/0603450}.

\bibitem[{\citenamefont{{Montroy} et~al.}(2006)\citenamefont{{Montroy}, {Ade},
  {Bock}, {Bond}, {Borrill}, {Boscaleri}, {Cabella}, {Contaldi}, {Crill}, {de
  Bernardis} et~al.}}]{boom_ee}
\bibinfo{author}{\bibfnamefont{T.~E.} \bibnamefont{{Montroy}}},
  \bibinfo{author}{\bibfnamefont{P.~A.~R.} \bibnamefont{{Ade}}},
  \bibinfo{author}{\bibfnamefont{J.~J.} \bibnamefont{{Bock}}},
  \bibinfo{author}{\bibfnamefont{J.~R.} \bibnamefont{{Bond}}},
  \bibinfo{author}{\bibfnamefont{J.}~\bibnamefont{{Borrill}}},
  \bibinfo{author}{\bibfnamefont{A.}~\bibnamefont{{Boscaleri}}},
  \bibinfo{author}{\bibfnamefont{P.}~\bibnamefont{{Cabella}}},
  \bibinfo{author}{\bibfnamefont{C.~R.} \bibnamefont{{Contaldi}}},
  \bibinfo{author}{\bibfnamefont{B.~P.} \bibnamefont{{Crill}}},
  \bibinfo{author}{\bibfnamefont{P.}~\bibnamefont{{de Bernardis}}},
  \bibnamefont{et~al.}, \bibinfo{journal}{Astrophys. J.}
  \textbf{\bibinfo{volume}{647}}, \bibinfo{pages}{813} (\bibinfo{year}{2006}),
  \eprint{astro-ph/0507514}.

\bibitem[{\citenamefont{{J. ~Khoury et al.}}(2001)}]{Khoury_et_01}
\bibinfo{author}{\bibnamefont{{J. ~Khoury et al.}}}, \bibinfo{journal}{Phys.
  Rev. D.} \textbf{\bibinfo{volume}{64}}, \bibinfo{pages}{123522}
  (\bibinfo{year}{2001}).

\bibitem[{\citenamefont{Steinhardt and
  Turok}(2002{\natexlab{a}})}]{Steinhardt_Turok_02a}
\bibinfo{author}{\bibfnamefont{P.}~\bibnamefont{Steinhardt}} \bibnamefont{and}
  \bibinfo{author}{\bibfnamefont{N.}~\bibnamefont{Turok}},
  \bibinfo{journal}{Science} \textbf{\bibinfo{volume}{296}},
  \bibinfo{pages}{1436} (\bibinfo{year}{2002}{\natexlab{a}}).

\bibitem[{\citenamefont{Steinhardt and
  Turok}(2002{\natexlab{b}})}]{Steinhardt_Turok_02b}
\bibinfo{author}{\bibfnamefont{P.}~\bibnamefont{Steinhardt}} \bibnamefont{and}
  \bibinfo{author}{\bibfnamefont{N.}~\bibnamefont{Turok}},
  \bibinfo{journal}{Phys. Rev. D.} \textbf{\bibinfo{volume}{65}},
  \bibinfo{pages}{126003} (\bibinfo{year}{2002}{\natexlab{b}}).

\bibitem[{\citenamefont{{Baumann}}(2009)}]{2009arXiv0907.5424B}
\bibinfo{author}{\bibfnamefont{D.}~\bibnamefont{{Baumann}}},
  \bibinfo{journal}{ArXiv e-prints}  (\bibinfo{year}{2009}),
  \eprint{0907.5424}.

\bibitem[{\citenamefont{Mukhanov
  et~al.}(1992{\natexlab{b}})\citenamefont{Mukhanov, Feldman, A., and
  Brandenberger}}]{MFB92}
\bibinfo{author}{\bibfnamefont{V.~F.} \bibnamefont{Mukhanov}},
  \bibinfo{author}{\bibnamefont{Feldman}},
  \bibinfo{author}{\bibfnamefont{H.}~\bibnamefont{A.}}, \bibnamefont{and}
  \bibinfo{author}{\bibnamefont{Brandenberger}}, \bibinfo{journal}{Phys. Rep.}
  \textbf{\bibinfo{volume}{215}}, \bibinfo{pages}{203}
  (\bibinfo{year}{1992}{\natexlab{b}}).

\bibitem[{\citenamefont{{Bardeen}}(1980)}]{Bardeen80}
\bibinfo{author}{\bibfnamefont{J.~M.} \bibnamefont{{Bardeen}}},
  \bibinfo{journal}{Phys. Rev. D.}  (\bibinfo{year}{1980}).

\bibitem[{\citenamefont{Kodama and Sasaki}(1984)}]{KS84}
\bibinfo{author}{\bibfnamefont{H.}~\bibnamefont{Kodama}} \bibnamefont{and}
  \bibinfo{author}{\bibfnamefont{M.}~\bibnamefont{Sasaki}},
  \bibinfo{journal}{Prog. Theor. Phys. Suppl.}  (\bibinfo{year}{1984}).

\bibitem[{\citenamefont{Sachs and Wolfe}(1967)}]{SW}
\bibinfo{author}{\bibfnamefont{R.~K.} \bibnamefont{Sachs}} \bibnamefont{and}
  \bibinfo{author}{\bibfnamefont{A.~M.} \bibnamefont{Wolfe}},
  \bibinfo{journal}{ApJ} \textbf{\bibinfo{volume}{147}}, \bibinfo{pages}{73}
  (\bibinfo{year}{1967}).

\bibitem[{\citenamefont{Bond and Efstathiou}(1987)}]{Bond_Efstathiou87}
\bibinfo{author}{\bibfnamefont{J.~R.} \bibnamefont{Bond}} \bibnamefont{and}
  \bibinfo{author}{\bibfnamefont{G.}~\bibnamefont{Efstathiou}},
  \bibinfo{journal}{Mon. Not. Roy. Astron. Soc.}  (\bibinfo{year}{1987}).

\bibitem[{\citenamefont{Peebles and Yu}(1970)}]{Peebles_Yu70}
\bibinfo{author}{\bibfnamefont{P.~J.~E.} \bibnamefont{Peebles}}
  \bibnamefont{and} \bibinfo{author}{\bibfnamefont{J.~T.} \bibnamefont{Yu}},
  \bibinfo{journal}{Astrophys. J.} \textbf{\bibinfo{volume}{162}}
  (\bibinfo{year}{1970}).

\bibitem[{\citenamefont{Seljak and Zaldarriaga}(1996)}]{cmbfast}
\bibinfo{author}{\bibfnamefont{U.}~\bibnamefont{Seljak}} \bibnamefont{and}
  \bibinfo{author}{\bibfnamefont{M.}~\bibnamefont{Zaldarriaga}},
  \bibinfo{journal}{ApJ.} \textbf{\bibinfo{volume}{469}}, \bibinfo{pages}{437}
  (\bibinfo{year}{1996}).

\bibitem[{\citenamefont{{Komatsu} et~al.}(2008)\citenamefont{{Komatsu},
  {Dunkley}, {Nolta}, {Bennett}, {Gold}, {Hinshaw}, {Jarosik}, {Larson},
  {Limon}, {Page} et~al.}}]{wmap5_cosmol}
\bibinfo{author}{\bibfnamefont{E.}~\bibnamefont{{Komatsu}}},
  \bibinfo{author}{\bibfnamefont{J.}~\bibnamefont{{Dunkley}}},
  \bibinfo{author}{\bibfnamefont{M.~R.} \bibnamefont{{Nolta}}},
  \bibinfo{author}{\bibfnamefont{C.~L.} \bibnamefont{{Bennett}}},
  \bibinfo{author}{\bibfnamefont{B.}~\bibnamefont{{Gold}}},
  \bibinfo{author}{\bibfnamefont{G.}~\bibnamefont{{Hinshaw}}},
  \bibinfo{author}{\bibfnamefont{N.}~\bibnamefont{{Jarosik}}},
  \bibinfo{author}{\bibfnamefont{D.}~\bibnamefont{{Larson}}},
  \bibinfo{author}{\bibfnamefont{M.}~\bibnamefont{{Limon}}},
  \bibinfo{author}{\bibfnamefont{L.}~\bibnamefont{{Page}}},
  \bibnamefont{et~al.}, \bibinfo{journal}{ArXiv e-prints}
  \textbf{\bibinfo{volume}{803}} (\bibinfo{year}{2008}), \eprint{0803.0547}.

\bibitem[{\citenamefont{Baumann et~al.}(2009)}]{cmbpol_Baumann08}
\bibinfo{author}{\bibfnamefont{D.}~\bibnamefont{Baumann}} \bibnamefont{et~al.}
  (\bibinfo{collaboration}{CMBPol Study Team}), \bibinfo{journal}{AIP Conf.
  Proc.} \textbf{\bibinfo{volume}{1141}}, \bibinfo{pages}{10}
  (\bibinfo{year}{2009}), \eprint{0811.3919}.

\bibitem[{\citenamefont{{Collaboration, The LIGO Scientific and Collaboration,
  the Virgo}}(2009)}]{Collaboration:2009ws}
\bibinfo{author}{\bibnamefont{{Collaboration, The LIGO Scientific and
  Collaboration, the Virgo}}}, \bibinfo{journal}{Nature}
  \textbf{\bibinfo{volume}{460}}, \bibinfo{pages}{990} (\bibinfo{year}{2009}),
  \eprint{0910.5772}.

\bibitem[{\citenamefont{Bartolo et~al.}(2007)\citenamefont{Bartolo, Matarrese,
  and Riotto}}]{Bartolo:2006fj}
\bibinfo{author}{\bibfnamefont{N.}~\bibnamefont{Bartolo}},
  \bibinfo{author}{\bibfnamefont{S.}~\bibnamefont{Matarrese}},
  \bibnamefont{and} \bibinfo{author}{\bibfnamefont{A.}~\bibnamefont{Riotto}},
  \bibinfo{journal}{JCAP} \textbf{\bibinfo{volume}{0701}}, \bibinfo{pages}{019}
  (\bibinfo{year}{2007}), \eprint{astro-ph/0610110}.

\bibitem[{\citenamefont{Baumann et~al.}(2007)\citenamefont{Baumann, Steinhardt,
  Takahashi, and Ichiki}}]{Baumann:2007zm}
\bibinfo{author}{\bibfnamefont{D.}~\bibnamefont{Baumann}},
  \bibinfo{author}{\bibfnamefont{P.~J.} \bibnamefont{Steinhardt}},
  \bibinfo{author}{\bibfnamefont{K.}~\bibnamefont{Takahashi}},
  \bibnamefont{and} \bibinfo{author}{\bibfnamefont{K.}~\bibnamefont{Ichiki}},
  \bibinfo{journal}{Phys. Rev.} \textbf{\bibinfo{volume}{D76}},
  \bibinfo{pages}{084019} (\bibinfo{year}{2007}), \eprint{hep-th/0703290}.

\bibitem[{\citenamefont{{Seljak} and
  {Zaldarriaga}}(1997)}]{SeljakZaldarriaga97}
\bibinfo{author}{\bibfnamefont{U.}~\bibnamefont{{Seljak}}} \bibnamefont{and}
  \bibinfo{author}{\bibfnamefont{M.}~\bibnamefont{{Zaldarriaga}}},
  \bibinfo{journal}{Phys. Rev. Lett.} \textbf{\bibinfo{volume}{78}},
  \bibinfo{pages}{2054} (\bibinfo{year}{1997}).

\bibitem[{\citenamefont{{Kamionkowski}
  et~al.}(1997{\natexlab{a}})\citenamefont{{Kamionkowski}, {Kosowsky}, and
  {Stebbins}}}]{1997PhRvD..55.7368K}
\bibinfo{author}{\bibfnamefont{M.}~\bibnamefont{{Kamionkowski}}},
  \bibinfo{author}{\bibfnamefont{A.}~\bibnamefont{{Kosowsky}}},
  \bibnamefont{and}
  \bibinfo{author}{\bibfnamefont{A.}~\bibnamefont{{Stebbins}}},
  \bibinfo{journal}{\prd} \textbf{\bibinfo{volume}{55}}, \bibinfo{pages}{7368}
  (\bibinfo{year}{1997}{\natexlab{a}}).

\bibitem[{\citenamefont{{Kamionkowski}
  et~al.}(1997{\natexlab{b}})\citenamefont{{Kamionkowski}, {Kosowsky}, and
  {Stebbins}}}]{1997PhRvL..78.2058K}
\bibinfo{author}{\bibfnamefont{M.}~\bibnamefont{{Kamionkowski}}},
  \bibinfo{author}{\bibfnamefont{A.}~\bibnamefont{{Kosowsky}}},
  \bibnamefont{and}
  \bibinfo{author}{\bibfnamefont{A.}~\bibnamefont{{Stebbins}}},
  \bibinfo{journal}{Phys. Rev. Lett.} \textbf{\bibinfo{volume}{78}},
  \bibinfo{pages}{2058} (\bibinfo{year}{1997}{\natexlab{b}}).

\bibitem[{\citenamefont{{ACTPol , BICEP, CAPMAP, CBI, Clover, CMBPol, EBEX,
  POLARBEAR, PIPER, PIQUE, QUaD, QUIET, SPIDER}}(current and upcomming
  experimets)}]{cmb_polarization_experiments}
\bibinfo{author}{\bibnamefont{{ACTPol , BICEP, CAPMAP, CBI, Clover, CMBPol,
  EBEX, POLARBEAR, PIPER, PIQUE, QUaD, QUIET, SPIDER}}} (\bibinfo{year}{current
  and upcomming experimets}).

\bibitem[{\citenamefont{{Seljak}}(1996)}]{1996ApJ...463....1S}
\bibinfo{author}{\bibfnamefont{U.}~\bibnamefont{{Seljak}}},
  \bibinfo{journal}{\apj} \textbf{\bibinfo{volume}{463}}, \bibinfo{pages}{1}
  (\bibinfo{year}{1996}), \eprint{arXiv:astro-ph/9505109}.

\bibitem[{\citenamefont{{Zaldarriaga} and
  {Seljak}}(1998)}]{1998PhRvD..58b3003Z}
\bibinfo{author}{\bibfnamefont{M.}~\bibnamefont{{Zaldarriaga}}}
  \bibnamefont{and} \bibinfo{author}{\bibfnamefont{U.}~\bibnamefont{{Seljak}}},
  \bibinfo{journal}{\prd} \textbf{\bibinfo{volume}{58}},
  \bibinfo{pages}{023003} (\bibinfo{year}{1998}),
  \eprint{arXiv:astro-ph/9803150}.


\bibitem[{\citenamefont{{Lue} et~al.}(1999)\citenamefont{{Lue}, {Wang}, and
  {Kamionkowski}}}]{LueWangKamionkowski}
\bibinfo{author}{\bibfnamefont{A.}~\bibnamefont{{Lue}}},
  \bibinfo{author}{\bibfnamefont{L.}~\bibnamefont{{Wang}}}, \bibnamefont{and}
  \bibinfo{author}{\bibfnamefont{M.}~\bibnamefont{{Kamionkowski}}},
  \bibinfo{journal}{Physical Review Letters} \textbf{\bibinfo{volume}{83}},
  \bibinfo{pages}{1506} (\bibinfo{year}{1999}),
  \eprint{arXiv:astro-ph/9812088}.


\bibitem[{\citenamefont{{Kamionkowski}}(2009)}]{Kamionkowski_rotation09}
\bibinfo{author}{\bibfnamefont{M.}~\bibnamefont{{Kamionkowski}}},
  \bibinfo{journal}{Physical Review Letters} \textbf{\bibinfo{volume}{102}},
  \bibinfo{pages}{111302} (\bibinfo{year}{2009}), \eprint{0810.1286}.



\bibitem[{\citenamefont{{Yadav} et~al.}(2009)\citenamefont{{Yadav}, {Biswas},
  {Su}, and {Zaldarriaga}}}]{Yadav_etal_09_rotation}
\bibinfo{author}{\bibfnamefont{A.~P.~S.} \bibnamefont{{Yadav}}},
  \bibinfo{author}{\bibfnamefont{R.}~\bibnamefont{{Biswas}}},
  \bibinfo{author}{\bibfnamefont{M.}~\bibnamefont{{Su}}}, \bibnamefont{and}
  \bibinfo{author}{\bibfnamefont{M.}~\bibnamefont{{Zaldarriaga}}},
  \bibinfo{journal}{\prd} \textbf{\bibinfo{volume}{79}},
  \bibinfo{pages}{123009} (\bibinfo{year}{2009}), \eprint{0902.4466}.

\bibitem[{\citenamefont{{Gluscevic} et~al.}(2009)\citenamefont{{Gluscevic},
  {Kamionkowski}, and {Cooray}}}]{GluscevicKamionkowskiCooray09}
\bibinfo{author}{\bibfnamefont{V.}~\bibnamefont{{Gluscevic}}},
  \bibinfo{author}{\bibfnamefont{M.}~\bibnamefont{{Kamionkowski}}},
  \bibnamefont{and} \bibinfo{author}{\bibfnamefont{A.}~\bibnamefont{{Cooray}}},
  \bibinfo{journal}{\prd} \textbf{\bibinfo{volume}{80}},
  \bibinfo{pages}{023510} (\bibinfo{year}{2009}), \eprint{0905.1687}.


\bibitem[{\citenamefont{Hu et~al.}(2003)\citenamefont{Hu, Hedman, and
  Zaldarriaga}}]{HHZ}
\bibinfo{author}{\bibfnamefont{W.}~\bibnamefont{Hu}},
  \bibinfo{author}{\bibfnamefont{M.~M.} \bibnamefont{Hedman}},
  \bibnamefont{and}
  \bibinfo{author}{\bibfnamefont{M.}~\bibnamefont{Zaldarriaga}},
  \bibinfo{journal}{Phys. Rev.} \textbf{\bibinfo{volume}{D67}},
  \bibinfo{pages}{043004} (\bibinfo{year}{2003}), \eprint{astro-ph/0210096}.

\bibitem[{\citenamefont{{Shimon} et~al.}(2008)\citenamefont{{Shimon},
  {Keating}, {Ponthieu}, and {Hivon}}}]{Shimon_etal08}
\bibinfo{author}{\bibfnamefont{M.}~\bibnamefont{{Shimon}}},
  \bibinfo{author}{\bibfnamefont{B.}~\bibnamefont{{Keating}}},
  \bibinfo{author}{\bibfnamefont{N.}~\bibnamefont{{Ponthieu}}},
  \bibnamefont{and} \bibinfo{author}{\bibfnamefont{E.}~\bibnamefont{{Hivon}}},
  \bibinfo{journal}{\prd} \textbf{\bibinfo{volume}{77}},
  \bibinfo{pages}{083003} (\bibinfo{year}{2008}), \eprint{0709.1513}.

\bibitem[{\citenamefont{Yadav et~al.}(2009)\citenamefont{Yadav, Su, and
  Zaldarriaga}}]{YSZ09}
\bibinfo{author}{\bibfnamefont{A.~P.~S.} \bibnamefont{Yadav}},
  \bibinfo{author}{\bibfnamefont{M.}~\bibnamefont{Su}}, \bibnamefont{and}
  \bibinfo{author}{\bibfnamefont{M.}~\bibnamefont{Zaldarriaga}}
  (\bibinfo{year}{2009}), \eprint{0912.3532}.

\bibitem[{\citenamefont{{Linde A. D.}}(1984)}]{Linde1984}
\bibinfo{author}{\bibnamefont{{Linde A. D.}}}, \bibinfo{journal}{JETP Lett.}
  \textbf{\bibinfo{volume}{40}}, \bibinfo{pages}{1333} (\bibinfo{year}{1984}).

\bibitem[{\citenamefont{{Kofman L. A. and Linde A. D.}}(1987)}]{Kofman1987}
\bibinfo{author}{\bibnamefont{{Kofman L. A. and Linde A. D.}}},
  \bibinfo{journal}{Nucl. Phys.} \textbf{\bibinfo{volume}{B282}},
  \bibinfo{pages}{555} (\bibinfo{year}{1987}).

\bibitem[{\citenamefont{{Polarski, David and Starobinsky, Alexei
  A.}}(1994)}]{Polarski:1994rz}
\bibinfo{author}{\bibnamefont{{Polarski, David and Starobinsky, Alexei A.}}},
  \bibinfo{journal}{Phys. Rev.} \textbf{\bibinfo{volume}{D50}},
  \bibinfo{pages}{6123} (\bibinfo{year}{1994}), \eprint{astro-ph/9404061}.

\bibitem[{\citenamefont{{Gordon, Christopher and Wands, David and Bassett,
  Bruce A. and Maartens, Roy}}(2001)}]{Gordon:2000hv}
\bibinfo{author}{\bibnamefont{{Gordon, Christopher and Wands, David and
  Bassett, Bruce A. and Maartens, Roy}}}, \bibinfo{journal}{Phys. Rev.}
  \textbf{\bibinfo{volume}{D63}}, \bibinfo{pages}{023506}
  (\bibinfo{year}{2001}), \eprint{astro-ph/0009131}.

\bibitem[{\citenamefont{{Linde A. D.}}(1985)}]{Linde1985}
\bibinfo{author}{\bibnamefont{{Linde A. D.}}}, \bibinfo{journal}{Phys. Lett.}
  \textbf{\bibinfo{volume}{B158}}, \bibinfo{pages}{375} (\bibinfo{year}{1985}).

\bibitem[{\citenamefont{{Efstathiou G. and Bond J.
  R.}}(1986)}]{Efstathiou_Bond_1986}
\bibinfo{author}{\bibnamefont{{Efstathiou G. and Bond J. R.}}},
  \bibinfo{journal}{Mon. Not. Roy. Astron. Soc.}
  \textbf{\bibinfo{volume}{218}}, \bibinfo{pages}{103} (\bibinfo{year}{1986}).

\bibitem[{\citenamefont{{Peebles P. J. E.}}(1987)}]{Peebles1987}
\bibinfo{author}{\bibnamefont{{Peebles P. J. E.}}}, \bibinfo{journal}{Nature}
  \textbf{\bibinfo{volume}{327}}, \bibinfo{pages}{210} (\bibinfo{year}{1987}).

\bibitem[{\citenamefont{{Kodama H. and Sasaki M.}}(1986)}]{Kodama_Sasaki_1986}
\bibinfo{author}{\bibnamefont{{Kodama H. and Sasaki M.}}},
  \bibinfo{journal}{Int. J. Mod. Phys.} \textbf{\bibinfo{volume}{A1}},
  \bibinfo{pages}{265} (\bibinfo{year}{1986}).

\bibitem[{\citenamefont{{Weinberg S.}}(2004)}]{Weinberg:2004kr}
\bibinfo{author}{\bibnamefont{{Weinberg S.}}}, \bibinfo{journal}{Phys. Rev.}
  \textbf{\bibinfo{volume}{D70}}, \bibinfo{pages}{043541}
  (\bibinfo{year}{2004}), \eprint{astro-ph/0401313}.

\bibitem[{\citenamefont{{Garcia-Bellido, Juan and Wands,
  David}}(1996)}]{GarciaBellido:1995qq}
\bibinfo{author}{\bibnamefont{{Garcia-Bellido, Juan and Wands, David}}},
  \bibinfo{journal}{Phys. Rev.} \textbf{\bibinfo{volume}{D53}},
  \bibinfo{pages}{5437} (\bibinfo{year}{1996}), \eprint{astro-ph/9511029}.

\bibitem[{\citenamefont{{Sasaki, Misao and Stewart, Ewan
  D.}}(1996)}]{Sasaki:1995aw}
\bibinfo{author}{\bibnamefont{{Sasaki, Misao and Stewart, Ewan D.}}},
  \bibinfo{journal}{Prog. Theor. Phys.} \textbf{\bibinfo{volume}{95}},
  \bibinfo{pages}{71} (\bibinfo{year}{1996}), \eprint{astro-ph/9507001}.

\bibitem[{\citenamefont{Sasaki and Tanaka}(1998)}]{Sasaki:1998ug}
\bibinfo{author}{\bibfnamefont{M.}~\bibnamefont{Sasaki}} \bibnamefont{and}
  \bibinfo{author}{\bibfnamefont{T.}~\bibnamefont{Tanaka}},
  \bibinfo{journal}{Prog. Theor. Phys.} \textbf{\bibinfo{volume}{99}},
  \bibinfo{pages}{763} (\bibinfo{year}{1998}), \eprint{gr-qc/9801017}.

\bibitem[{\citenamefont{Bartolo et~al.}(2001)\citenamefont{Bartolo, Matarrese,
  and Riotto}}]{Bartolo:2001rt}
\bibinfo{author}{\bibfnamefont{N.}~\bibnamefont{Bartolo}},
  \bibinfo{author}{\bibfnamefont{S.}~\bibnamefont{Matarrese}},
  \bibnamefont{and} \bibinfo{author}{\bibfnamefont{A.}~\bibnamefont{Riotto}},
  \bibinfo{journal}{Phys. Rev.} \textbf{\bibinfo{volume}{D64}},
  \bibinfo{pages}{123504} (\bibinfo{year}{2001}), \eprint{astro-ph/0107502}.

\bibitem[{\citenamefont{Bucher et~al.}(2002)\citenamefont{Bucher, Moodley, and
  Turok}}]{Bucher:2000kb}
\bibinfo{author}{\bibfnamefont{M.}~\bibnamefont{Bucher}},
  \bibinfo{author}{\bibfnamefont{K.}~\bibnamefont{Moodley}}, \bibnamefont{and}
  \bibinfo{author}{\bibfnamefont{N.}~\bibnamefont{Turok}},
  \bibinfo{journal}{Phys. Rev.} \textbf{\bibinfo{volume}{D66}},
  \bibinfo{pages}{023528} (\bibinfo{year}{2002}), \eprint{astro-ph/0007360}.

\bibitem[{\citenamefont{Bucher et~al.}(2001)\citenamefont{Bucher, Moodley, and
  Turok}}]{Bucher_Moodley_Turok_01}
\bibinfo{author}{\bibfnamefont{M.}~\bibnamefont{Bucher}},
  \bibinfo{author}{\bibfnamefont{K.}~\bibnamefont{Moodley}}, \bibnamefont{and}
  \bibinfo{author}{\bibfnamefont{N.}~\bibnamefont{Turok}},
  \bibinfo{journal}{Phys. Rev. Lett.} \textbf{\bibinfo{volume}{87}},
  \bibinfo{pages}{191301} (\bibinfo{year}{2001}).

\bibitem[{\citenamefont{{Bean} et~al.}(2006)\citenamefont{{Bean}, {Dunkley},
  and {Pierpaoli}}}]{Bean_etal_06}
\bibinfo{author}{\bibfnamefont{R.}~\bibnamefont{{Bean}}},
  \bibinfo{author}{\bibfnamefont{J.}~\bibnamefont{{Dunkley}}},
  \bibnamefont{and}
  \bibinfo{author}{\bibfnamefont{E.}~\bibnamefont{{Pierpaoli}}},
  \bibinfo{journal}{Phys. Rev. D.} \textbf{\bibinfo{volume}{74}},
  \bibinfo{pages}{063503} (\bibinfo{year}{2006}), \eprint{astro-ph/0606685}.

\bibitem[{\citenamefont{Gupta et~al.}(2002)\citenamefont{Gupta, Berera,
  Heavens, and Matarrese}}]{Gupta_et02}
\bibinfo{author}{\bibfnamefont{S.}~\bibnamefont{Gupta}},
  \bibinfo{author}{\bibfnamefont{A.}~\bibnamefont{Berera}},
  \bibinfo{author}{\bibfnamefont{A.~F.} \bibnamefont{Heavens}},
  \bibnamefont{and}
  \bibinfo{author}{\bibfnamefont{S.}~\bibnamefont{Matarrese}},
  \bibinfo{journal}{Phys. Rev. D.} \textbf{\bibinfo{volume}{66}},
  \bibinfo{pages}{043510} (\bibinfo{year}{2002}).

\bibitem[{\citenamefont{Allen et~al.}(1987)\citenamefont{Allen, Grinstein, and
  Wise}}]{Allen_et87}
\bibinfo{author}{\bibfnamefont{T.~J.} \bibnamefont{Allen}},
  \bibinfo{author}{\bibfnamefont{B.}~\bibnamefont{Grinstein}},
  \bibnamefont{and} \bibinfo{author}{\bibfnamefont{M.~B.} \bibnamefont{Wise}},
  \bibinfo{journal}{Phys. Lett. B.} \textbf{\bibinfo{volume}{197}},
  \bibinfo{pages}{66} (\bibinfo{year}{1987}).

\bibitem[{\citenamefont{Lesgourgues et~al.}(1997)\citenamefont{Lesgourgues,
  Polarski, and Starobinsky}}]{Lesgourgues_et97}
\bibinfo{author}{\bibfnamefont{J.}~\bibnamefont{Lesgourgues}},
  \bibinfo{author}{\bibfnamefont{D.}~\bibnamefont{Polarski}}, \bibnamefont{and}
  \bibinfo{author}{\bibfnamefont{A.~A.} \bibnamefont{Starobinsky}},
  \bibinfo{journal}{Nucl. Phys. B.} \textbf{\bibinfo{volume}{497}},
  \bibinfo{pages}{479} (\bibinfo{year}{1997}).

\bibitem[{\citenamefont{Martin et~al.}(2000)\citenamefont{Martin, Riazuelo, and
  Sakellariadou}}]{Martin_et2000}
\bibinfo{author}{\bibfnamefont{J.}~\bibnamefont{Martin}},
  \bibinfo{author}{\bibfnamefont{A.}~\bibnamefont{Riazuelo}}, \bibnamefont{and}
  \bibinfo{author}{\bibfnamefont{M.}~\bibnamefont{Sakellariadou}},
  \bibinfo{journal}{Phys. Rev. D.} \textbf{\bibinfo{volume}{61}},
  \bibinfo{pages}{083518} (\bibinfo{year}{2000}).

\bibitem[{\citenamefont{Creminelli}(2003)}]{Creminelli03}
\bibinfo{author}{\bibfnamefont{P.}~\bibnamefont{Creminelli}},
  \bibinfo{journal}{J. Cosmol. Astropart. Phys.} \textbf{\bibinfo{volume}{10}},
  \bibinfo{pages}{003} (\bibinfo{year}{2003}).

\bibitem[{\citenamefont{{Arkani-Hamed}
  et~al.}(2004)\citenamefont{{Arkani-Hamed}, {Creminelli}, {Mukohyama}, and
  {Zaldarriaga}}}]{Arkani_et_04}
\bibinfo{author}{\bibfnamefont{N.}~\bibnamefont{{Arkani-Hamed}}},
  \bibinfo{author}{\bibfnamefont{P.}~\bibnamefont{{Creminelli}}},
  \bibinfo{author}{\bibfnamefont{S.}~\bibnamefont{{Mukohyama}}},
  \bibnamefont{and}
  \bibinfo{author}{\bibfnamefont{M.}~\bibnamefont{{Zaldarriaga}}},
  \bibinfo{journal}{Journal of Cosmology and Astro-Particle Physics}
  \textbf{\bibinfo{volume}{4}}, \bibinfo{pages}{1} (\bibinfo{year}{2004}).

\bibitem[{\citenamefont{Dvali et~al.}(2003)\citenamefont{Dvali, Gruzinov, and
  Zaldarriaga}}]{Dvali_Gruzinov_Zaldarriaga_03}
\bibinfo{author}{\bibfnamefont{G.}~\bibnamefont{Dvali}},
  \bibinfo{author}{\bibfnamefont{A.}~\bibnamefont{Gruzinov}}, \bibnamefont{and}
  \bibinfo{author}{\bibfnamefont{M.}~\bibnamefont{Zaldarriaga}},
  \bibinfo{journal}{Phys. Rev. D} \textbf{\bibinfo{volume}{69}},
  \bibinfo{pages}{023505} (\bibinfo{year}{2003}).

\bibitem[{\citenamefont{Zaldarriaga}(2004)}]{Zaldarriaga04}
\bibinfo{author}{\bibfnamefont{M.}~\bibnamefont{Zaldarriaga}},
  \bibinfo{journal}{Phys. Rev. D} \textbf{\bibinfo{volume}{69}},
  \bibinfo{pages}{043508} (\bibinfo{year}{2004}).

\bibitem[{\citenamefont{{Lyth} et~al.}(2003)\citenamefont{{Lyth}, {Ungarelli},
  and {Wands}}}]{Lyth_etal03}
\bibinfo{author}{\bibfnamefont{D.~H.} \bibnamefont{{Lyth}}},
  \bibinfo{author}{\bibfnamefont{C.}~\bibnamefont{{Ungarelli}}},
  \bibnamefont{and} \bibinfo{author}{\bibfnamefont{D.}~\bibnamefont{{Wands}}},
  \bibinfo{journal}{Phys. Rev. D.} \textbf{\bibinfo{volume}{67}},
  \bibinfo{pages}{023503} (\bibinfo{year}{2003}), \eprint{astro-ph/0208055}.

\bibitem[{\citenamefont{Komatsu et~al.}(2009)}]{Komatsu:2009kd}
\bibinfo{author}{\bibfnamefont{E.}~\bibnamefont{Komatsu}} \bibnamefont{et~al.}
  (\bibinfo{year}{2009}), \eprint{0902.4759}.

\bibitem[{\citenamefont{{Babich} et~al.}(2004)\citenamefont{{Babich},
  {Creminelli}, and {Zaldarriaga}}}]{Babich_etal_04}
\bibinfo{author}{\bibfnamefont{D.}~\bibnamefont{{Babich}}},
  \bibinfo{author}{\bibfnamefont{P.}~\bibnamefont{{Creminelli}}},
  \bibnamefont{and}
  \bibinfo{author}{\bibfnamefont{M.}~\bibnamefont{{Zaldarriaga}}},
  \bibinfo{journal}{Journal of Cosmology and Astro-Particle Physics}
  \textbf{\bibinfo{volume}{8}}, \bibinfo{pages}{9} (\bibinfo{year}{2004}).

\bibitem[{\citenamefont{Bartolo et~al.}(2002)\citenamefont{Bartolo, Matarrese,
  and Riotto}}]{Bartolo:2001cw}
\bibinfo{author}{\bibfnamefont{N.}~\bibnamefont{Bartolo}},
  \bibinfo{author}{\bibfnamefont{S.}~\bibnamefont{Matarrese}},
  \bibnamefont{and} \bibinfo{author}{\bibfnamefont{A.}~\bibnamefont{Riotto}},
  \bibinfo{journal}{Phys. Rev.} \textbf{\bibinfo{volume}{D65}},
  \bibinfo{pages}{103505} (\bibinfo{year}{2002}), \eprint{hep-ph/0112261}.

\bibitem[{\citenamefont{Bernardeau and Uzan}(2002)}]{Bernardeau:2002jy}
\bibinfo{author}{\bibfnamefont{F.}~\bibnamefont{Bernardeau}} \bibnamefont{and}
  \bibinfo{author}{\bibfnamefont{J.-P.} \bibnamefont{Uzan}},
  \bibinfo{journal}{Phys. Rev.} \textbf{\bibinfo{volume}{D66}},
  \bibinfo{pages}{103506} (\bibinfo{year}{2002}), \eprint{hep-ph/0207295}.

\bibitem[{\citenamefont{Bernardeau and Uzan}(2003)}]{Bernardeau:2002jf}
\bibinfo{author}{\bibfnamefont{F.}~\bibnamefont{Bernardeau}} \bibnamefont{and}
  \bibinfo{author}{\bibfnamefont{J.-P.} \bibnamefont{Uzan}},
  \bibinfo{journal}{Phys. Rev.} \textbf{\bibinfo{volume}{D67}},
  \bibinfo{pages}{121301} (\bibinfo{year}{2003}), \eprint{astro-ph/0209330}.

\bibitem[{\citenamefont{Sasaki}(2008)}]{Sasaki:2008uc}
\bibinfo{author}{\bibfnamefont{M.}~\bibnamefont{Sasaki}},
  \bibinfo{journal}{Prog. Theor. Phys.} \textbf{\bibinfo{volume}{120}},
  \bibinfo{pages}{159} (\bibinfo{year}{2008}), \eprint{0805.0974}.

\bibitem[{\citenamefont{Naruko and Sasaki}(2009)}]{Naruko:2008sq}
\bibinfo{author}{\bibfnamefont{A.}~\bibnamefont{Naruko}} \bibnamefont{and}
  \bibinfo{author}{\bibfnamefont{M.}~\bibnamefont{Sasaki}},
  \bibinfo{journal}{Prog. Theor. Phys.} \textbf{\bibinfo{volume}{121}},
  \bibinfo{pages}{193} (\bibinfo{year}{2009}), \eprint{0807.0180}.

\bibitem[{\citenamefont{Byrnes et~al.}(2008)\citenamefont{Byrnes, Choi, and
  Hall}}]{Byrnes:2008wi}
\bibinfo{author}{\bibfnamefont{C.~T.} \bibnamefont{Byrnes}},
  \bibinfo{author}{\bibfnamefont{K.-Y.} \bibnamefont{Choi}}, \bibnamefont{and}
  \bibinfo{author}{\bibfnamefont{L.~M.~H.} \bibnamefont{Hall}},
  \bibinfo{journal}{JCAP} \textbf{\bibinfo{volume}{0810}}, \bibinfo{pages}{008}
  (\bibinfo{year}{2008}), \eprint{0807.1101}.

\bibitem[{\citenamefont{Byrnes and Wands}(2006)}]{Byrnes:2006fr}
\bibinfo{author}{\bibfnamefont{C.~T.} \bibnamefont{Byrnes}} \bibnamefont{and}
  \bibinfo{author}{\bibfnamefont{D.}~\bibnamefont{Wands}},
  \bibinfo{journal}{Phys. Rev.} \textbf{\bibinfo{volume}{D74}},
  \bibinfo{pages}{043529} (\bibinfo{year}{2006}), \eprint{astro-ph/0605679}.

\bibitem[{\citenamefont{Langlois et~al.}(2008)\citenamefont{Langlois, Vernizzi,
  and Wands}}]{Langlois:2008vk}
\bibinfo{author}{\bibfnamefont{D.}~\bibnamefont{Langlois}},
  \bibinfo{author}{\bibfnamefont{F.}~\bibnamefont{Vernizzi}}, \bibnamefont{and}
  \bibinfo{author}{\bibfnamefont{D.}~\bibnamefont{Wands}},
  \bibinfo{journal}{JCAP} \textbf{\bibinfo{volume}{0812}}, \bibinfo{pages}{004}
  (\bibinfo{year}{2008}), \eprint{0809.4646}.

\bibitem[{\citenamefont{Valiviita et~al.}(2008)\citenamefont{Valiviita,
  Assadullahi, and Wands}}]{Valiviita:2008zb}
\bibinfo{author}{\bibfnamefont{J.}~\bibnamefont{Valiviita}},
  \bibinfo{author}{\bibfnamefont{H.}~\bibnamefont{Assadullahi}},
  \bibnamefont{and} \bibinfo{author}{\bibfnamefont{D.}~\bibnamefont{Wands}}
  (\bibinfo{year}{2008}), \eprint{0806.0623}.

\bibitem[{\citenamefont{Assadullahi et~al.}(2007)\citenamefont{Assadullahi,
  Valiviita, and Wands}}]{Assadullahi:2007uw}
\bibinfo{author}{\bibfnamefont{H.}~\bibnamefont{Assadullahi}},
  \bibinfo{author}{\bibfnamefont{J.}~\bibnamefont{Valiviita}},
  \bibnamefont{and} \bibinfo{author}{\bibfnamefont{D.}~\bibnamefont{Wands}},
  \bibinfo{journal}{Phys. Rev.} \textbf{\bibinfo{volume}{D76}},
  \bibinfo{pages}{103003} (\bibinfo{year}{2007}), \eprint{0708.0223}.

\bibitem[{\citenamefont{Valiviita et~al.}(2006)\citenamefont{Valiviita, Sasaki,
  and Wands}}]{Valiviita:2006mz}
\bibinfo{author}{\bibfnamefont{J.}~\bibnamefont{Valiviita}},
  \bibinfo{author}{\bibfnamefont{M.}~\bibnamefont{Sasaki}}, \bibnamefont{and}
  \bibinfo{author}{\bibfnamefont{D.}~\bibnamefont{Wands}}
  (\bibinfo{year}{2006}), \eprint{astro-ph/0610001}.

\bibitem[{\citenamefont{Vernizzi and Wands}(2006)}]{Vernizzi:2006ve}
\bibinfo{author}{\bibfnamefont{F.}~\bibnamefont{Vernizzi}} \bibnamefont{and}
  \bibinfo{author}{\bibfnamefont{D.}~\bibnamefont{Wands}},
  \bibinfo{journal}{JCAP} \textbf{\bibinfo{volume}{0605}}, \bibinfo{pages}{019}
  (\bibinfo{year}{2006}), \eprint{astro-ph/0603799}.

\bibitem[{\citenamefont{Allen et~al.}(2006)\citenamefont{Allen, Gupta, and
  Wands}}]{Allen:2005ye}
\bibinfo{author}{\bibfnamefont{L.~E.} \bibnamefont{Allen}},
  \bibinfo{author}{\bibfnamefont{S.}~\bibnamefont{Gupta}}, \bibnamefont{and}
  \bibinfo{author}{\bibfnamefont{D.}~\bibnamefont{Wands}},
  \bibinfo{journal}{JCAP} \textbf{\bibinfo{volume}{0601}}, \bibinfo{pages}{006}
  (\bibinfo{year}{2006}), \eprint{astro-ph/0509719}.

\bibitem[{\citenamefont{{Linde} and {Mukhanov}}(1997)}]{Linde_Mukhanov1997}
\bibinfo{author}{\bibfnamefont{A.}~\bibnamefont{{Linde}}} \bibnamefont{and}
  \bibinfo{author}{\bibfnamefont{V.}~\bibnamefont{{Mukhanov}}},
  \bibinfo{journal}{Phys. Rev. D.} \textbf{\bibinfo{volume}{56}},
  \bibinfo{pages}{535} (\bibinfo{year}{1997}), \eprint{arXiv:astro-ph/9610219}.

\bibitem[{\citenamefont{Kofman}(2003)}]{Kofman:2003nx}
\bibinfo{author}{\bibfnamefont{L.}~\bibnamefont{Kofman}}
  (\bibinfo{year}{2003}), \eprint{astro-ph/0303614}.

\bibitem[{\citenamefont{Lehners and
  Steinhardt}(2008{\natexlab{a}})}]{Lehners:2007wc}
\bibinfo{author}{\bibfnamefont{J.-L.} \bibnamefont{Lehners}} \bibnamefont{and}
  \bibinfo{author}{\bibfnamefont{P.~J.} \bibnamefont{Steinhardt}},
  \bibinfo{journal}{Phys. Rev.} \textbf{\bibinfo{volume}{D77}},
  \bibinfo{pages}{063533} (\bibinfo{year}{2008}{\natexlab{a}}),
  \eprint{0712.3779}.

\bibitem[{\citenamefont{Lehners and
  Steinhardt}(2008{\natexlab{b}})}]{Lehners:2008my}
\bibinfo{author}{\bibfnamefont{J.-L.} \bibnamefont{Lehners}} \bibnamefont{and}
  \bibinfo{author}{\bibfnamefont{P.~J.} \bibnamefont{Steinhardt}},
  \bibinfo{journal}{Phys. Rev.} \textbf{\bibinfo{volume}{D78}},
  \bibinfo{pages}{023506} (\bibinfo{year}{2008}{\natexlab{b}}),
  \eprint{0804.1293}.

\bibitem[{\citenamefont{Koyama et~al.}(2007)\citenamefont{Koyama, Mizuno, and
  Wands}}]{Koyama:2007ag}
\bibinfo{author}{\bibfnamefont{K.}~\bibnamefont{Koyama}},
  \bibinfo{author}{\bibfnamefont{S.}~\bibnamefont{Mizuno}}, \bibnamefont{and}
  \bibinfo{author}{\bibfnamefont{D.}~\bibnamefont{Wands}},
  \bibinfo{journal}{Class. Quant. Grav.} \textbf{\bibinfo{volume}{24}},
  \bibinfo{pages}{3919} (\bibinfo{year}{2007}), \eprint{0704.1152}.

\bibitem[{\citenamefont{Holman and Tolley}(2008)}]{Holman:2007na}
\bibinfo{author}{\bibfnamefont{R.}~\bibnamefont{Holman}} \bibnamefont{and}
  \bibinfo{author}{\bibfnamefont{A.~J.} \bibnamefont{Tolley}},
  \bibinfo{journal}{JCAP} \textbf{\bibinfo{volume}{0805}}, \bibinfo{pages}{001}
  (\bibinfo{year}{2008}), \eprint{0710.1302}.

\bibitem[{\citenamefont{{Chen} et~al.}(2007)\citenamefont{{Chen}, {Huang},
  {Kachru}, and {Shiu}}}]{Chen_etal07}
\bibinfo{author}{\bibfnamefont{X.}~\bibnamefont{{Chen}}},
  \bibinfo{author}{\bibfnamefont{M.-x.} \bibnamefont{{Huang}}},
  \bibinfo{author}{\bibfnamefont{S.}~\bibnamefont{{Kachru}}}, \bibnamefont{and}
  \bibinfo{author}{\bibfnamefont{G.}~\bibnamefont{{Shiu}}},
  \bibinfo{journal}{Journal of Cosmology and Astro-Particle Physics}
  \textbf{\bibinfo{volume}{1}}, \bibinfo{pages}{2} (\bibinfo{year}{2007}).

\bibitem[{\citenamefont{{Verde} et~al.}(2000)\citenamefont{{Verde}, {Wang},
  {Heavens}, and {Kamionkowski}}}]{verde00}
\bibinfo{author}{\bibfnamefont{L.}~\bibnamefont{{Verde}}},
  \bibinfo{author}{\bibfnamefont{L.}~\bibnamefont{{Wang}}},
  \bibinfo{author}{\bibfnamefont{A.~F.} \bibnamefont{{Heavens}}},
  \bibnamefont{and}
  \bibinfo{author}{\bibfnamefont{M.}~\bibnamefont{{Kamionkowski}}},
  \bibinfo{journal}{MNRAS} \textbf{\bibinfo{volume}{313}}, \bibinfo{pages}{141}
  (\bibinfo{year}{2000}).

\bibitem[{\citenamefont{Komatsu and Spergel}(2001)}]{KS2001}
\bibinfo{author}{\bibfnamefont{E.~N.} \bibnamefont{Komatsu}} \bibnamefont{and}
  \bibinfo{author}{\bibfnamefont{D.~N.} \bibnamefont{Spergel}},
  \bibinfo{journal}{Phys. Rev. D} \textbf{\bibinfo{volume}{63}},
  \bibinfo{pages}{063002} (\bibinfo{year}{2001}).

\bibitem[{\citenamefont{{Koyama} et~al.}(2007)\citenamefont{{Koyama}, {Mizuno},
  {Vernizzi}, and {Wands}}}]{Koyama_etal07}
\bibinfo{author}{\bibfnamefont{K.}~\bibnamefont{{Koyama}}},
  \bibinfo{author}{\bibfnamefont{S.}~\bibnamefont{{Mizuno}}},
  \bibinfo{author}{\bibfnamefont{F.}~\bibnamefont{{Vernizzi}}},
  \bibnamefont{and} \bibinfo{author}{\bibfnamefont{D.}~\bibnamefont{{Wands}}},
  \bibinfo{journal}{arXiv:astro-ph/0708.4321}  (\bibinfo{year}{2007}).

\bibitem[{\citenamefont{{Buchbinder} et~al.}(2007)\citenamefont{{Buchbinder},
  {Khoury}, and {Ovrut}}}]{Buchbinder_etal07}
\bibinfo{author}{\bibfnamefont{E.~I.} \bibnamefont{{Buchbinder}}},
  \bibinfo{author}{\bibfnamefont{J.}~\bibnamefont{{Khoury}}}, \bibnamefont{and}
  \bibinfo{author}{\bibfnamefont{B.~A.} \bibnamefont{{Ovrut}}},
  \bibinfo{journal}{arXiv:astro-ph/0710.5172}  (\bibinfo{year}{2007}).

\bibitem[{\citenamefont{Maldacena}(2003{\natexlab{b}})}]{Maldacena02}
\bibinfo{author}{\bibfnamefont{J.}~\bibnamefont{Maldacena}},
  \bibinfo{journal}{JHEP} \textbf{\bibinfo{volume}{0305}}, \bibinfo{pages}{013}
  (\bibinfo{year}{2003}{\natexlab{b}}).

\bibitem[{\citenamefont{{Creminelli}}(2003)}]{2003JCAP...10..003C}
\bibinfo{author}{\bibfnamefont{P.}~\bibnamefont{{Creminelli}}},
  \bibinfo{journal}{Journal of Cosmology and Astro-Particle Physics}
  \textbf{\bibinfo{volume}{10}}, \bibinfo{pages}{3} (\bibinfo{year}{2003}),
  \eprint{arXiv:astro-ph/0306122}.

\bibitem[{\citenamefont{{Alishahiha} et~al.}(2004)\citenamefont{{Alishahiha},
  {Silverstein}, and {Tong}}}]{Alishahiha_etal04}
\bibinfo{author}{\bibfnamefont{M.}~\bibnamefont{{Alishahiha}}},
  \bibinfo{author}{\bibfnamefont{E.}~\bibnamefont{{Silverstein}}},
  \bibnamefont{and} \bibinfo{author}{\bibfnamefont{D.}~\bibnamefont{{Tong}}},
  \bibinfo{journal}{Phys. Rev. D.} \textbf{\bibinfo{volume}{70}},
  \bibinfo{pages}{123505} (\bibinfo{year}{2004}).

\bibitem[{\citenamefont{{Seery} and {Lidsey}}(2005)}]{2005JCAP...06..003S}
\bibinfo{author}{\bibfnamefont{D.}~\bibnamefont{{Seery}}} \bibnamefont{and}
  \bibinfo{author}{\bibfnamefont{J.~E.} \bibnamefont{{Lidsey}}},
  \bibinfo{journal}{Journal of Cosmology and Astro-Particle Physics}
  \textbf{\bibinfo{volume}{6}}, \bibinfo{pages}{3} (\bibinfo{year}{2005}),
  \eprint{arXiv:astro-ph/0503692}.

\bibitem[{\citenamefont{{Cheung} et~al.}(2007)\citenamefont{{Cheung},
  {Creminelli}, {Fitzpatrick}, {Kaplan}, and
  {Senatore}}}]{Cheung_Creminelli_etal07}
\bibinfo{author}{\bibfnamefont{C.}~\bibnamefont{{Cheung}}},
  \bibinfo{author}{\bibfnamefont{P.}~\bibnamefont{{Creminelli}}},
  \bibinfo{author}{\bibfnamefont{A.~L.} \bibnamefont{{Fitzpatrick}}},
  \bibinfo{author}{\bibfnamefont{J.}~\bibnamefont{{Kaplan}}}, \bibnamefont{and}
  \bibinfo{author}{\bibfnamefont{L.}~\bibnamefont{{Senatore}}},
  \bibinfo{journal}{arXiv:astro-ph/0709.0293}  (\bibinfo{year}{2007}).

\bibitem[{\citenamefont{{Meerburg} et~al.}(2009)\citenamefont{{Meerburg}, {van
  der Schaar}, and {Stefano Corasaniti}}}]{2009JCAP...05..018M}
\bibinfo{author}{\bibfnamefont{P.~D.} \bibnamefont{{Meerburg}}},
  \bibinfo{author}{\bibfnamefont{J.~P.} \bibnamefont{{van der Schaar}}},
  \bibnamefont{and} \bibinfo{author}{\bibfnamefont{P.}~\bibnamefont{{Stefano
  Corasaniti}}}, \bibinfo{journal}{Journal of Cosmology and Astro-Particle
  Physics} \textbf{\bibinfo{volume}{5}}, \bibinfo{pages}{18}
  (\bibinfo{year}{2009}), \eprint{0901.4044}.

\bibitem[{\citenamefont{Penzias and Wilson}(1965)}]{PW69}
\bibinfo{author}{\bibfnamefont{A.}~\bibnamefont{Penzias}} \bibnamefont{and}
  \bibinfo{author}{\bibfnamefont{R.}~\bibnamefont{Wilson}},
  \bibinfo{journal}{ApJ.} \textbf{\bibinfo{volume}{142}}, \bibinfo{pages}{419}
  (\bibinfo{year}{1965}).

\bibitem[{\citenamefont{{Smoot} et~al.}(1992)\citenamefont{{Smoot}, {Bennett},
  {Kogut}, {Wright}, {Aymon}, {Boggess}, {Cheng}, {de Amici}, {Gulkis},
  {Hauser} et~al.}}]{COBE_first_det_1992}
\bibinfo{author}{\bibfnamefont{G.~F.} \bibnamefont{{Smoot}}},
  \bibinfo{author}{\bibfnamefont{C.~L.} \bibnamefont{{Bennett}}},
  \bibinfo{author}{\bibfnamefont{A.}~\bibnamefont{{Kogut}}},
  \bibinfo{author}{\bibfnamefont{E.~L.} \bibnamefont{{Wright}}},
  \bibinfo{author}{\bibfnamefont{J.}~\bibnamefont{{Aymon}}},
  \bibinfo{author}{\bibfnamefont{N.~W.} \bibnamefont{{Boggess}}},
  \bibinfo{author}{\bibfnamefont{E.~S.} \bibnamefont{{Cheng}}},
  \bibinfo{author}{\bibfnamefont{G.}~\bibnamefont{{de Amici}}},
  \bibinfo{author}{\bibfnamefont{S.}~\bibnamefont{{Gulkis}}},
  \bibinfo{author}{\bibfnamefont{M.~G.} \bibnamefont{{Hauser}}},
  \bibnamefont{et~al.}, \bibinfo{journal}{\apjl}
  \textbf{\bibinfo{volume}{396}}, \bibinfo{pages}{L1} (\bibinfo{year}{1992}).

\bibitem[{\citenamefont{Lewis et~al.}(2000)\citenamefont{Lewis, Challinor, and
  Lasenby}}]{camb}
\bibinfo{author}{\bibfnamefont{A.}~\bibnamefont{Lewis}},
  \bibinfo{author}{\bibfnamefont{A.}~\bibnamefont{Challinor}},
  \bibnamefont{and} \bibinfo{author}{\bibfnamefont{A.}~\bibnamefont{Lasenby}},
  \bibinfo{journal}{Astrophys. J.} \textbf{\bibinfo{volume}{538}},
  \bibinfo{pages}{473} (\bibinfo{year}{2000}), \eprint{astro-ph/9911177}.

\bibitem[{\citenamefont{Senatore
  et~al.}(2009{\natexlab{a}})\citenamefont{Senatore, Smith, and
  Zaldarriaga}}]{Senatore:2009gt}
\bibinfo{author}{\bibfnamefont{L.}~\bibnamefont{Senatore}},
  \bibinfo{author}{\bibfnamefont{K.~M.} \bibnamefont{Smith}}, \bibnamefont{and}
  \bibinfo{author}{\bibfnamefont{M.}~\bibnamefont{Zaldarriaga}}
  (\bibinfo{year}{2009}{\natexlab{a}}), \eprint{0905.3746}.

\bibitem[{\citenamefont{{Fergusson} and
  {Shellard}}(2009)}]{2009PhRvD..80d3510F}
\bibinfo{author}{\bibfnamefont{J.~R.} \bibnamefont{{Fergusson}}}
  \bibnamefont{and} \bibinfo{author}{\bibfnamefont{E.~P.~S.}
  \bibnamefont{{Shellard}}}, \bibinfo{journal}{\prd}
  \textbf{\bibinfo{volume}{80}}, \bibinfo{pages}{043510}
  (\bibinfo{year}{2009}), \eprint{0812.3413}.

\bibitem[{\citenamefont{Smith et~al.}(2009)\citenamefont{Smith, Senatore, and
  Zaldarriaga}}]{Smith:2009jr}
\bibinfo{author}{\bibfnamefont{K.~M.} \bibnamefont{Smith}},
  \bibinfo{author}{\bibfnamefont{L.}~\bibnamefont{Senatore}}, \bibnamefont{and}
  \bibinfo{author}{\bibfnamefont{M.}~\bibnamefont{Zaldarriaga}},
  \bibinfo{journal}{JCAP} \textbf{\bibinfo{volume}{0909}}, \bibinfo{pages}{006}
  (\bibinfo{year}{2009}), \eprint{0901.2572}.

\bibitem[{\citenamefont{Komatsu et~al.}(2010)}]{Komatsu:2010fb}
\bibinfo{author}{\bibfnamefont{E.}~\bibnamefont{Komatsu}} \bibnamefont{et~al.}
  (\bibinfo{year}{2010}), \eprint{1001.4538}.

\bibitem[{\citenamefont{{Creminelli}
  et~al.}(2007{\natexlab{b}})\citenamefont{{Creminelli}, {Senatore}, and
  {Zaldarriaga}}}]{creminelli_estimators_nong}
\bibinfo{author}{\bibfnamefont{P.}~\bibnamefont{{Creminelli}}},
  \bibinfo{author}{\bibfnamefont{L.}~\bibnamefont{{Senatore}}},
  \bibnamefont{and}
  \bibinfo{author}{\bibfnamefont{M.}~\bibnamefont{{Zaldarriaga}}},
  \bibinfo{journal}{Journal of Cosmology and Astro-Particle Physics}
  \textbf{\bibinfo{volume}{3}}, \bibinfo{pages}{19}
  (\bibinfo{year}{2007}{\natexlab{b}}), \eprint{arXiv:astro-ph/0606001}.

\bibitem[{\citenamefont{{Liguori} et~al.}(2007)\citenamefont{{Liguori},
  {Yadav}, {Hansen}, {Komatsu}, {Matarrese}, and
  {Wandelt}}}]{Liguori_Yadav_etal07}
\bibinfo{author}{\bibfnamefont{M.}~\bibnamefont{{Liguori}}},
  \bibinfo{author}{\bibfnamefont{A.}~\bibnamefont{{Yadav}}},
  \bibinfo{author}{\bibfnamefont{F.~K.} \bibnamefont{{Hansen}}},
  \bibinfo{author}{\bibfnamefont{E.}~\bibnamefont{{Komatsu}}},
  \bibinfo{author}{\bibfnamefont{S.}~\bibnamefont{{Matarrese}}},
  \bibnamefont{and}
  \bibinfo{author}{\bibfnamefont{B.}~\bibnamefont{{Wandelt}}},
  \bibinfo{journal}{Phys. Rev. D.} \textbf{\bibinfo{volume}{76}},
  \bibinfo{pages}{105016} (\bibinfo{year}{2007}).

\bibitem[{\citenamefont{Bartolo and Riotto}(2009)}]{Bartolo:2008sg}
\bibinfo{author}{\bibfnamefont{N.}~\bibnamefont{Bartolo}} \bibnamefont{and}
  \bibinfo{author}{\bibfnamefont{A.}~\bibnamefont{Riotto}},
  \bibinfo{journal}{JCAP} \textbf{\bibinfo{volume}{0903}}, \bibinfo{pages}{017}
  (\bibinfo{year}{2009}), \eprint{0811.4584}.

\bibitem[{\citenamefont{Sefusatti et~al.}(2009)\citenamefont{Sefusatti,
  Liguori, Yadav, Jackson, and Pajer}}]{Sefusatti:2009xu}
\bibinfo{author}{\bibfnamefont{E.}~\bibnamefont{Sefusatti}},
  \bibinfo{author}{\bibfnamefont{M.}~\bibnamefont{Liguori}},
  \bibinfo{author}{\bibfnamefont{A.~P.~S.} \bibnamefont{Yadav}},
  \bibinfo{author}{\bibfnamefont{M.~G.} \bibnamefont{Jackson}},
  \bibnamefont{and} \bibinfo{author}{\bibfnamefont{E.}~\bibnamefont{Pajer}}
  (\bibinfo{year}{2009}), \eprint{0906.0232}.

\bibitem[{\citenamefont{{Serra} and {Cooray}}(2008)}]{SerraCooray08}
\bibinfo{author}{\bibfnamefont{P.}~\bibnamefont{{Serra}}} \bibnamefont{and}
  \bibinfo{author}{\bibfnamefont{A.}~\bibnamefont{{Cooray}}},
  \bibinfo{journal}{ArXiv e-prints} \textbf{\bibinfo{volume}{801}}
  (\bibinfo{year}{2008}), \eprint{0801.3276}.

\bibitem[{\citenamefont{Goldberg and Spergel}(1999)}]{Goldberg_spergel_1999}
\bibinfo{author}{\bibfnamefont{D.~M.} \bibnamefont{Goldberg}} \bibnamefont{and}
  \bibinfo{author}{\bibfnamefont{D.~N.} \bibnamefont{Spergel}},
  \bibinfo{journal}{Phys. Rev.} \textbf{\bibinfo{volume}{D59}},
  \bibinfo{pages}{103002} (\bibinfo{year}{1999}), \eprint{astro-ph/9811251}.

\bibitem[{\citenamefont{Spergel and Goldberg}(1999)}]{Spergel_Goldberg_1999}
\bibinfo{author}{\bibfnamefont{D.~N.} \bibnamefont{Spergel}} \bibnamefont{and}
  \bibinfo{author}{\bibfnamefont{D.~M.} \bibnamefont{Goldberg}},
  \bibinfo{journal}{Phys. Rev.} \textbf{\bibinfo{volume}{D59}},
  \bibinfo{pages}{103001} (\bibinfo{year}{1999}), \eprint{astro-ph/9811252}.

\bibitem[{\citenamefont{Cooray and Hu}(2000)}]{Cooray_Hu_2000}
\bibinfo{author}{\bibfnamefont{A.~R.} \bibnamefont{Cooray}} \bibnamefont{and}
  \bibinfo{author}{\bibfnamefont{W.}~\bibnamefont{Hu}},
  \bibinfo{journal}{Astrophys. J.} \textbf{\bibinfo{volume}{534}},
  \bibinfo{pages}{533} (\bibinfo{year}{2000}), \eprint{astro-ph/9910397}.

\bibitem[{\citenamefont{{Smith} and
  {Zaldarriaga}}(2006{\natexlab{b}})}]{SmithZaldarriaga}
\bibinfo{author}{\bibfnamefont{K.~M.} \bibnamefont{{Smith}}} \bibnamefont{and}
  \bibinfo{author}{\bibfnamefont{M.}~\bibnamefont{{Zaldarriaga}}},
  \bibinfo{journal}{ArXiv Astrophysics e-prints, astro-ph/0612571}
  (\bibinfo{year}{2006}{\natexlab{b}}), \eprint{astro-ph/0612571}.

\bibitem[{\citenamefont{Mangilli and Verde}(2009)}]{Mangilli:2009dr}
\bibinfo{author}{\bibfnamefont{A.}~\bibnamefont{Mangilli}} \bibnamefont{and}
  \bibinfo{author}{\bibfnamefont{L.}~\bibnamefont{Verde}},
  \bibinfo{journal}{Phys. Rev.} \textbf{\bibinfo{volume}{D80}},
  \bibinfo{pages}{123007} (\bibinfo{year}{2009}), \eprint{0906.2317}.

\bibitem[{\citenamefont{Babich and Pierpaoli}(2008)}]{Babich:2008uw}
\bibinfo{author}{\bibfnamefont{D.}~\bibnamefont{Babich}} \bibnamefont{and}
  \bibinfo{author}{\bibfnamefont{E.}~\bibnamefont{Pierpaoli}},
  \bibinfo{journal}{Phys. Rev.} \textbf{\bibinfo{volume}{D77}},
  \bibinfo{pages}{123011} (\bibinfo{year}{2008}), \eprint{0803.1161}.

\bibitem[{\citenamefont{{Pitrou} et~al.}(2010)\citenamefont{{Pitrou}, {Uzan},
  and {Bernardeau}}}]{2010arXiv1003.0481P}
\bibinfo{author}{\bibfnamefont{C.}~\bibnamefont{{Pitrou}}},
  \bibinfo{author}{\bibfnamefont{J.}~\bibnamefont{{Uzan}}}, \bibnamefont{and}
  \bibinfo{author}{\bibfnamefont{F.}~\bibnamefont{{Bernardeau}}},
  \bibinfo{journal}{ArXiv e-prints}  (\bibinfo{year}{2010}),
  \eprint{1003.0481}.

\bibitem[{\citenamefont{Hu et~al.}(1994)\citenamefont{Hu, Scott, and
  Silk}}]{Hu:1993tc}
\bibinfo{author}{\bibfnamefont{W.}~\bibnamefont{Hu}},
  \bibinfo{author}{\bibfnamefont{D.}~\bibnamefont{Scott}}, \bibnamefont{and}
  \bibinfo{author}{\bibfnamefont{J.}~\bibnamefont{Silk}},
  \bibinfo{journal}{Phys. Rev.} \textbf{\bibinfo{volume}{D49}},
  \bibinfo{pages}{648} (\bibinfo{year}{1994}), \eprint{astro-ph/9305038}.

\bibitem[{\citenamefont{Dodelson and Jubas}(1995)}]{Dodelson:1993xz}
\bibinfo{author}{\bibfnamefont{S.}~\bibnamefont{Dodelson}} \bibnamefont{and}
  \bibinfo{author}{\bibfnamefont{J.~M.} \bibnamefont{Jubas}},
  \bibinfo{journal}{Astrophys. J.} \textbf{\bibinfo{volume}{439}},
  \bibinfo{pages}{503} (\bibinfo{year}{1995}), \eprint{astro-ph/9308019}.

\bibitem[{\citenamefont{Pyne and Carroll}(1996)}]{Pyne:1995bs}
\bibinfo{author}{\bibfnamefont{T.}~\bibnamefont{Pyne}} \bibnamefont{and}
  \bibinfo{author}{\bibfnamefont{S.~M.} \bibnamefont{Carroll}},
  \bibinfo{journal}{Phys. Rev.} \textbf{\bibinfo{volume}{D53}},
  \bibinfo{pages}{2920} (\bibinfo{year}{1996}), \eprint{astro-ph/9510041}.

\bibitem[{\citenamefont{Mollerach and Matarrese}(1997)}]{Mollerach:1997up}
\bibinfo{author}{\bibfnamefont{S.}~\bibnamefont{Mollerach}} \bibnamefont{and}
  \bibinfo{author}{\bibfnamefont{S.}~\bibnamefont{Matarrese}},
  \bibinfo{journal}{Phys. Rev.} \textbf{\bibinfo{volume}{D56}},
  \bibinfo{pages}{4494} (\bibinfo{year}{1997}), \eprint{astro-ph/9702234}.

\bibitem[{\citenamefont{Matarrese et~al.}(1998)\citenamefont{Matarrese,
  Mollerach, and Bruni}}]{Matarrese:1997ay}
\bibinfo{author}{\bibfnamefont{S.}~\bibnamefont{Matarrese}},
  \bibinfo{author}{\bibfnamefont{S.}~\bibnamefont{Mollerach}},
  \bibnamefont{and} \bibinfo{author}{\bibfnamefont{M.}~\bibnamefont{Bruni}},
  \bibinfo{journal}{Phys. Rev.} \textbf{\bibinfo{volume}{D58}},
  \bibinfo{pages}{043504} (\bibinfo{year}{1998}), \eprint{astro-ph/9707278}.

\bibitem[{\citenamefont{{Bartolo}
  et~al.}(2004{\natexlab{b}})\citenamefont{{Bartolo}, {Matarrese}, and
  {Riotto}}}]{2004JHEP...04..006B}
\bibinfo{author}{\bibfnamefont{N.}~\bibnamefont{{Bartolo}}},
  \bibinfo{author}{\bibfnamefont{S.}~\bibnamefont{{Matarrese}}},
  \bibnamefont{and} \bibinfo{author}{\bibfnamefont{A.}~\bibnamefont{{Riotto}}},
  \emph{\bibinfo{title}{{Enhancement of non-gaussianity after inflation}}}
  (\bibinfo{year}{2004}{\natexlab{b}}), \eprint{arXiv:astro-ph/0308088}.

\bibitem[{\citenamefont{{Bartolo}
  et~al.}(2004{\natexlab{c}})\citenamefont{{Bartolo}, {Matarrese}, and
  {Riotto}}}]{2004JCAP...01..003B}
\bibinfo{author}{\bibfnamefont{N.}~\bibnamefont{{Bartolo}}},
  \bibinfo{author}{\bibfnamefont{S.}~\bibnamefont{{Matarrese}}},
  \bibnamefont{and} \bibinfo{author}{\bibfnamefont{A.}~\bibnamefont{{Riotto}}},
  \bibinfo{journal}{Journal of Cosmology and Astro-Particle Physics}
  \textbf{\bibinfo{volume}{1}}, \bibinfo{pages}{3}
  (\bibinfo{year}{2004}{\natexlab{c}}), \eprint{arXiv:astro-ph/0309692}.

\bibitem[{\citenamefont{Creminelli and Zaldarriaga}(2004)}]{Creminelli:2004pv}
\bibinfo{author}{\bibfnamefont{P.}~\bibnamefont{Creminelli}} \bibnamefont{and}
  \bibinfo{author}{\bibfnamefont{M.}~\bibnamefont{Zaldarriaga}},
  \bibinfo{journal}{Phys. Rev.} \textbf{\bibinfo{volume}{D70}},
  \bibinfo{pages}{083532} (\bibinfo{year}{2004}), \eprint{astro-ph/0405428}.

\bibitem[{\citenamefont{Bartolo et~al.}(2004)\citenamefont{Bartolo, Matarrese,
  and Riotto}}]{Bartolo:2004ty}
\bibinfo{author}{\bibfnamefont{N.}~\bibnamefont{Bartolo}},
  \bibinfo{author}{\bibfnamefont{S.}~\bibnamefont{Matarrese}},
  \bibnamefont{and} \bibinfo{author}{\bibfnamefont{A.}~\bibnamefont{Riotto}},
  \bibinfo{journal}{Phys. Rev. Lett.} \textbf{\bibinfo{volume}{93}},
  \bibinfo{pages}{231301} (\bibinfo{year}{2004}), \eprint{astro-ph/0407505}.

\bibitem[{\citenamefont{Tomita}(2005)}]{Tomita:2005et}
\bibinfo{author}{\bibfnamefont{K.}~\bibnamefont{Tomita}},
  \bibinfo{journal}{Phys. Rev.} \textbf{\bibinfo{volume}{D71}},
  \bibinfo{pages}{083504} (\bibinfo{year}{2005}), \eprint{astro-ph/0501663}.

\bibitem[{\citenamefont{Bartolo et~al.}(2005)\citenamefont{Bartolo, Matarrese,
  and Riotto}}]{Bartolo:2005fp}
\bibinfo{author}{\bibfnamefont{N.}~\bibnamefont{Bartolo}},
  \bibinfo{author}{\bibfnamefont{S.}~\bibnamefont{Matarrese}},
  \bibnamefont{and} \bibinfo{author}{\bibfnamefont{A.}~\bibnamefont{Riotto}},
  \bibinfo{journal}{JCAP} \textbf{\bibinfo{volume}{0508}}, \bibinfo{pages}{010}
  (\bibinfo{year}{2005}), \eprint{astro-ph/0506410}.

\bibitem[{\citenamefont{Bartolo
  et~al.}(2006{\natexlab{a}})\citenamefont{Bartolo, Matarrese, and
  Riotto}}]{Bartolo:2005kv}
\bibinfo{author}{\bibfnamefont{N.}~\bibnamefont{Bartolo}},
  \bibinfo{author}{\bibfnamefont{S.}~\bibnamefont{Matarrese}},
  \bibnamefont{and} \bibinfo{author}{\bibfnamefont{A.}~\bibnamefont{Riotto}},
  \bibinfo{journal}{JCAP} \textbf{\bibinfo{volume}{0605}}, \bibinfo{pages}{010}
  (\bibinfo{year}{2006}{\natexlab{a}}), \eprint{astro-ph/0512481}.

\bibitem[{\citenamefont{Tomita}(2008)}]{Tomita:2007kc}
\bibinfo{author}{\bibfnamefont{K.}~\bibnamefont{Tomita}},
  \bibinfo{journal}{Phys. Rev.} \textbf{\bibinfo{volume}{D77}},
  \bibinfo{pages}{103521} (\bibinfo{year}{2008}), \eprint{0712.2511}.

\bibitem[{\citenamefont{Pitrou et~al.}(2008)\citenamefont{Pitrou, Uzan, and
  Bernardeau}}]{Pitrou:2008ak}
\bibinfo{author}{\bibfnamefont{C.}~\bibnamefont{Pitrou}},
  \bibinfo{author}{\bibfnamefont{J.-P.} \bibnamefont{Uzan}}, \bibnamefont{and}
  \bibinfo{author}{\bibfnamefont{F.}~\bibnamefont{Bernardeau}},
  \bibinfo{journal}{Phys. Rev.} \textbf{\bibinfo{volume}{D78}},
  \bibinfo{pages}{063526} (\bibinfo{year}{2008}), \eprint{0807.0341}.

\bibitem[{\citenamefont{Bartolo
  et~al.}(2006{\natexlab{b}})\citenamefont{Bartolo, Matarrese, and
  Riotto}}]{Bartolo:2006cu}
\bibinfo{author}{\bibfnamefont{N.}~\bibnamefont{Bartolo}},
  \bibinfo{author}{\bibfnamefont{S.}~\bibnamefont{Matarrese}},
  \bibnamefont{and} \bibinfo{author}{\bibfnamefont{A.}~\bibnamefont{Riotto}},
  \bibinfo{journal}{JCAP} \textbf{\bibinfo{volume}{0606}}, \bibinfo{pages}{024}
  (\bibinfo{year}{2006}{\natexlab{b}}), \eprint{astro-ph/0604416}.

\bibitem[{\citenamefont{Pitrou}(2009)}]{Pitrou:2008hy}
\bibinfo{author}{\bibfnamefont{C.}~\bibnamefont{Pitrou}},
  \bibinfo{journal}{Class. Quant. Grav.} \textbf{\bibinfo{volume}{26}},
  \bibinfo{pages}{065006} (\bibinfo{year}{2009}), \eprint{0809.3036}.

\bibitem[{\citenamefont{Senatore
  et~al.}(2009{\natexlab{b}})\citenamefont{Senatore, Tassev, and
  Zaldarriaga}}]{Senatore:2008vi}
\bibinfo{author}{\bibfnamefont{L.}~\bibnamefont{Senatore}},
  \bibinfo{author}{\bibfnamefont{S.}~\bibnamefont{Tassev}}, \bibnamefont{and}
  \bibinfo{author}{\bibfnamefont{M.}~\bibnamefont{Zaldarriaga}},
  \bibinfo{journal}{JCAP} \textbf{\bibinfo{volume}{0908}}, \bibinfo{pages}{031}
  (\bibinfo{year}{2009}{\natexlab{b}}), \eprint{0812.3652}.

\bibitem[{\citenamefont{{Boubekeur} et~al.}(2009)\citenamefont{{Boubekeur},
  {Creminelli}, {D'Amico}, {Nore{\~n}a}, and {Vernizzi}}}]{2009JCAP...08..029B}
\bibinfo{author}{\bibfnamefont{L.}~\bibnamefont{{Boubekeur}}},
  \bibinfo{author}{\bibfnamefont{P.}~\bibnamefont{{Creminelli}}},
  \bibinfo{author}{\bibfnamefont{G.}~\bibnamefont{{D'Amico}}},
  \bibinfo{author}{\bibfnamefont{J.}~\bibnamefont{{Nore{\~n}a}}},
  \bibnamefont{and}
  \bibinfo{author}{\bibfnamefont{F.}~\bibnamefont{{Vernizzi}}},
  \bibinfo{journal}{Journal of Cosmology and Astro-Particle Physics}
  \textbf{\bibinfo{volume}{8}}, \bibinfo{pages}{29} (\bibinfo{year}{2009}),
  \eprint{0906.0980}.

\bibitem[{\citenamefont{{Khatri} and
  {Wandelt}}(2010{\natexlab{a}})}]{2010ApJ...711.1310K}
\bibinfo{author}{\bibfnamefont{R.}~\bibnamefont{{Khatri}}} \bibnamefont{and}
  \bibinfo{author}{\bibfnamefont{B.~D.} \bibnamefont{{Wandelt}}},
  \bibinfo{journal}{\apj} \textbf{\bibinfo{volume}{711}}, \bibinfo{pages}{1310}
  (\bibinfo{year}{2010}{\natexlab{a}}), \eprint{0910.5218}.

\bibitem[{\citenamefont{{Bartolo} et~al.}(2010)\citenamefont{{Bartolo},
  {Matarrese}, and {Riotto}}}]{2010arXiv1001.3957B}
\bibinfo{author}{\bibfnamefont{N.}~\bibnamefont{{Bartolo}}},
  \bibinfo{author}{\bibfnamefont{S.}~\bibnamefont{{Matarrese}}},
  \bibnamefont{and} \bibinfo{author}{\bibfnamefont{A.}~\bibnamefont{{Riotto}}},
  \bibinfo{journal}{ArXiv e-prints}  (\bibinfo{year}{2010}),
  \eprint{1001.3957}.

\bibitem[{\citenamefont{Novosyadlyj}(2006)}]{Novosyadlyj:2006fw}
\bibinfo{author}{\bibfnamefont{B.}~\bibnamefont{Novosyadlyj}},
  \bibinfo{journal}{Mon. Not. Roy. Astron. Soc.}
  \textbf{\bibinfo{volume}{370}}, \bibinfo{pages}{1771} (\bibinfo{year}{2006}),
  \eprint{astro-ph/0603674}.

\bibitem[{\citenamefont{Khatri and Wandelt}(2009)}]{Khatri:2008kb}
\bibinfo{author}{\bibfnamefont{R.}~\bibnamefont{Khatri}} \bibnamefont{and}
  \bibinfo{author}{\bibfnamefont{B.~D.} \bibnamefont{Wandelt}},
  \bibinfo{journal}{Phys. Rev.} \textbf{\bibinfo{volume}{D79}},
  \bibinfo{pages}{023501} (\bibinfo{year}{2009}), \eprint{0810.4370}.

\bibitem[{\citenamefont{Senatore
  et~al.}(2009{\natexlab{c}})\citenamefont{Senatore, Tassev, and
  Zaldarriaga}}]{Senatore:2008wk}
\bibinfo{author}{\bibfnamefont{L.}~\bibnamefont{Senatore}},
  \bibinfo{author}{\bibfnamefont{S.}~\bibnamefont{Tassev}}, \bibnamefont{and}
  \bibinfo{author}{\bibfnamefont{M.}~\bibnamefont{Zaldarriaga}},
  \bibinfo{journal}{JCAP} \textbf{\bibinfo{volume}{0909}}, \bibinfo{pages}{038}
  (\bibinfo{year}{2009}{\natexlab{c}}), \eprint{0812.3658}.

\bibitem[{\citenamefont{{Khatri} and
  {Wandelt}}(2010{\natexlab{b}})}]{2010PhRvD..81j3518K}
\bibinfo{author}{\bibfnamefont{R.}~\bibnamefont{{Khatri}}} \bibnamefont{and}
  \bibinfo{author}{\bibfnamefont{B.~D.} \bibnamefont{{Wandelt}}},
  \bibinfo{journal}{\prd} \textbf{\bibinfo{volume}{81}},
  \bibinfo{pages}{103518} (\bibinfo{year}{2010}{\natexlab{b}}),
  \eprint{0903.0871}.

\bibitem[{\citenamefont{Nitta et~al.}(2009)\citenamefont{Nitta, Komatsu,
  Bartolo, Matarrese, and Riotto}}]{Nitta:2009jp}
\bibinfo{author}{\bibfnamefont{D.}~\bibnamefont{Nitta}},
  \bibinfo{author}{\bibfnamefont{E.}~\bibnamefont{Komatsu}},
  \bibinfo{author}{\bibfnamefont{N.}~\bibnamefont{Bartolo}},
  \bibinfo{author}{\bibfnamefont{S.}~\bibnamefont{Matarrese}},
  \bibnamefont{and} \bibinfo{author}{\bibfnamefont{A.}~\bibnamefont{Riotto}},
  \bibinfo{journal}{JCAP} \textbf{\bibinfo{volume}{0905}}, \bibinfo{pages}{014}
  (\bibinfo{year}{2009}), \eprint{0903.0894}.

\bibitem[{\citenamefont{Liguori and Riotto}(2008)}]{Liguori:2008vf}
\bibinfo{author}{\bibfnamefont{M.}~\bibnamefont{Liguori}} \bibnamefont{and}
  \bibinfo{author}{\bibfnamefont{A.}~\bibnamefont{Riotto}},
  \bibinfo{journal}{Phys. Rev.} \textbf{\bibinfo{volume}{D78}},
  \bibinfo{pages}{123004} (\bibinfo{year}{2008}), \eprint{0808.3255}.

\bibitem[{\citenamefont{{Su} et~al.}(2009)\citenamefont{{Su}, {Yadav}, and
  {Zaldarriaga}}}]{SYZ09}
\bibinfo{author}{\bibfnamefont{M.}~\bibnamefont{{Su}}},
  \bibinfo{author}{\bibfnamefont{A.~P.~S.} \bibnamefont{{Yadav}}},
  \bibnamefont{and}
  \bibinfo{author}{\bibfnamefont{M.}~\bibnamefont{{Zaldarriaga}}},
  \bibinfo{journal}{\prd} \textbf{\bibinfo{volume}{79}},
  \bibinfo{pages}{123002} (\bibinfo{year}{2009}), \eprint{0901.0285}.

\bibitem[{\citenamefont{Hanson et~al.}(2009)\citenamefont{Hanson, Smith,
  Challinor, and Liguori}}]{Hanson:2009kg}
\bibinfo{author}{\bibfnamefont{D.}~\bibnamefont{Hanson}},
  \bibinfo{author}{\bibfnamefont{K.~M.} \bibnamefont{Smith}},
  \bibinfo{author}{\bibfnamefont{A.}~\bibnamefont{Challinor}},
  \bibnamefont{and} \bibinfo{author}{\bibfnamefont{M.}~\bibnamefont{Liguori}},
  \bibinfo{journal}{Phys. Rev.} \textbf{\bibinfo{volume}{D80}},
  \bibinfo{pages}{083004} (\bibinfo{year}{2009}), \eprint{0905.4732}.

\bibitem[{\citenamefont{Cooray et~al.}(2008)\citenamefont{Cooray, Sarkar, and
  Serra}}]{Cooray:2008xz}
\bibinfo{author}{\bibfnamefont{A.}~\bibnamefont{Cooray}},
  \bibinfo{author}{\bibfnamefont{D.}~\bibnamefont{Sarkar}}, \bibnamefont{and}
  \bibinfo{author}{\bibfnamefont{P.}~\bibnamefont{Serra}},
  \bibinfo{journal}{Phys. Rev.} \textbf{\bibinfo{volume}{D77}},
  \bibinfo{pages}{123006} (\bibinfo{year}{2008}), \eprint{0803.4194}.

\bibitem[{\citenamefont{Donzelli et~al.}(2009)\citenamefont{Donzelli, Hansen,
  Liguori, and Maino}}]{Donzelli:2009ya}
\bibinfo{author}{\bibfnamefont{S.}~\bibnamefont{Donzelli}},
  \bibinfo{author}{\bibfnamefont{F.~K.} \bibnamefont{Hansen}},
  \bibinfo{author}{\bibfnamefont{M.}~\bibnamefont{Liguori}}, \bibnamefont{and}
  \bibinfo{author}{\bibfnamefont{D.}~\bibnamefont{Maino}},
  \bibinfo{journal}{Astrophys. J.} \textbf{\bibinfo{volume}{706}},
  \bibinfo{pages}{1226} (\bibinfo{year}{2009}), \eprint{0907.4650}.

\bibitem[{\citenamefont{Okamoto and Hu}(2002)}]{Okamoto:2002ik}
\bibinfo{author}{\bibfnamefont{T.}~\bibnamefont{Okamoto}} \bibnamefont{and}
  \bibinfo{author}{\bibfnamefont{W.}~\bibnamefont{Hu}}, \bibinfo{journal}{Phys.
  Rev.} \textbf{\bibinfo{volume}{D66}}, \bibinfo{pages}{063008}
  (\bibinfo{year}{2002}), \eprint{astro-ph/0206155}.

\bibitem[{\citenamefont{Kogo and Komatsu}(2006)}]{Kogo:2006kh}
\bibinfo{author}{\bibfnamefont{N.}~\bibnamefont{Kogo}} \bibnamefont{and}
  \bibinfo{author}{\bibfnamefont{E.}~\bibnamefont{Komatsu}},
  \bibinfo{journal}{Phys. Rev.} \textbf{\bibinfo{volume}{D73}},
  \bibinfo{pages}{083007} (\bibinfo{year}{2006}), \eprint{astro-ph/0602099}.

\bibitem[{\citenamefont{Creminelli et~al.}(2007)\citenamefont{Creminelli,
  Senatore, and Zaldarriaga}}]{Creminelli:2006gc}
\bibinfo{author}{\bibfnamefont{P.}~\bibnamefont{Creminelli}},
  \bibinfo{author}{\bibfnamefont{L.}~\bibnamefont{Senatore}}, \bibnamefont{and}
  \bibinfo{author}{\bibfnamefont{M.}~\bibnamefont{Zaldarriaga}},
  \bibinfo{journal}{JCAP} \textbf{\bibinfo{volume}{0703}}, \bibinfo{pages}{019}
  (\bibinfo{year}{2007}), \eprint{astro-ph/0606001}.

\bibitem[{\citenamefont{Munshi et~al.}(2009)}]{Munshi:2009wy}
\bibinfo{author}{\bibfnamefont{D.}~\bibnamefont{Munshi}} \bibnamefont{et~al.}
  (\bibinfo{year}{2009}), \eprint{0910.3693}.

\bibitem[{\citenamefont{{Regan} et~al.}(2010)\citenamefont{{Regan}, {Shellard},
  and {Fergusson}}}]{2010arXiv1004.2915R}
\bibinfo{author}{\bibfnamefont{D.~M.} \bibnamefont{{Regan}}},
  \bibinfo{author}{\bibfnamefont{E.~P.~S.} \bibnamefont{{Shellard}}},
  \bibnamefont{and} \bibinfo{author}{\bibfnamefont{J.~R.}
  \bibnamefont{{Fergusson}}}, \bibinfo{journal}{ArXiv e-prints}
  (\bibinfo{year}{2010}), \eprint{1004.2915}.

\bibitem[{\citenamefont{Seery et~al.}(2007)\citenamefont{Seery, Lidsey, and
  Sloth}}]{Seery:2006vu}
\bibinfo{author}{\bibfnamefont{D.}~\bibnamefont{Seery}},
  \bibinfo{author}{\bibfnamefont{J.~E.} \bibnamefont{Lidsey}},
  \bibnamefont{and} \bibinfo{author}{\bibfnamefont{M.~S.} \bibnamefont{Sloth}},
  \bibinfo{journal}{JCAP} \textbf{\bibinfo{volume}{0701}}, \bibinfo{pages}{027}
  (\bibinfo{year}{2007}), \eprint{astro-ph/0610210}.

\bibitem[{\citenamefont{{Huang} and {Shiu}}(2006)}]{2006PhRvD..74l1301H}
\bibinfo{author}{\bibfnamefont{M.}~\bibnamefont{{Huang}}} \bibnamefont{and}
  \bibinfo{author}{\bibfnamefont{G.}~\bibnamefont{{Shiu}}},
  \bibinfo{journal}{\prd} \textbf{\bibinfo{volume}{74}},
  \bibinfo{pages}{121301} (\bibinfo{year}{2006}),
  \eprint{arXiv:hep-th/0610235}.

\bibitem[{\citenamefont{{Chen} et~al.}(2009)\citenamefont{{Chen}, {Hu},
  {Huang}, {Shiu}, and {Wang}}}]{2009JCAP...08..008C}
\bibinfo{author}{\bibfnamefont{X.}~\bibnamefont{{Chen}}},
  \bibinfo{author}{\bibfnamefont{B.}~\bibnamefont{{Hu}}},
  \bibinfo{author}{\bibfnamefont{M.}~\bibnamefont{{Huang}}},
  \bibinfo{author}{\bibfnamefont{G.}~\bibnamefont{{Shiu}}}, \bibnamefont{and}
  \bibinfo{author}{\bibfnamefont{Y.}~\bibnamefont{{Wang}}},
  \bibinfo{journal}{Journal of Cosmology and Astro-Particle Physics}
  \textbf{\bibinfo{volume}{8}}, \bibinfo{pages}{8} (\bibinfo{year}{2009}),
  \eprint{0905.3494}.

\bibitem[{\citenamefont{{Arroja} et~al.}(2009)\citenamefont{{Arroja}, {Mizuno},
  {Koyama}, and {Tanaka}}}]{2009PhRvD..80d3527A}
\bibinfo{author}{\bibfnamefont{F.}~\bibnamefont{{Arroja}}},
  \bibinfo{author}{\bibfnamefont{S.}~\bibnamefont{{Mizuno}}},
  \bibinfo{author}{\bibfnamefont{K.}~\bibnamefont{{Koyama}}}, \bibnamefont{and}
  \bibinfo{author}{\bibfnamefont{T.}~\bibnamefont{{Tanaka}}},
  \bibinfo{journal}{\prd} \textbf{\bibinfo{volume}{80}},
  \bibinfo{pages}{043527} (\bibinfo{year}{2009}), \eprint{0905.3641}.

\bibitem[{\citenamefont{Engel et~al.}(2009)\citenamefont{Engel, Lee, and
  Wise}}]{Engel:2008fu}
\bibinfo{author}{\bibfnamefont{K.~T.} \bibnamefont{Engel}},
  \bibinfo{author}{\bibfnamefont{K.~S.~M.} \bibnamefont{Lee}},
  \bibnamefont{and} \bibinfo{author}{\bibfnamefont{M.~B.} \bibnamefont{Wise}},
  \bibinfo{journal}{Phys. Rev.} \textbf{\bibinfo{volume}{D79}},
  \bibinfo{pages}{103530} (\bibinfo{year}{2009}), \eprint{0811.3964}.

\bibitem[{\citenamefont{Byrnes et~al.}(2006)\citenamefont{Byrnes, Sasaki, and
  Wands}}]{Byrnes:2006vq}
\bibinfo{author}{\bibfnamefont{C.~T.} \bibnamefont{Byrnes}},
  \bibinfo{author}{\bibfnamefont{M.}~\bibnamefont{Sasaki}}, \bibnamefont{and}
  \bibinfo{author}{\bibfnamefont{D.}~\bibnamefont{Wands}},
  \bibinfo{journal}{Phys. Rev.} \textbf{\bibinfo{volume}{D74}},
  \bibinfo{pages}{123519} (\bibinfo{year}{2006}), \eprint{astro-ph/0611075}.

\bibitem[{\citenamefont{{Lehners} and
  {Steinhardt}}(2009)}]{2009PhRvD..80j3520L}
\bibinfo{author}{\bibfnamefont{J.}~\bibnamefont{{Lehners}}} \bibnamefont{and}
  \bibinfo{author}{\bibfnamefont{P.~J.} \bibnamefont{{Steinhardt}}},
  \bibinfo{journal}{\prd} \textbf{\bibinfo{volume}{80}},
  \bibinfo{pages}{103520} (\bibinfo{year}{2009}), \eprint{0909.2558}.

\bibitem[{\citenamefont{Smidt et~al.}(2010)}]{Smidt:2010sv}
\bibinfo{author}{\bibfnamefont{J.}~\bibnamefont{Smidt}} \bibnamefont{et~al.}
  (\bibinfo{year}{2010}), \eprint{1001.5026}.

\bibitem[{\citenamefont{Mecke et~al.}(1994)\citenamefont{Mecke, Buchert, and
  Wagner}}]{Mecke_etal1994}
\bibinfo{author}{\bibfnamefont{K.~R.} \bibnamefont{Mecke}},
  \bibinfo{author}{\bibfnamefont{T.}~\bibnamefont{Buchert}}, \bibnamefont{and}
  \bibinfo{author}{\bibfnamefont{H.}~\bibnamefont{Wagner}},
  \bibinfo{journal}{Astronomy and Astrophysics} \textbf{\bibinfo{volume}{288}},
  \bibinfo{pages}{697} (\bibinfo{year}{1994}).

\bibitem[{\citenamefont{Schmalzing and Buchert}(1997)}]{Schmalzing:1997aj}
\bibinfo{author}{\bibfnamefont{J.}~\bibnamefont{Schmalzing}} \bibnamefont{and}
  \bibinfo{author}{\bibfnamefont{T.}~\bibnamefont{Buchert}},
  \bibinfo{journal}{Astrophys. J.} \textbf{\bibinfo{volume}{482}},
  \bibinfo{pages}{L1} (\bibinfo{year}{1997}), \eprint{astro-ph/9702130}.

\bibitem[{\citenamefont{Schmalzing and Gorski}(1997)}]{Schmalzing:1997uc}
\bibinfo{author}{\bibfnamefont{J.}~\bibnamefont{Schmalzing}} \bibnamefont{and}
  \bibinfo{author}{\bibfnamefont{K.~M.} \bibnamefont{Gorski}}
  (\bibinfo{year}{1997}), \eprint{astro-ph/9710185}.

\bibitem[{\citenamefont{Winitzki and Kosowsky}(1998)}]{Winitzki:1997jj}
\bibinfo{author}{\bibfnamefont{S.}~\bibnamefont{Winitzki}} \bibnamefont{and}
  \bibinfo{author}{\bibfnamefont{A.}~\bibnamefont{Kosowsky}},
  \bibinfo{journal}{New Astron.} \textbf{\bibinfo{volume}{3}},
  \bibinfo{pages}{75} (\bibinfo{year}{1998}), \eprint{astro-ph/9710164}.

\bibitem[{\citenamefont{Matsubara}(1994)}]{Matsubara:1994wn}
\bibinfo{author}{\bibfnamefont{T.}~\bibnamefont{Matsubara}},
  \bibinfo{journal}{\apj} \textbf{\bibinfo{volume}{434}}, \bibinfo{pages}{L43}
  (\bibinfo{year}{1994}), \eprint{astro-ph/9405037}.

\bibitem[{\citenamefont{Matsubara}(2003)}]{Matsubara2003}
\bibinfo{author}{\bibfnamefont{T.}~\bibnamefont{Matsubara}},
  \bibinfo{journal}{\apj} \textbf{\bibinfo{volume}{584}}, \bibinfo{pages}{1}
  (\bibinfo{year}{2003}).

\bibitem[{\citenamefont{{Hikage} et~al.}(2006)\citenamefont{{Hikage},
  {Komatsu}, and {Matsubara}}}]{Hikage_Komatsu_etal06}
\bibinfo{author}{\bibfnamefont{C.}~\bibnamefont{{Hikage}}},
  \bibinfo{author}{\bibfnamefont{E.}~\bibnamefont{{Komatsu}}},
  \bibnamefont{and}
  \bibinfo{author}{\bibfnamefont{T.}~\bibnamefont{{Matsubara}}},
  \bibinfo{journal}{Astrophys. J.} \textbf{\bibinfo{volume}{653}},
  \bibinfo{pages}{11} (\bibinfo{year}{2006}), \eprint{arXiv:astro-ph/0607284}.

\bibitem[{\citenamefont{Hikage et~al.}(2008)}]{Hikage:2008gy}
\bibinfo{author}{\bibfnamefont{C.}~\bibnamefont{Hikage}} \bibnamefont{et~al.},
  \bibinfo{journal}{Mon. Not. Roy. Astron. Soc.}
  \textbf{\bibinfo{volume}{389}}, \bibinfo{pages}{1439} (\bibinfo{year}{2008}),
  \eprint{0802.3677}.

\bibitem[{\citenamefont{Hikage et~al.}(2006)\citenamefont{Hikage, Komatsu, and
  Matsubara}}]{Hikage:2006fe}
\bibinfo{author}{\bibfnamefont{C.}~\bibnamefont{Hikage}},
  \bibinfo{author}{\bibfnamefont{E.}~\bibnamefont{Komatsu}}, \bibnamefont{and}
  \bibinfo{author}{\bibfnamefont{T.}~\bibnamefont{Matsubara}},
  \bibinfo{journal}{Astrophys. J.} \textbf{\bibinfo{volume}{653}},
  \bibinfo{pages}{11} (\bibinfo{year}{2006}), \eprint{astro-ph/0607284}.

\bibitem[{\citenamefont{{Aghanim} and {Forni}}(1999)}]{1999A&A...347..409A}
\bibinfo{author}{\bibfnamefont{N.}~\bibnamefont{{Aghanim}}} \bibnamefont{and}
  \bibinfo{author}{\bibfnamefont{O.}~\bibnamefont{{Forni}}},
  \bibinfo{journal}{\aap} \textbf{\bibinfo{volume}{347}}, \bibinfo{pages}{409}
  (\bibinfo{year}{1999}), \eprint{arXiv:astro-ph/9905124}.

\bibitem[{\citenamefont{{Hobson} et~al.}(1999)\citenamefont{{Hobson}, {Jones},
  and {Lasenby}}}]{1999MNRAS.309..125H}
\bibinfo{author}{\bibfnamefont{M.~P.} \bibnamefont{{Hobson}}},
  \bibinfo{author}{\bibfnamefont{A.~W.} \bibnamefont{{Jones}}},
  \bibnamefont{and} \bibinfo{author}{\bibfnamefont{A.~N.}
  \bibnamefont{{Lasenby}}}, \bibinfo{journal}{\mnras}
  \textbf{\bibinfo{volume}{309}}, \bibinfo{pages}{125} (\bibinfo{year}{1999}),
  \eprint{arXiv:astro-ph/9810200}.

\bibitem[{\citenamefont{{Vielva} et~al.}(2004)\citenamefont{{Vielva},
  {Mart{\'{\i}}nez-Gonz{\'a}lez}, {Barreiro}, {Sanz}, and
  {Cay{\'o}n}}}]{2004ApJ...609...22V}
\bibinfo{author}{\bibfnamefont{P.}~\bibnamefont{{Vielva}}},
  \bibinfo{author}{\bibfnamefont{E.}~\bibnamefont{{Mart{\'{\i}}nez-Gonz{\'a}le%
z}}}, \bibinfo{author}{\bibfnamefont{R.~B.} \bibnamefont{{Barreiro}}},
  \bibinfo{author}{\bibfnamefont{J.~L.} \bibnamefont{{Sanz}}},
  \bibnamefont{and}
  \bibinfo{author}{\bibfnamefont{L.}~\bibnamefont{{Cay{\'o}n}}},
  \bibinfo{journal}{\apj} \textbf{\bibinfo{volume}{609}}, \bibinfo{pages}{22}
  (\bibinfo{year}{2004}), \eprint{arXiv:astro-ph/0310273}.

\bibitem[{\citenamefont{{Starck} et~al.}(2004)\citenamefont{{Starck},
  {Aghanim}, and {Forni}}}]{2004A&A...416....9S}
\bibinfo{author}{\bibfnamefont{J.}~\bibnamefont{{Starck}}},
  \bibinfo{author}{\bibfnamefont{N.}~\bibnamefont{{Aghanim}}},
  \bibnamefont{and} \bibinfo{author}{\bibfnamefont{O.}~\bibnamefont{{Forni}}},
  \bibinfo{journal}{\aap} \textbf{\bibinfo{volume}{416}}, \bibinfo{pages}{9}
  (\bibinfo{year}{2004}), \eprint{arXiv:astro-ph/0311577}.

\bibitem[{\citenamefont{{Jin} et~al.}(2005)\citenamefont{{Jin}, {Starck},
  {Donoho}, {Aghanim}, and {Forni}}}]{2005JASP...15.2470J}
\bibinfo{author}{\bibfnamefont{J.}~\bibnamefont{{Jin}}},
  \bibinfo{author}{\bibfnamefont{J.}~\bibnamefont{{Starck}}},
  \bibinfo{author}{\bibfnamefont{D.~L.} \bibnamefont{{Donoho}}},
  \bibinfo{author}{\bibfnamefont{N.}~\bibnamefont{{Aghanim}}},
  \bibnamefont{and} \bibinfo{author}{\bibfnamefont{O.}~\bibnamefont{{Forni}}},
  \bibinfo{journal}{EURASIP Journal on Applied Signal Processing, Vol.~2005, No
  15, page 2470} \textbf{\bibinfo{volume}{15}}, \bibinfo{pages}{2470}
  (\bibinfo{year}{2005}), \eprint{arXiv:astro-ph/0503374}.

\bibitem[{\citenamefont{{Vielva} et~al.}(2006)\citenamefont{{Vielva}, {Wiaux},
  {Mart{\'{\i}}nez-Gonz{\'a}lez}, and {Vandergheynst}}}]{2006NewAR..50..880V}
\bibinfo{author}{\bibfnamefont{P.}~\bibnamefont{{Vielva}}},
  \bibinfo{author}{\bibfnamefont{Y.}~\bibnamefont{{Wiaux}}},
  \bibinfo{author}{\bibfnamefont{E.}~\bibnamefont{{Mart{\'{\i}}nez-Gonz{\'a}le%
z}}}, \bibnamefont{and}
  \bibinfo{author}{\bibfnamefont{P.}~\bibnamefont{{Vandergheynst}}},
  \bibinfo{journal}{New Astronomy Review} \textbf{\bibinfo{volume}{50}},
  \bibinfo{pages}{880} (\bibinfo{year}{2006}), \eprint{arXiv:astro-ph/0609147}.

\bibitem[{\citenamefont{{McEwen} et~al.}(2007)\citenamefont{{McEwen}, {Vielva},
  {Wiaux}, {Barreiro}, {Cayon}, {Hobson}, {Lasenby}, {Martinez-Gonzalez}, and
  {Sanz}}}]{2007JFAA...13..495M}
\bibinfo{author}{\bibfnamefont{J.~D.} \bibnamefont{{McEwen}}},
  \bibinfo{author}{\bibfnamefont{P.}~\bibnamefont{{Vielva}}},
  \bibinfo{author}{\bibfnamefont{Y.}~\bibnamefont{{Wiaux}}},
  \bibinfo{author}{\bibfnamefont{R.~B.} \bibnamefont{{Barreiro}}},
  \bibinfo{author}{\bibfnamefont{L.}~\bibnamefont{{Cayon}}},
  \bibinfo{author}{\bibfnamefont{M.~P.} \bibnamefont{{Hobson}}},
  \bibinfo{author}{\bibfnamefont{A.~N.} \bibnamefont{{Lasenby}}},
  \bibinfo{author}{\bibfnamefont{E.}~\bibnamefont{{Martinez-Gonzalez}}},
  \bibnamefont{and} \bibinfo{author}{\bibfnamefont{J.~L.}
  \bibnamefont{{Sanz}}}, \bibinfo{journal}{Journal of Fourier Analysis and
  Applications} \textbf{\bibinfo{volume}{13}}, \bibinfo{pages}{495}
  (\bibinfo{year}{2007}), \eprint{0704.3158}.

\bibitem[{\citenamefont{{Cay{\'o}n} et~al.}(2003)\citenamefont{{Cay{\'o}n},
  {Mart{\'{\i}}nez-Gonz{\'a}lez}, {Arg{\"u}eso}, {Banday}, and
  {G{\'o}rski}}}]{2003MNRAS.339.1189C}
\bibinfo{author}{\bibfnamefont{L.}~\bibnamefont{{Cay{\'o}n}}},
  \bibinfo{author}{\bibfnamefont{E.}~\bibnamefont{{Mart{\'{\i}}nez-Gonz{\'a}le%
z}}}, \bibinfo{author}{\bibfnamefont{F.}~\bibnamefont{{Arg{\"u}eso}}},
  \bibinfo{author}{\bibfnamefont{A.~J.} \bibnamefont{{Banday}}},
  \bibnamefont{and} \bibinfo{author}{\bibfnamefont{K.~M.}
  \bibnamefont{{G{\'o}rski}}}, \bibinfo{journal}{\mnras}
  \textbf{\bibinfo{volume}{339}}, \bibinfo{pages}{1189} (\bibinfo{year}{2003}),
  \eprint{arXiv:astro-ph/0211399}.

\bibitem[{\citenamefont{{Curto}
  et~al.}(2009{\natexlab{a}})\citenamefont{{Curto},
  {Mart{\'{\i}}nez-Gonz{\'a}lez}, {Mukherjee}, {Barreiro}, {Hansen}, {Liguori},
  and {Matarrese}}}]{2009MNRAS.393..615C}
\bibinfo{author}{\bibfnamefont{A.}~\bibnamefont{{Curto}}},
  \bibinfo{author}{\bibfnamefont{E.}~\bibnamefont{{Mart{\'{\i}}nez-Gonz{\'a}le%
z}}}, \bibinfo{author}{\bibfnamefont{P.}~\bibnamefont{{Mukherjee}}},
  \bibinfo{author}{\bibfnamefont{R.~B.} \bibnamefont{{Barreiro}}},
  \bibinfo{author}{\bibfnamefont{F.~K.} \bibnamefont{{Hansen}}},
  \bibinfo{author}{\bibfnamefont{M.}~\bibnamefont{{Liguori}}},
  \bibnamefont{and}
  \bibinfo{author}{\bibfnamefont{S.}~\bibnamefont{{Matarrese}}},
  \bibinfo{journal}{\mnras} \textbf{\bibinfo{volume}{393}},
  \bibinfo{pages}{615} (\bibinfo{year}{2009}{\natexlab{a}}),
  \eprint{0807.0231}.

\bibitem[{\citenamefont{{Curto}
  et~al.}(2009{\natexlab{b}})\citenamefont{{Curto},
  {Mart{\'{\i}}nez-Gonz{\'a}lez}, and {Barreiro}}}]{2009ApJ...706..399C}
\bibinfo{author}{\bibfnamefont{A.}~\bibnamefont{{Curto}}},
  \bibinfo{author}{\bibfnamefont{E.}~\bibnamefont{{Mart{\'{\i}}nez-Gonz{\'a}le%
z}}}, \bibnamefont{and} \bibinfo{author}{\bibfnamefont{R.~B.}
  \bibnamefont{{Barreiro}}}, \bibinfo{journal}{\apj}
  \textbf{\bibinfo{volume}{706}}, \bibinfo{pages}{399}
  (\bibinfo{year}{2009}{\natexlab{b}}), \eprint{0902.1523}.

\bibitem[{\citenamefont{Narcowich
  et~al.}(2006{\natexlab{a}})\citenamefont{Narcowich, Petrushev, and
  Ward}}]{MR2237162}
\bibinfo{author}{\bibfnamefont{F.~J.} \bibnamefont{Narcowich}},
  \bibinfo{author}{\bibfnamefont{P.}~\bibnamefont{Petrushev}},
  \bibnamefont{and} \bibinfo{author}{\bibfnamefont{J.~D.} \bibnamefont{Ward}},
  \bibinfo{journal}{SIAM J. Math. Anal.} \textbf{\bibinfo{volume}{38}},
  \bibinfo{pages}{574} (\bibinfo{year}{2006}{\natexlab{a}}), ISSN
  \bibinfo{issn}{0036-1410},
  \urlprefix\url{http://dx.doi.org/10.1137/040614359}.

\bibitem[{\citenamefont{Narcowich
  et~al.}(2006{\natexlab{b}})\citenamefont{Narcowich, Petrushev, and
  Ward}}]{MR2253732}
\bibinfo{author}{\bibfnamefont{F.}~\bibnamefont{Narcowich}},
  \bibinfo{author}{\bibfnamefont{P.}~\bibnamefont{Petrushev}},
  \bibnamefont{and} \bibinfo{author}{\bibfnamefont{J.}~\bibnamefont{Ward}},
  \bibinfo{journal}{J. Funct. Anal.} \textbf{\bibinfo{volume}{238}},
  \bibinfo{pages}{530} (\bibinfo{year}{2006}{\natexlab{b}}), ISSN
  \bibinfo{issn}{0022-1236}.

\bibitem[{\citenamefont{Baldi et~al.}(2008)\citenamefont{Baldi, Kerkyacharian,
  Marinucci, and Picard}}]{baldi-2008-99}
\bibinfo{author}{\bibfnamefont{P.}~\bibnamefont{Baldi}},
  \bibinfo{author}{\bibfnamefont{G.}~\bibnamefont{Kerkyacharian}},
  \bibinfo{author}{\bibfnamefont{D.}~\bibnamefont{Marinucci}},
  \bibnamefont{and} \bibinfo{author}{\bibfnamefont{D.}~\bibnamefont{Picard}},
  \bibinfo{journal}{Journal of Multivariate Analysis}
  \textbf{\bibinfo{volume}{99}}, \bibinfo{pages}{606} (\bibinfo{year}{2008}),
  \urlprefix\url{http://www.citebase.org/abstract?id=oai:arXiv.org:math/060615%
4}.

\bibitem[{\citenamefont{Baldi et~al.}(2009{\natexlab{a}})\citenamefont{Baldi,
  Kerkyacharian, Marinucci, and Picard}}]{baldi-2009-37}
\bibinfo{author}{\bibfnamefont{P.}~\bibnamefont{Baldi}},
  \bibinfo{author}{\bibfnamefont{G.}~\bibnamefont{Kerkyacharian}},
  \bibinfo{author}{\bibfnamefont{D.}~\bibnamefont{Marinucci}},
  \bibnamefont{and} \bibinfo{author}{\bibfnamefont{D.}~\bibnamefont{Picard}},
  \bibinfo{journal}{ANNALS OF STATISTICS} \textbf{\bibinfo{volume}{37}},
  \bibinfo{pages}{1150} (\bibinfo{year}{2009}{\natexlab{a}}),
  \urlprefix\url{doi:10.1214/08-AOS601}.

\bibitem[{\citenamefont{Baldi et~al.}(2009{\natexlab{b}})\citenamefont{Baldi,
  Kerkyacharian, Marinucci, and Picard}}]{baldi-2009-15}
\bibinfo{author}{\bibfnamefont{P.}~\bibnamefont{Baldi}},
  \bibinfo{author}{\bibfnamefont{G.}~\bibnamefont{Kerkyacharian}},
  \bibinfo{author}{\bibfnamefont{D.}~\bibnamefont{Marinucci}},
  \bibnamefont{and} \bibinfo{author}{\bibfnamefont{D.}~\bibnamefont{Picard}},
  \bibinfo{journal}{BERNOULLI} \textbf{\bibinfo{volume}{15}},
  \bibinfo{pages}{438} (\bibinfo{year}{2009}{\natexlab{b}}),
  \urlprefix\url{doi:10.3150/08-BEJ164}.

\bibitem[{\citenamefont{{Lan} and {Marinucci}}(2008)}]{2008EJSta...2..332L}
\bibinfo{author}{\bibfnamefont{X.}~\bibnamefont{{Lan}}} \bibnamefont{and}
  \bibinfo{author}{\bibfnamefont{D.}~\bibnamefont{{Marinucci}}},
  \bibinfo{journal}{Electronic Journal of Statistics}
  \textbf{\bibinfo{volume}{2}}, \bibinfo{pages}{332} (\bibinfo{year}{2008}),
  \eprint{0802.4020}.

\bibitem[{\citenamefont{{Pietrobon} et~al.}(2009)\citenamefont{{Pietrobon},
  {Cabella}, {Balbi}, {de Gasperis}, and {Vittorio}}}]{2009MNRAS.396.1682P}
\bibinfo{author}{\bibfnamefont{D.}~\bibnamefont{{Pietrobon}}},
  \bibinfo{author}{\bibfnamefont{P.}~\bibnamefont{{Cabella}}},
  \bibinfo{author}{\bibfnamefont{A.}~\bibnamefont{{Balbi}}},
  \bibinfo{author}{\bibfnamefont{G.}~\bibnamefont{{de Gasperis}}},
  \bibnamefont{and}
  \bibinfo{author}{\bibfnamefont{N.}~\bibnamefont{{Vittorio}}},
  \bibinfo{journal}{\mnras} \textbf{\bibinfo{volume}{396}},
  \bibinfo{pages}{1682} (\bibinfo{year}{2009}), \eprint{0812.2478}.

\bibitem[{\citenamefont{{Pietrobon} et~al.}(2010)\citenamefont{{Pietrobon},
  {Cabella}, {Balbi}, {Crittenden}, {de Gasperis}, and
  {Vittorio}}}]{2010MNRAS.402L..34P}
\bibinfo{author}{\bibfnamefont{D.}~\bibnamefont{{Pietrobon}}},
  \bibinfo{author}{\bibfnamefont{P.}~\bibnamefont{{Cabella}}},
  \bibinfo{author}{\bibfnamefont{A.}~\bibnamefont{{Balbi}}},
  \bibinfo{author}{\bibfnamefont{R.}~\bibnamefont{{Crittenden}}},
  \bibinfo{author}{\bibfnamefont{G.}~\bibnamefont{{de Gasperis}}},
  \bibnamefont{and}
  \bibinfo{author}{\bibfnamefont{N.}~\bibnamefont{{Vittorio}}},
  \bibinfo{journal}{\mnras} \textbf{\bibinfo{volume}{402}},
  \bibinfo{pages}{L34} (\bibinfo{year}{2010}), \eprint{0905.3702}.

\bibitem[{\citenamefont{{Cabella} et~al.}(2010)\citenamefont{{Cabella},
  {Pietrobon}, {Veneziani}, {Balbi}, {Crittenden}, {de Gasperis},
  {Quercellini}, and {Vittorio}}}]{2010MNRAS.tmp..504C}
\bibinfo{author}{\bibfnamefont{P.}~\bibnamefont{{Cabella}}},
  \bibinfo{author}{\bibfnamefont{D.}~\bibnamefont{{Pietrobon}}},
  \bibinfo{author}{\bibfnamefont{M.}~\bibnamefont{{Veneziani}}},
  \bibinfo{author}{\bibfnamefont{A.}~\bibnamefont{{Balbi}}},
  \bibinfo{author}{\bibfnamefont{R.}~\bibnamefont{{Crittenden}}},
  \bibinfo{author}{\bibfnamefont{G.}~\bibnamefont{{de Gasperis}}},
  \bibinfo{author}{\bibfnamefont{C.}~\bibnamefont{{Quercellini}}},
  \bibnamefont{and}
  \bibinfo{author}{\bibfnamefont{N.}~\bibnamefont{{Vittorio}}},
  \bibinfo{journal}{\mnras} pp. \bibinfo{pages}{504--+} (\bibinfo{year}{2010}),
  \eprint{0910.4362}.

\bibitem[{\citenamefont{Rudjord et~al.}(2009)}]{Rudjord:2009mh}
\bibinfo{author}{\bibfnamefont{O.}~\bibnamefont{Rudjord}} \bibnamefont{et~al.},
  \bibinfo{journal}{Astrophys. J.} \textbf{\bibinfo{volume}{701}},
  \bibinfo{pages}{369} (\bibinfo{year}{2009}), \eprint{0901.3154}.

\bibitem[{\citenamefont{{Rudjord} et~al.}(2010)\citenamefont{{Rudjord},
  {Hansen}, {Lan}, {Liguori}, {Marinucci}, and
  {Matarrese}}}]{2010ApJ...708.1321R}
\bibinfo{author}{\bibfnamefont{{\O}.}~\bibnamefont{{Rudjord}}},
  \bibinfo{author}{\bibfnamefont{F.~K.} \bibnamefont{{Hansen}}},
  \bibinfo{author}{\bibfnamefont{X.}~\bibnamefont{{Lan}}},
  \bibinfo{author}{\bibfnamefont{M.}~\bibnamefont{{Liguori}}},
  \bibinfo{author}{\bibfnamefont{D.}~\bibnamefont{{Marinucci}}},
  \bibnamefont{and}
  \bibinfo{author}{\bibfnamefont{S.}~\bibnamefont{{Matarrese}}},
  \bibinfo{journal}{\apj} \textbf{\bibinfo{volume}{708}}, \bibinfo{pages}{1321}
  (\bibinfo{year}{2010}), \eprint{0906.3232}.

\bibitem[{\citenamefont{{Rocha} et~al.}(2001)\citenamefont{{Rocha}, {Magueijo},
  {Hobson}, and {Lasenby}}}]{2001PhRvD..64f3512R}
\bibinfo{author}{\bibfnamefont{G.}~\bibnamefont{{Rocha}}},
  \bibinfo{author}{\bibfnamefont{J.}~\bibnamefont{{Magueijo}}},
  \bibinfo{author}{\bibfnamefont{M.}~\bibnamefont{{Hobson}}}, \bibnamefont{and}
  \bibinfo{author}{\bibfnamefont{A.}~\bibnamefont{{Lasenby}}},
  \bibinfo{journal}{\prd} \textbf{\bibinfo{volume}{64}},
  \bibinfo{pages}{063512} (\bibinfo{year}{2001}),
  \eprint{arXiv:astro-ph/0008070}.

\bibitem[{\citenamefont{{Vielva} and {Sanz}}(2008)}]{2008arXiv0812.1756V}
\bibinfo{author}{\bibfnamefont{P.}~\bibnamefont{{Vielva}}} \bibnamefont{and}
  \bibinfo{author}{\bibfnamefont{J.~L.} \bibnamefont{{Sanz}}},
  \bibinfo{journal}{ArXiv e-prints}  (\bibinfo{year}{2008}),
  \eprint{0812.1756}.

\bibitem[{\citenamefont{{En{\ss}lin} et~al.}(2009)\citenamefont{{En{\ss}lin},
  {Frommert}, and {Kitaura}}}]{2009PhRvD..80j5005E}
\bibinfo{author}{\bibfnamefont{T.~A.} \bibnamefont{{En{\ss}lin}}},
  \bibinfo{author}{\bibfnamefont{M.}~\bibnamefont{{Frommert}}},
  \bibnamefont{and} \bibinfo{author}{\bibfnamefont{F.~S.}
  \bibnamefont{{Kitaura}}}, \bibinfo{journal}{\prd}
  \textbf{\bibinfo{volume}{80}}, \bibinfo{pages}{105005}
  (\bibinfo{year}{2009}), \eprint{0806.3474}.

\bibitem[{\citenamefont{{Elsner} et~al.}(2010)\citenamefont{{Elsner},
  {Wandelt}, and {Schneider}}}]{2010arXiv1002.1713E}
\bibinfo{author}{\bibfnamefont{F.}~\bibnamefont{{Elsner}}},
  \bibinfo{author}{\bibfnamefont{B.~D.} \bibnamefont{{Wandelt}}},
  \bibnamefont{and} \bibinfo{author}{\bibfnamefont{M.~D.}
  \bibnamefont{{Schneider}}}, \bibinfo{journal}{ArXiv e-prints}
  (\bibinfo{year}{2010}), \eprint{1002.1713}.

\bibitem[{\citenamefont{{Komatsu} et~al.}(2002)\citenamefont{{Komatsu},
  {Wandelt}, {Spergel}, {Banday}, and {G{\'o}rski}}}]{2002ApJ...566...19K}
\bibinfo{author}{\bibfnamefont{E.}~\bibnamefont{{Komatsu}}},
  \bibinfo{author}{\bibfnamefont{B.~D.} \bibnamefont{{Wandelt}}},
  \bibinfo{author}{\bibfnamefont{D.~N.} \bibnamefont{{Spergel}}},
  \bibinfo{author}{\bibfnamefont{A.~J.} \bibnamefont{{Banday}}},
  \bibnamefont{and} \bibinfo{author}{\bibfnamefont{K.~M.}
  \bibnamefont{{G{\'o}rski}}}, \bibinfo{journal}{\apj}
  \textbf{\bibinfo{volume}{566}}, \bibinfo{pages}{19} (\bibinfo{year}{2002}),
  \eprint{arXiv:astro-ph/0107605}.

\bibitem[{\citenamefont{Santos et~al.}(2003)}]{Santos:2002df}
\bibinfo{author}{\bibfnamefont{M.~G.} \bibnamefont{Santos}}
  \bibnamefont{et~al.}, \bibinfo{journal}{Mon. Not. Roy. Astron. Soc.}
  \textbf{\bibinfo{volume}{341}}, \bibinfo{pages}{623} (\bibinfo{year}{2003}),
  \eprint{astro-ph/0211123}.

\bibitem[{\citenamefont{{Smith} et~al.}(2004)\citenamefont{{Smith}, {Rocha},
  {Challinor}, {Battye}, {Carreira}, {Cleary}, {Davies}, {Davis}, {Dickinson},
  {Genova-Santos} et~al.}}]{2004MNRAS.352..887S}
\bibinfo{author}{\bibfnamefont{S.}~\bibnamefont{{Smith}}},
  \bibinfo{author}{\bibfnamefont{G.}~\bibnamefont{{Rocha}}},
  \bibinfo{author}{\bibfnamefont{A.}~\bibnamefont{{Challinor}}},
  \bibinfo{author}{\bibfnamefont{R.~A.} \bibnamefont{{Battye}}},
  \bibinfo{author}{\bibfnamefont{P.}~\bibnamefont{{Carreira}}},
  \bibinfo{author}{\bibfnamefont{K.}~\bibnamefont{{Cleary}}},
  \bibinfo{author}{\bibfnamefont{R.~D.} \bibnamefont{{Davies}}},
  \bibinfo{author}{\bibfnamefont{R.~J.} \bibnamefont{{Davis}}},
  \bibinfo{author}{\bibfnamefont{C.}~\bibnamefont{{Dickinson}}},
  \bibinfo{author}{\bibfnamefont{R.}~\bibnamefont{{Genova-Santos}}},
  \bibnamefont{et~al.}, \bibinfo{journal}{\mnras}
  \textbf{\bibinfo{volume}{352}}, \bibinfo{pages}{887} (\bibinfo{year}{2004}),
  \eprint{arXiv:astro-ph/0401618}.

\bibitem[{\citenamefont{{de Troia} et~al.}(2007)\citenamefont{{de Troia},
  {Ade}, {Bock}, {Bond}, {Borrill}, {Boscaleri}, {Cabella}, {Contaldi},
  {Crill}, {de Bernardis} et~al.}}]{2007NewAR..51..250D}
\bibinfo{author}{\bibfnamefont{G.}~\bibnamefont{{de Troia}}},
  \bibinfo{author}{\bibfnamefont{P.~A.~R.} \bibnamefont{{Ade}}},
  \bibinfo{author}{\bibfnamefont{J.~J.} \bibnamefont{{Bock}}},
  \bibinfo{author}{\bibfnamefont{J.~R.} \bibnamefont{{Bond}}},
  \bibinfo{author}{\bibfnamefont{J.}~\bibnamefont{{Borrill}}},
  \bibinfo{author}{\bibfnamefont{A.}~\bibnamefont{{Boscaleri}}},
  \bibinfo{author}{\bibfnamefont{P.}~\bibnamefont{{Cabella}}},
  \bibinfo{author}{\bibfnamefont{C.~R.} \bibnamefont{{Contaldi}}},
  \bibinfo{author}{\bibfnamefont{B.~P.} \bibnamefont{{Crill}}},
  \bibinfo{author}{\bibfnamefont{P.}~\bibnamefont{{de Bernardis}}},
  \bibnamefont{et~al.}, \bibinfo{journal}{New Astronomy Review}
  \textbf{\bibinfo{volume}{51}}, \bibinfo{pages}{250} (\bibinfo{year}{2007}),
  \eprint{0705.1615}.

\bibitem[{\citenamefont{Curto et~al.}(2008)}]{Curto:2008gg}
\bibinfo{author}{\bibfnamefont{A.}~\bibnamefont{Curto}} \bibnamefont{et~al.}
  (\bibinfo{year}{2008}), \eprint{0804.0136}.

\bibitem[{\citenamefont{Natoli et~al.}(2009)}]{Natoli:2009wk}
\bibinfo{author}{\bibfnamefont{P.}~\bibnamefont{Natoli}} \bibnamefont{et~al.}
  (\bibinfo{year}{2009}), \eprint{0905.4301}.

\bibitem[{\citenamefont{Smidt et~al.}(2009)\citenamefont{Smidt, Amblard, Serra,
  and Cooray}}]{Smidt:2009ir}
\bibinfo{author}{\bibfnamefont{J.}~\bibnamefont{Smidt}},
  \bibinfo{author}{\bibfnamefont{A.}~\bibnamefont{Amblard}},
  \bibinfo{author}{\bibfnamefont{P.}~\bibnamefont{Serra}}, \bibnamefont{and}
  \bibinfo{author}{\bibfnamefont{A.}~\bibnamefont{Cooray}},
  \bibinfo{journal}{Phys. Rev.} \textbf{\bibinfo{volume}{D80}},
  \bibinfo{pages}{123005} (\bibinfo{year}{2009}), \eprint{0907.4051}.

\end{thebibliography}

%

\end{document}